\documentclass[cernpreprint,atlasdraft=true,texlive=2019,UKenglish,texmf]{atlasdoc}
 
\usepackage[default, subfig]{atlaspackage}

\usepackage[acronym,shortcuts]{glossaries}
\glsdisablehyper
 
\usepackage{todonotes}
\usepackage{ulem}
\usepackage{atlasbiblatex}
 
\usepackage{atlasphysics}
 
\graphicspath{{logos/}{figures/}}
 
\usepackage{ANA-TDAQ-2019-02-PAPER-defs}

\AtlasTitle{The ATLAS Fast TracKer system}
 
\AtlasJournalRef{JINST 16 (2021) P07006}
\AtlasDOI{10.1088/1748-0221/16/07/P07006}
\AtlasAbstract{
The ATLAS Fast TracKer (FTK) was designed to provide full tracking for the ATLAS  high-level trigger by using
pattern recognition based on Associative Memory (AM) chips and fitting in high-speed field programmable gate arrays.
The tracks found by the FTK are based on inputs from all modules of the pixel and silicon microstrip trackers.
The as-built FTK system and components are described, as is the online software used to control them while running in the ATLAS data acquisition system.
Also described is the simulation of the FTK hardware and the optimization of the AM pattern banks.
An optimization for long-lived particles with large impact parameter values is included.
A test of the FTK system with the data playback facility that allowed the FTK to be commissioned during the shutdown between Run~2 and Run~3 of the LHC is reported.
The resulting tracks from part of the FTK system covering a limited $\eta$--$\phi$ region of the detector
are compared with the output from the FTK simulation.
It is shown that FTK performance is in good agreement with the simulation.
 
}
 
\author{The ATLAS Collaboration}
 
\AtlasRefCode{TDAQ-2019-02}
 
\PreprintIdNumber{CERN-EP-2020-232}

\hypersetup{pdftitle={ATLAS document},pdfauthor={The ATLAS Collaboration}}
 
\begin{document}
 
\maketitle
 
\tableofcontents
 
\clearpage
 
\section{Introduction}
\label{sec:intro}
\newcommand{\AtlasCoordFootnote}{
ATLAS uses a right-handed coordinate system with its origin at the nominal interaction point (IP)
in the centre of the detector and the \(z\)-axis along the beam pipe.
The \(x\)-axis points from the IP to the centre of the LHC ring,
and the \(y\)-axis points upwards.
Cylindrical coordinates \((r,\phi)\) are used in the transverse plane,
\(\phi\) being the azimuthal angle around the \(z\)-axis.
The pseudorapidity is defined in terms of the polar angle \(\theta\) as \(\eta = -\ln \tan(\theta/2)\).
Angular distance is measured in units of \(\Delta R \equiv \sqrt{(\Delta\eta)^{2} + (\Delta\phi)^{2}}\).
}
 
The \ac{FTK} system~\cite{FTKTDR} was foreseen as an evolution of the ATLAS experiment's~\cite{PERF-2007-01} trigger strategy, designed to bring the tracking of charged particles into trigger decisions at the earliest possible stage in the trigger algorithms.
The ATLAS trigger~\cite{PERF-2011-02,TRIG-2016-01} was originally designed around initial signatures based on energy deposits in the ATLAS calorimeters and the tracking of muons in the \ac{MS}, which surrounds the calorimeters.
In Runs\,1--3 of the \ac{LHC} the ATLAS first-level (L1) trigger system uses hardware-based algorithms to reduce the 40\,MHz rate from the beam-crossings to approximately 100\,kHz, which can then be processed in the \ac{HLT} to reduce the rate by another factor of 50 or more before recording events.
In 2018, for example, the average event-recording rate was 1200~Hz~\cite{ATL-DAQ-PUB-2019-001}.
Initially, in Run 1, the commodity-computing-based HLT was designed with a first step to process small areas around each of the L1 trigger objects.
Within these \ac{RoIs}, software-based tracking occurred as late as possible in the trigger algorithms to minimize the use of HLT resources.
In many cases, the RoI approach led to significant inefficiencies, for example in jet finding, because of differences between the online and offline strategies~\cite{TAMSETT2013253,ATLAS:2016qun}.
In addition, the instantaneous luminosity of the \ac{LHC} increased to $2 \times 10^{34}\,\text{cm}^{-2} \text{s}^{-1}$  and
the number of overlapping collisions, pile-up, frequently exceeded 50, leading to large inefficiencies.
The FTK was designed to mitigate this by using custom hardware to provide tracks to the \ac{HLT} at the full L1 trigger rate for events with up to 80 pile-up collisions.

By reconstructing all charged particles in an event with transverse momentum (\pt) down to 1~\GeV\ with good efficiency, it is possible to use algorithms such as particle flow~\cite{PERF-2015-09} to accurately separate the energy deposited in the target event from that of the pile-up interactions.
Such techniques for pile-up mitigation  are especially important for final states that require jets to be reconstructed and in events with signatures involving \ac{MET}~\cite{Aad:2020les}.
Examples of areas where ATLAS triggers would benefit from full tracking down to low \pt\ include di-Higgs production with decays to the four-$b$-quark final state~\cite{EXOT-2016-31} (tracks are also used to tag jets containing $b$-hadrons) and exotic~\cite{Aaboud:2019trc} and/or supersymmetry~\cite{compressed-susy} signatures involving \met.
In addition, full reconstruction of charged tracks in an event, such as the \ac{FTK} provides, allows for opportunistic use of these tracks in other signatures, e.g.\ $B$-physics, outside of \ac{RoIs} or in other future scenarios.

The following sections present the as-built FTK system, along with the simulation and control software that was used for system commissioning at the end of Run 2 data-taking and during \ac{LS2} before the project was stopped in 2019.
The \ac{FTK} system will not be used during Run 3.
Contributors to this decision were the lower than expected pile-up due to cryogenic limits of the LHC \cite{Petersen:2750301}, significant gains from optimization of the HLT software-tracking algorithm, and potential resource shortages.
Instead, it is envisioned to perform more tracking in the \ac{HLT} for signatures that benefit the most, e.g.\ jets and \met.
 
A comparison of the performance of the hardware with simulation is presented.
The paper is organized as follows:
an overview of the FTK system is described in Section~\ref{sec:ftk_system_overview};
the components of the FTK hardware, including those used for the comparison, are described in more detail in Section~\ref{sec:hardware};
the online software as well as FTK control and monitoring are described in Section~\ref{sec:online};
the FTK configuration, its optimization using simulation and its performance in a vertical slice, a minimal set of electronics boards to process tracks, are described in Section~\ref{sec:simulation}; and
Section~\ref{sec:conclusion} contains the conclusion.
 
\section{FTK system overview}
\label{sec:ftk_system_overview}
 
The \ac{FTK} system was developed to provide tracking of charged particles over the full acceptance of the ATLAS inner detector at the full L1 trigger rate of 100\,kHz.
It was designed to provide tracking information to the HLT, which operates asynchronously from the LHC clock. Tracking information from the FTK was provided to the HLT and the ATLAS data acquisition system \cite{TDAQ2016} in the same manner as is used for detector readout.
These systems do not require a fixed latency.
The FTK uses four layers of hits from the \ac{pixel} system (including the \ac{IBL}~\cite{ATLAS-TDR-19, PIX-2018-001} installed in 2015) and eight layers of hits from the four stereo pairs in the \ac{SCT} system for a total of twelve layers.
The inner detector is surrounded by a thin superconducting solenoid providing a \SI{2}{\tesla} axial magnetic field; it covers the pseudorapidity range \(|\eta| < 2.5\) and full range in azimuthal angle $\phi$.\footnote{\AtlasCoordFootnote}
Tracks in ATLAS are defined by five helix-parameters ($d_{0}$, $z_{0}$, $q/\pt$, $\eta$, and $\phi$), where $d_{0}$ and $z_{0}$ are the transverse and longitudinal impact parameters respectively.
Track quality can be assessed with a $\chi^2$ value from summing over the residuals of the hits and the track.
The \ac{FTK} system performs tracking in two steps.
First, initial track candidates are found by using pattern recognition with eight layers of hits.
Second, four additional layers of hits are added to the track and the track helix parameters are evaluated.
The \ac{FTK} system does not perform minimizations to fit tracks; instead, it evaluates a system of linear equations to perform a linearized track reconstruction.
 
Challenges faced by the \ac{FTK} system include handling, processing, buffering and synchronizing a high bandwidth of data on multiple parallel data paths (towers) in a large heterogeneous system with many interconnects; input hit organization across towers requiring duplication and sharing through multiple interconnects; creating, loading, and validating a large amount of configuration data for the pattern matching and track fitting; and managing and monitoring a large number of high-power-density electronics boards requiring active cooling.
 
\begin{figure}[htbp]
\centering
\includegraphics[width=0.7\textwidth]{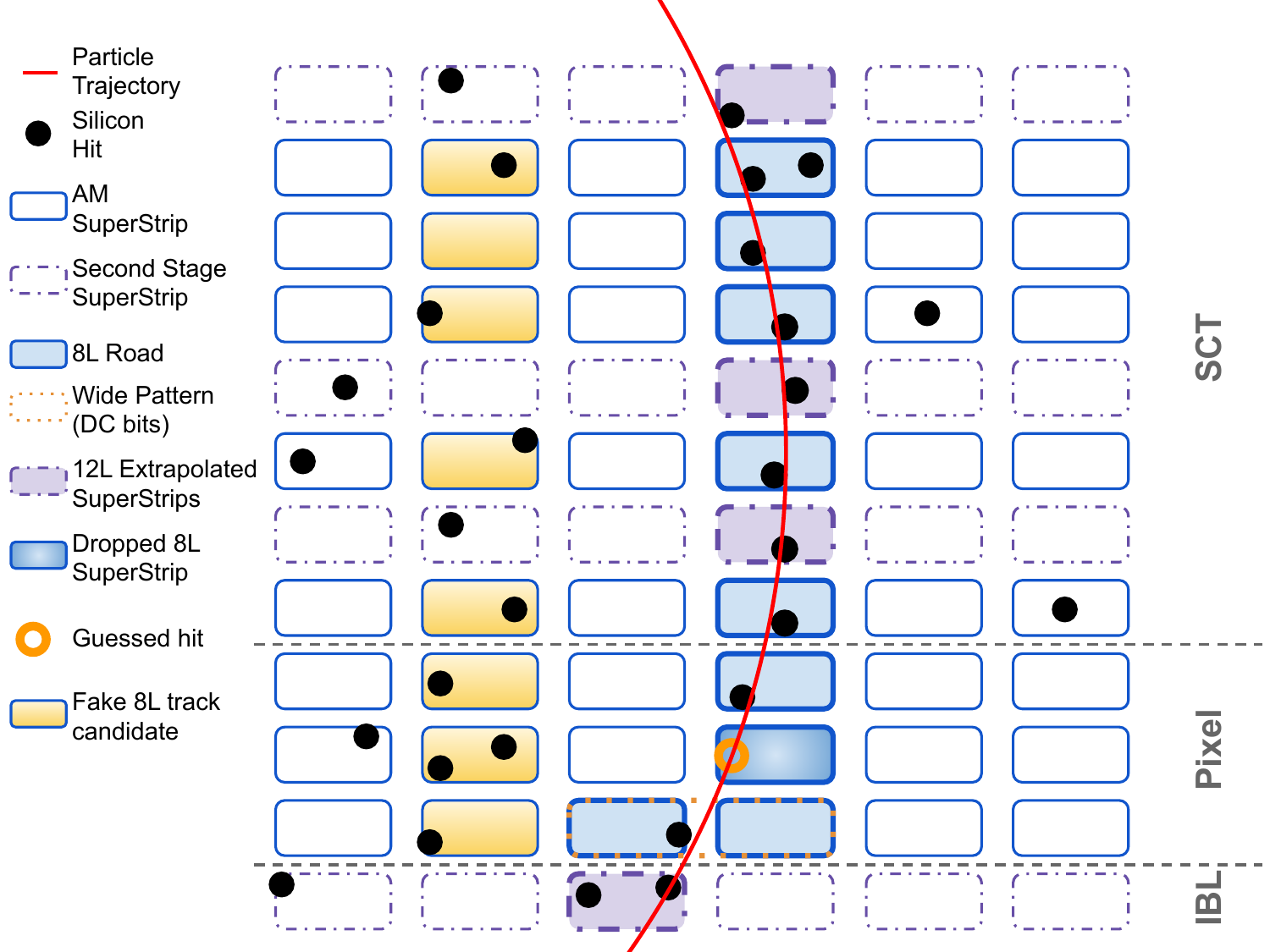}
\caption{Diagram of the track-finding method used in the FTK.
Rounded boxes represent the coarse resolution \aclp{SS} used to bin the hits from the silicon detectors.
Once eight layers (\acf{AM} SuperStrips) with hits match a stored pattern they become a road (solid filled boxes), which is then extrapolated into the remaining four layers (dash-dot filled boxes).
One layer may be dropped in the first eight layers used for pattern matching and another in the additional four layers.
Fake track candidates (e.g.\ the gradient-filled boxes to the left) can be rejected because they have a large $\chi^2$ value or fail extrapolation to the additional four layers.}
\label{fig:ftk_tracking}
\end{figure}

At the core of the \ac{FTK} is the content-addressable memory~\cite{DellOrso:1988gmu, Adelman:2007zz}, the \ac{AM}~\cite{Annovi2228284} that performs the initial pattern recognition and forms candidate tracks to be processed further.
The \ac{FTK} \ac{AM} is an ASIC specifically designed for the \ac{FTK} system using a 65~nm CMOS technology.
It is used to associate the data hits from the tracking detectors with stored patterns.
The \ac{AM} patterns use up to eight layers of hits (three \ac{pixel} and five \ac{SCT}), which are binned with a variable granularity in the azimuthal angle and pseudorapidity.
The AM layers were chosen to provide candidate tracks with a sufficient purity to allow downstream processing.
The candidate tracks (roads) found by the AM pass a $\chi^2$ selection before proceeding to second-stage processing that extrapolates them to the innermost Pixel layer (the \ac{IBL}) and to the three additional \ac{SCT} layers.
Figure~\ref{fig:ftk_tracking} illustrates the track-finding algorithm used in the \ac{FTK}, with more details described at the end of this section, after an overview of the \ac{FTK} system components
 
The stages of processing in the \ac{FTK}, clustering, data transport and sharing, pattern matching, and evaluations of track parameters in two stages,
are summarized in Table~\ref{tab:overview}.
These functions are spread across several electronics boards.
 
\begin{table}[htbp]
\centering
\caption{ Overview of the FTK system in the order in which the data are processed.   The AUX appears in the table twice because of its dual functions.
The table shows the system as envisioned for Run\,3 of the LHC where a direct connection of the SSB to the ATLAS \acf{ROS} was foreseen.
In this case, the \acf{FLIC} would not be used.
\label{tab:overview}}
\vspace{0.2cm}
\begin{tabular}{lcp{4.5cm}cc}
\hline
&\multicolumn{1}{c}{Module}   & \multicolumn{1}{c}{Function}  &  Type & Number  \\
\hline
IM         &  \acl{IM} &  Cluster \ac{pixel} and \ac{SCT} hits, format module data                                      & Mezzanine  & 128 \\
DF        &  \acl{DF} & Transport and duplicate module hit data to $\eta$--$\phi$ towers    &  ATCA & ~~32 \\
AUX     &  \acl{AUX} &  Transport coarse-resolution 8-layer hit data to AMB                            & VME & 128 \\
AMB     &  AM Board &   Transport  hit data to AM                                              & VME & 128 \\
AM       &  \acl{AM} & Match hits to patterns                                       &  ASIC & 8192~~ \\
AUX     &  \acl{AUX}  & Evaluate track candidates in matched patterns & VME & 128 \\
SSB     &  \acl{SSB}  &  Add remaining hits to 8-layer tracks, fit, remove overlaps &  VME & ~~32 \\
FLIC     & HLT Interface Board  & Interface to ATLAS readout                             & ATCA   & ~~~~2 \\
\hline
\end{tabular}
\end{table}
 
The \acp{IM}, 128 cards mounted on the \ac{DF} modules, have a total of 512 inputs.
The \ac{IM} cards cluster the hits from the \ac{pixel} and \ac{SCT} sensors. The \ac{DF} system, which consists of 32 ATCA~\cite{ATCA}
boards, routes the packets of hits corresponding to individual \ac{pixel} and \ac{SCT} modules to an array of 4 ($\eta$) $\times$ 16 ($\phi$) overlapping regions in the $\eta$--$\phi$ plane called towers.
The \ac{DF} boards duplicate the module packets as needed to allow sharing between overlapping towers.
They transmit the packets across the ATCA backplane and over fibres connecting the ATCA shelves to each other.
Towers overlap in order to keep the track-finding efficiency high near the boundaries.
 
\begin{figure}[htbp]
\begin{center}
\subfloat[]{
\includegraphics[width=0.90\textwidth]{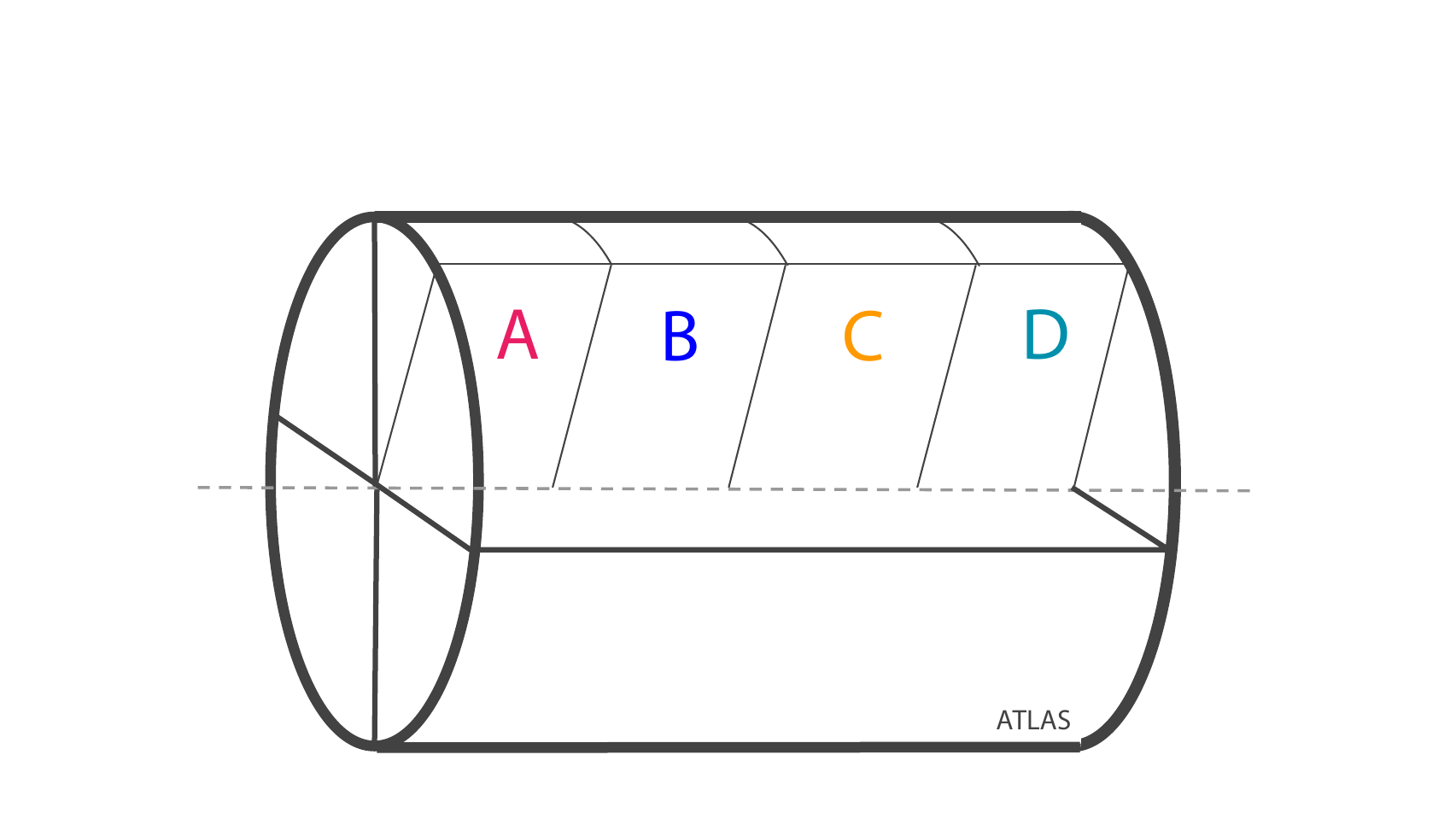}\label{fig:overview:segmantation}
}\\
\subfloat[]{
\includegraphics[width=0.90\textwidth]{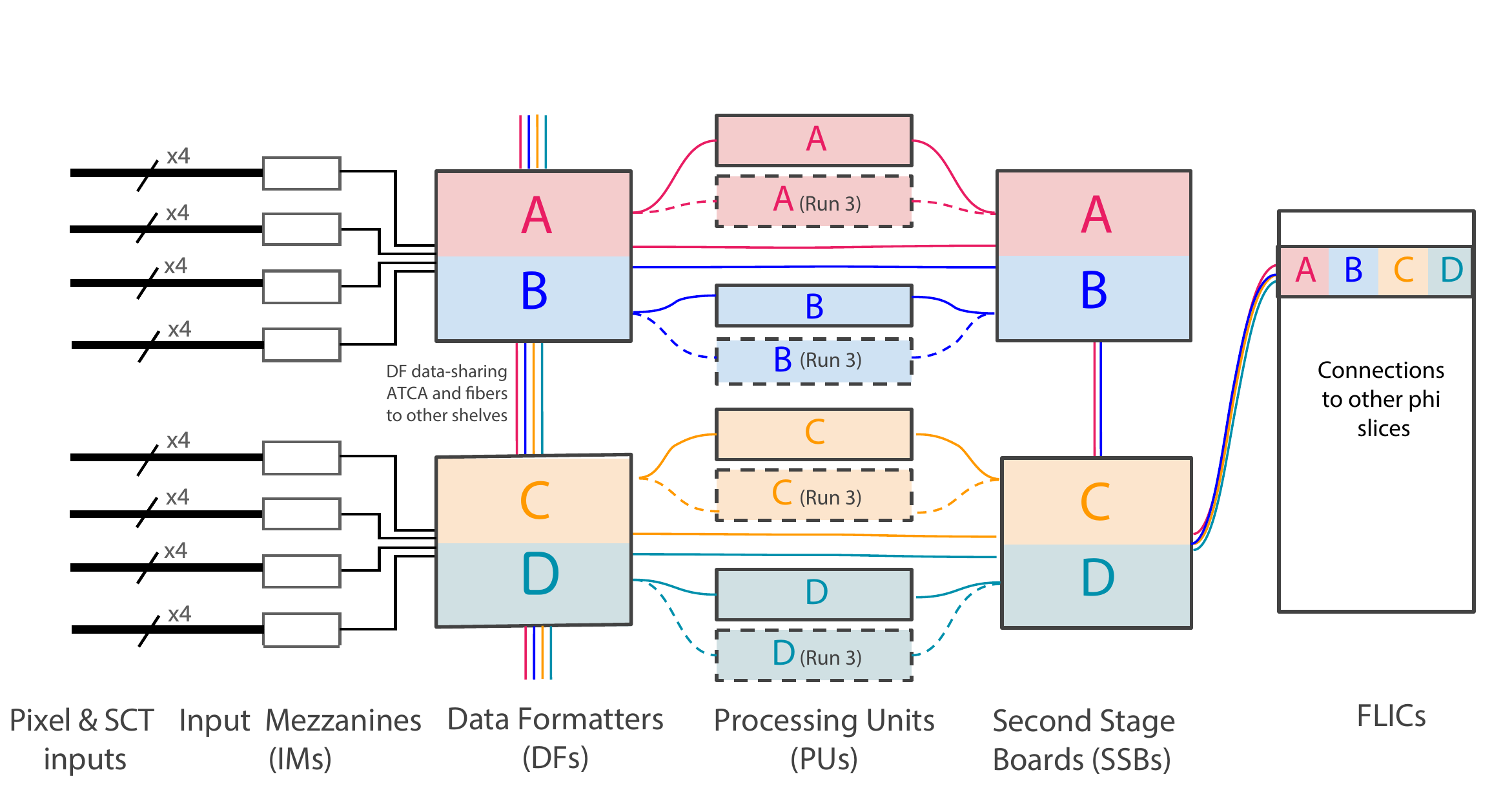}\label{fig:overview:four-tower-system}
}
\end{center}
\caption{
Figure \ref{fig:overview:segmantation} illustrates the tower geometry of the FTK.   Two adjacent towers, adjacent in $\eta$, are serviced by a single DF board, e.g.\ A and B.   An ATCA shelf containing eight DF boards serves a quadrant of the FTK.
Figure \ref{fig:overview:four-tower-system} shows the board layout and interconnections covering one slice in $\phi$ of the ATLAS detector.  Solid lines indicate connections used in the 64-PU configuration, and dashed lines indicate additional connections and boards used in the 128-PU configuration intended for use during Run\,3.  For Run\,3, the SSB could optionally (not shown) connect directly to the ATLAS  \acf{ROS} and have the connection between \acp{SSB} omitted.
\label{fig:overview}}
\end{figure}

Each of the 32 \ac{DF} boards serves two of the 64 towers and sorts the module hit data by layer
before transmitting them to four of the 128 \acp{PU}
and one of the 32 \acp{SSB}. The system can optionally be configured with two \acp{PU} per \ac{DF}.
The segmentation of connectivity of these towers is illustrated in Figure~\ref{fig:overview}.
 
Each \ac{PU} is composed of an \ac{AUX} and an \ac{AMB}.
The \ac{AUX} receives hits from eight layers in a given tower and converts them to coarse resolution hits, or \ac{SS}, which are sent to the \ac{AMB} to be matched against the stored patterns of \acp{SS} in the \ac{AM}.
The eight layers used in pattern matching are drawn in blue in Figure~\ref{fig:ftk_tracking}.
An \ac{SS} is a set of silicon strips or pixels, typically 128 strips or $30\times76$ pixels in size.
The strips have a typical pitch of 80~$\mu$m and the pixels are $400\times50~\mu$m in the outer layers, and $250\times50~\mu$m in the innermost layer.
The matching patterns from the \ac{AMB} correspond to 8-layer tracks.
These 8-layer tracks are processed using linearized fits with precomputed constants and filtered by the \ac{AUX} and then sent to the \ac{SSB}.

In the \ac{SSB}, the hits from the remaining four layers are matched to the 8-layer tracks, and tracks with up to 12 layers are formed and  processed using linearized fits.
Each \ac{SSB} receives tracks from two adjacent towers.
In addition to forming 12-layer tracks, the \acp{SSB} provide duplicate-track removal when tracks share more than an allowed number of hits.
 
The final tracks from the \ac{SSB} are forwarded to the \ac{FLIC}, which is in turn interfaced to the ATLAS \ac{ROS} \cite{Borga:1710776}.
Both the \ac{AUX} and  \ac{SSB} are capable of connecting directly to the ROS system.
In the case of the \ac{AUX}, this feature was used for testing;
in the case of the \ac{SSB}, it would have allowed inter-SSB track-overlap removal to be offloaded to the \ac{HLT}.
 
Using a linear system of equations in the hit coordinates allows fast computation of the track helix parameters and $\chi^2$ values on \acp{FPGA}.
The same underlying principle is used for both the 8-layer fit performed by the \ac{AUX} and the 12-layer fit performed by the \ac{SSB}.
The difference is that the \ac{AUX} computes the $\chi^2$ for the 8-layer track candidate, whereas the \ac{SSB} computes the same plus the helix parameters for 12-layer tracks.
In both cases, one hit may be dropped, or is allowed to be missing, from the fit in order to improve efficiency, i.e.\ the AUX may drop a hit from its eight considered layers and the SSB may drop a hit in the additional four layers.
If hits are found on every layer, the AUX and SSB each perform a nominal fit with hits on each layer and a recovery fit for each permutation of one dropped hit on their respective layers.
If the nominal fit does not pass an initial $\chi^2$ requirement, the recovery fit with the smallest $\chi^2$ is used instead.
In this case, the best expected hit position on the dropped layer is used in order to compute track parameters, represented by an orange circle as the `guessed hit' in Figure~\ref{fig:ftk_tracking}.
 
The \ac{AUX} uses 8 silicon hits (5 \ac{SCT} layers and 3 \ac{pixel} layers), with a total of 11 coordinates, and 5 helix parameters.
Thus, there are 6 degrees of freedom, each of which gives a function of the coordinates that should equal zero.
The $\chi^2$ is the sum of the squares of those functions:
\begin{equation}
\label{8-layer-chi}
\chi^2 = \sum_{i=1}^6 \left( \sum_{j=1}^{11} S_{ij}x_j+h_i\right)^2,
\end{equation}
where $x_j$ is a hit coordinate and $S_{ij}$ and $h_i$ are precalculated constants for each sector.
The constants are computed per sector, a small region of the detector containing one silicon module in each layer, typically a few centimeters in size.
There are on average 13\,000 8-layer and 28\,000 12-layer sectors per tower, which may share modules.
 
The \ac{SSB} extrapolates the 8-layer candidate track into the additional four layers and evaluates the track with all 12 layers, giving 16 coordinates, using a method similar to that in Eq.~(\ref{8-layer-chi}).
For the purpose of track fitting, a system of equations is built:
\begin{equation}
\label{lin-track-fit}
y_i = \sum_{j=1}^{16} C_{ij}x_j+q_i,
\end{equation}
where $C$ and $q$ are a $16\times16$ matrix and $16\times1$ array of precalculated constants respectively, $x_j$ is an array of the 16 hit coordinates, and $y_i=(d_{0},\ldots,\phi,\chi_1,\ldots,\chi_{11})^\text{T}$ is an array of the 5 helix parameters and 11 partial $\chi$-values to be summed as in the previous equation.
The \ac{SSB} extrapolates the 8-layer tracks to the expected hit positions in the additional layers by taking Eq.~(\ref{lin-track-fit}), dropping the helix parameter indices, assuming the $\chi$-values to be zero, and inverting the equation to solve for the hit positions.
Instead of performing all of these computations on the \ac{FPGA}, separate constants are produced for extrapolation, so that this too is a linear equation, which can be evaluated quickly.
Details of these procedures are described in Ref.~\cite{FTKTDR}.

\FloatBarrier

\section{Hardware and firmware}
\label{sec:hardware}
The \ac{FTK} infrastructure in the ATLAS underground counting room (USA15) is described in Section\,\ref{sec:input}.
This is followed by a description of the data playback system, QuestNP, in Section\,\ref{sec:questnp}.
The remaining sections, Sections~\ref{sec:im} to \ref{sec:flic}, describe the individual FTK components (previously summarized in Table\,\ref{tab:overview}): \ac{IM}, \ac{DF}, \ac{AUX}, \ac{AMB}, \ac{AM}, \ac{SSB}, and  \ac{FLIC}.
 
A common feature of the firmware for all components in the system is that it is data driven.
The system was designed so that the main bandwidth limitation occurs on the input links from the Pixel and \ac{SCT} systems.
Each stage of the processing has both  input and output buffers.
In case the input buffers are in danger of overflowing, it is possible for each component in the system to ask the previous component to pause data transmission by asserting back pressure.
Back pressure is asserted using the XOFF feature of the S-LINK \cite{slink,VanderBij:405075}, a signal sent to the upstream transceiver, or dedicated signal lines.
 
In much of the system,  data can be processed without regard to event synchronization.
There are various places where data are buffered until the relevant information for a given event is  available:
at the input of the \ac{DF} where data are received from the on-board \acp{IM};
at the output of the \ac{DF} before complete data from individual layers are sent to the \ac{AUX} and \ac{SSB};
at the output of the \ac{AUX} where data are merged from multiple FPGAs;
at the input of the \ac{SSB} where data are received from the \ac{AUX} and \ac{DF};
and at the output of the \ac{SSB} where data from two adjacent towers are combined before being sent to the \ac{FLIC} or \ac{ROS}.
These synchronizations use the packet structure of the data, specifically an event identifier in the header (packet metadata before the payload).
 
The firmware designs of the FPGAs used in the system have a shared concept of monitoring.
It is based on circular buffers (spybuffers) to sample data as they flow through the system, readout of monitoring registers, and the ability to freeze these monitoring quantities in case of errors for debugging.
Additionally, buffers are also used to store synchronized monitoring information and histograms of monitoring values.
These blocks use a common firmware design, with some exceptions due to design requirements and \ac{FPGA} chip architecture.
The design of the synchronization blocks in the previous paragraph is also shared between the FTK components.
 
The testing and commissioning of the FTK system involved a wide array of set-ups from individual components to the full data-processing chain.
Individual component testing, normally involving an entire electronics board, made use of the ability either to self-drive the firmware with test data or to directly drive the board's input connections in a stand-alone configuration.
This allowed independent commissioning without waiting for the debugging of upstream data-processing.
Other tests of the complete data-processing chain and scaling up of the system were made, some of which are described later.
Examples of these tests range from pairs of electronics boards to the set of electronics boards in Figure~\ref{fig:overview:four-tower-system}, to full shelves of \acp{DF} or full crates of \acp{PU} and \acp{SSB}.
Near the end of Run 2, a vertical slice of the FTK including each board needed for data-processing (a \ac{DF} with four \acp{IM}, \ac{PU}, \ac{SSB}, and \ac{FLIC}) was regularly included in ATLAS data-taking.
This set of boards was used in the collection of the data analysed in Section~\ref{sec:result}.
A second slice with a \ac{PU} writing directly to the readout system was also used for testing the upstream boards.
\subsection{Infrastructure and interface to the Pixel and SCT detectors}
\label{sec:input}
The final layout of the \ac{FTK} infrastructures is composed of seven racks.
Four racks are  dedicated to the custom VME~\cite{VME} core crates hosting the pattern matching and initial fitting, i.e.\ the \acp{PU}.
These VME crates also host the boards for the final fitting (\acp{SSB}) and their \acp{RTM}.
The remaining three racks are used by the five ATCA shelves which host \ac{DF} boards for data routing and sharing and the \acp{FLIC} for the interface to the \ac{HLT}.
All racks are located in USA15, in two consecutive rows of three racks each and a third row containing the remaining rack.
 
The electronics in these seven racks are air cooled with circulation provided by turbines on top of the racks~\cite{Thomas:38031} and by additional custom fan trays which are used around the VME crates.
As the air passes through and out of the racks, it is cooled by multiple heat-exchangers circulating chilled water.
 
The layout of the VME racks~\cite{Sotiropoulou:2295706} was optimized to keep the PUs' operating temperature below $80^{\circ}\mathrm{C}$, a temperature dictated by the expected lifetime of the \ac{AM} ASIC.
The custom fan trays were designed to minimize the temperature in the VME crates, which each consume about 6\,kW.
The fan tray design specifications required the usage of high-performance Sanyo-Denki fans (San Ace 9HV1248P1G001~\cite{Sanyo}) supplied by 48\,V~DC.
This in turn required a modification of the service voltage of the CAEN Power Supply A3488 \cite{CAEN}.
The speed of each fan row is automatically controlled by the \ac{DCS} via temperature sensors on the \acp{PU}.
In this configuration, the maximum temperature of the \acp{AMB} in a fully equipped crate was measured to remain within an acceptable range~\cite{Sottocornola:2741522}.
 
The two racks hosting \ac{DF} boards are equipped with two ATCA shelves each (ASIS Maxum 500 14-Slot~\cite{ASIS})
supplied by one AC--DC converter (UNIPOWER Guardian 3U \cite{UNIPOWER}) able to deliver up to 6\,kW.
Data are delivered to boards in the racks from the \ac{pixel} and \ac{SCT} \acp{ROD}  \cite{Balbi_2016,Abdesselam:2008zza}
by means of S-LINK over fibres.
 
The third ATCA rack hosts the \acp{FLIC} in an ASIS 6-slot horizontal 6U shelf (ATCA6u-010-3308) powered by a 1.5\,kW AC--DC supply.
Data are delivered out of the rack to the \acp{ROS} using S-LINK over fibres.

\subsection{QuestNP}
\label{sec:questnp}
The \ac{QNP} card is an S-LINK data source card based on the \ac{CRORC}~\cite{Borga_2015} hardware.
It can be used to replay data from the Pixel and SCT detectors.
The ability to replay ATLAS data to the FTK system, as in a data-taking run, is crucial in order to allow full commissioning during a shutdown and complements data-playback features available in individual FTK boards.
 
\begin{figure}[htbp]
\begin{center}
\includegraphics[width=0.75\textwidth]{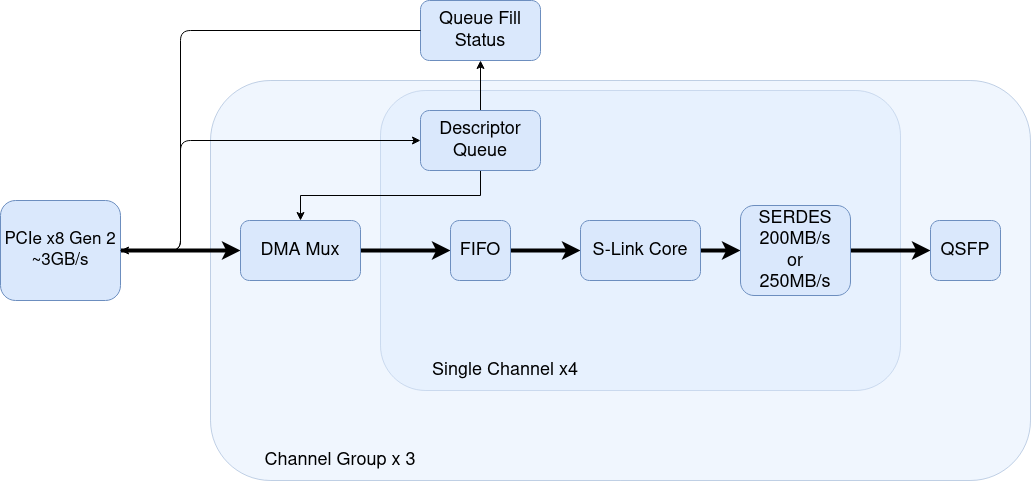}
\end{center}
\caption{\label{fig:FTH_Hardware:QNP:DMA} Diagram of the QuestNP firmware design used to input test data into the FTK via the QSFP connection.  The SERDES block is a \acl{SerDes}.  }
\end{figure}

The \ac{QNP} card is implemented by using a dedicated firmware design reusing elements from the firmware of the RobinNP~\cite{Borga:1710776} \ac{ROS}.
The \ac{CRORC} is a PCI express card using a Xilinx Virtex-6 \ac{FPGA} \cite{Xilinx:6series} as the main component handling the communication through a PCIe Gen 2 ($8 \times 5.0$\,Gb/s) interface and implementing the interfaces to twelve optical links via three \ac{QSFP} transceivers.
The \ac{QNP} card makes it possible for user-defined data to be transferred to remote S-LINK data receivers in parallel on each of the 12 S-LINKs.
The \ac{QNP} card has been successfully used to reliably transmit data to the FTK system at an event rate of 1~kHz and has been used for testing up to 19~kHz.
A version of the firmware including a \ac{DMA} structure for each of the channels in the card is shown in Figure~\ref{fig:FTH_Hardware:QNP:DMA}.
It has reached a fragment rate  corresponding to the 100~kHz required by the FTK system.
 
In a possible final implementation of the firmware, the \ac{QNP} card would be able to synchronize the event header between its 12 internal channels.
A daisy chain of the cards using \ac{LVDS} pairs routed on the eight-position, eight-contact socket (RJ45 connector following the IEC 60603-7 standard) and a dedicated switch would also allow the synchronization of the output of multiple \ac{QNP} cards.
During Run~2, an event arrival-time difference between data from the Pixel and SCT detectors was observed, and the aforementioned synchronization mechanism would have been used to emulate this skew.
 
Up to three \ac{QNP} cards can be installed in a 3U rack-mounted server.
Five dedicated servers were used to host the 13 \ac{QNP} cards used to feed data to half of the \ac{FTK} system.
A state-aware sender application, compatible with the ATLAS Run Control~\cite{TDAQ2016}, is used to configure and control the cards.
Monitoring of the application, of the cards' parameters and of the sent data is implemented through the \ac{IS}~\cite{Kolos_2012} and the information is displayed using Grafana \cite{Grafana}.


\subsection{\acl{IM}}
\label{sec:im}
The \ac{IM} module receives and processes data from the inner detector.
Each input channel, fed by an inner detector readout link, is processed independently of the others and propagated to the \ac{DF}.
The \ac{IM} forms inner detector hits into contiguous clusters and computes their centroid positions.
\ac{SCT} data split at the ABCD~\cite{842514} front-end readout chip boundary are merged into a single cluster.
 
The key performance requirement for the IM is maintaining \ac{pixel} clustering throughput at the full rate of incoming data.
In order to achieve this goal, the firmware is built with parallel clustering cores fed with a load balancing mechanism.
Pixel clustering reduces the data size by a factor of 2--2.5.
Simulation with Pixel data has shown that four cores are needed to achieve the required processing rate.
This performance was demonstrated using simulated data with a pile-up of 80 before the boards went into production, as reported in Ref.~\cite{TNS-Clustering}.
The implemented firmware instantiates eight cores, allowing for an additional safety margin.
Bit-level simulations and debugging were performed by comparing bit-by-bit the Pixel clusters from the IM with those produced within the FTK simulation in order to reach per-bit agreement.
 
\begin{figure}[htbp]
\begin{center}
\includegraphics[width=\textwidth]{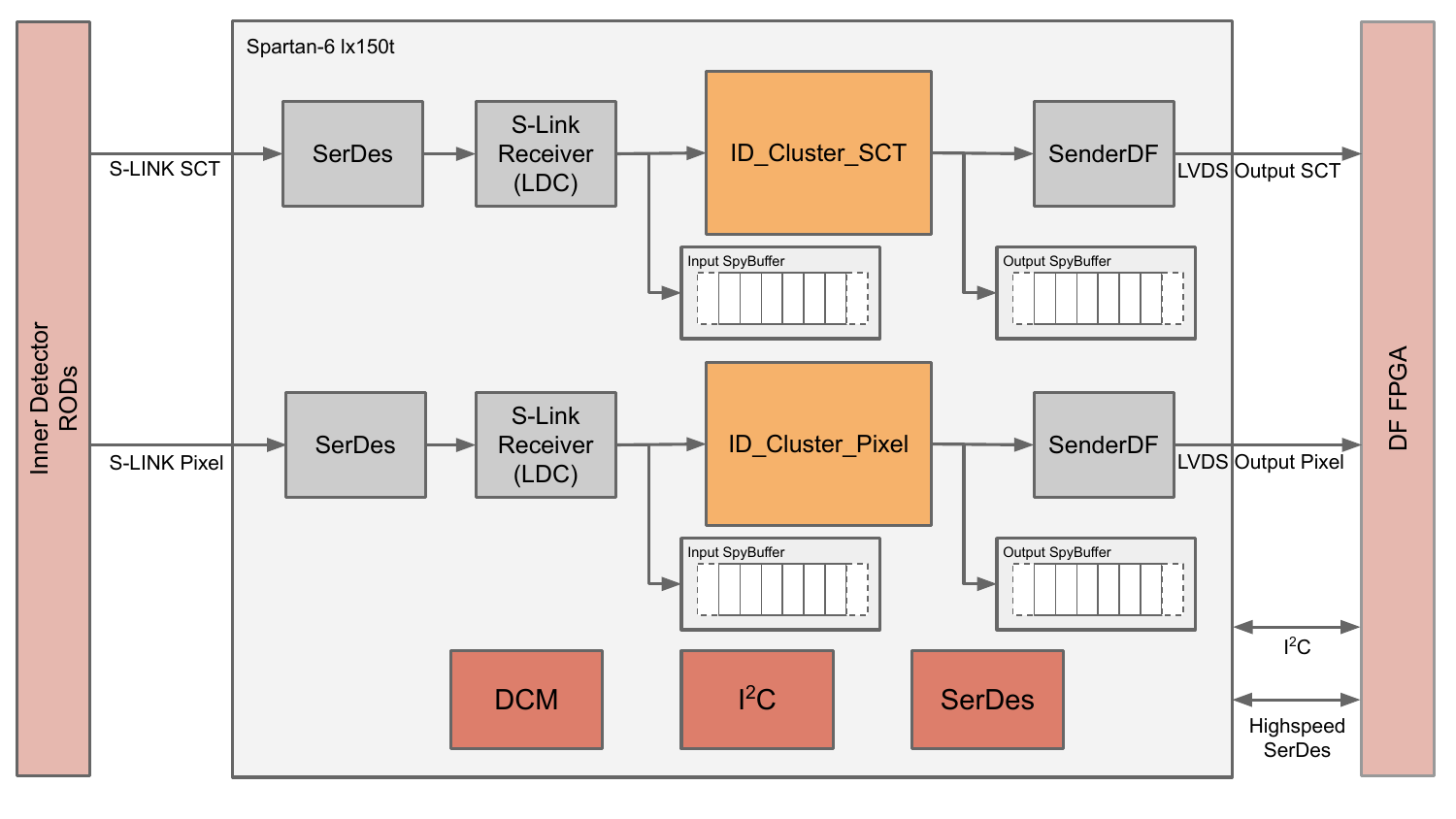}\\
\end{center}
\caption{Top-level diagram for each of the two FPGAs on the \acl{IM} card. The ID\_Cluster\_SCT and ID\_Cluster\_Pixel are the two main blocks performing SCT and Pixel clustering respectively. Data flows from the inner detector \acp{ROD} to the \acl{DF} FPGA. \label{fig:im-toplevel}}
\end{figure}

Each IM receives data from four readout fibres using S-LINK.
The fibres are distributed to two FPGAs, two links per FPGA, with the output sent to the \ac{DF} over a parallel DDR \ac{LVDS} bus, shown in Figure~\ref{fig:im-toplevel}.
The ID\_Cluster\_SCT and ID\_Cluster\_Pixel are the two main blocks performing SCT and Pixel clustering respectively.
The \ac{SerDes} plus S-LINK blocks receive data from the inner detector \acp{ROD}.
The SenderDF converts data to LVDS~DDR output for the \ac{DF}.
The $\mathrm{I^2C}$ protocol is used for configuration.
An additional \ac{SerDes} per FPGA is connected to the DF~FPGA as a spare 2~Gb/s line.
Each FPGA is also connected to an external RAM that can be used to store test data for IM and system testing purposes.
The firmware is modular, which allows different versions of the firmware to be compiled, allocating one FPGA for SCT--SCT, or SCT--Pixel, or Pixel--Pixel processing.
 
\begin{figure}[htbp]
\begin{center}
\includegraphics[width=0.85\textwidth]{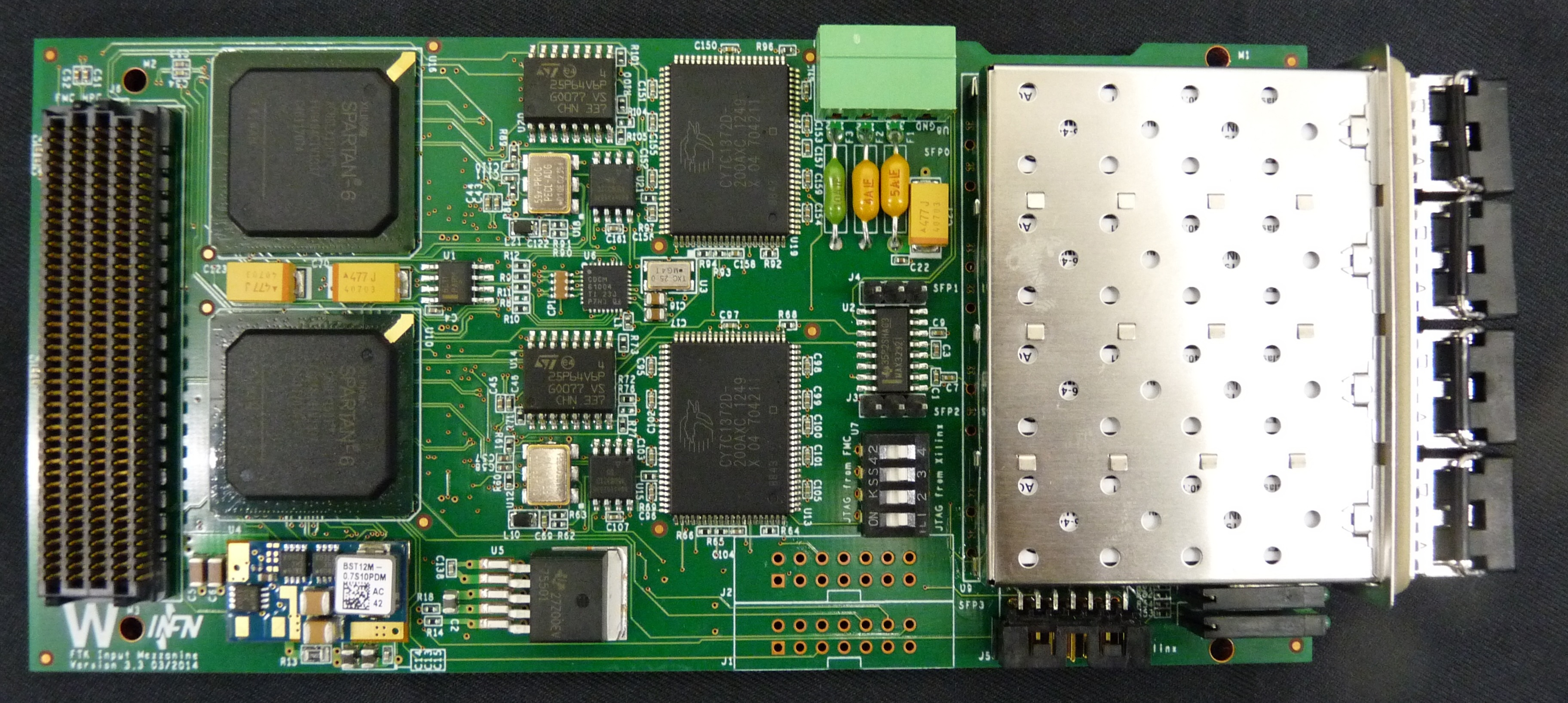}\\
\end{center}
\caption{Photo of the Spartan-6 \acl{IM} card, component side. \label{fig:im-top-photo}}
\end{figure}

Two versions of the IM cards were produced.
The first version is based on the Spartan-6 150T FPGA~\cite{Xilinx:Spartan6} and is shown in Figure~\ref{fig:im-top-photo}.
It was tuned for Pixel processing during the initial design.
The addition of the \ac{IBL} required a more powerful FPGA, because the effective input hit rate increased by a factor of 2--2.5.
The second half of the production used an updated design with Artix-7 200T FPGAs \cite{Xilinx:7series} to provide additional processing power.
 
The full FTK system requires 128 \acp{IM} installed on 32 \acp{DF}.
This allows for up to 256 Pixel+IBL input links and up to 256 SCT input links.
The 256 inputs for Pixel+IBL exceed the original design requirements in Ref.~\cite{FTKTDR} and were sufficient to cover increases in the number of Pixel links during Run~2, which were used to achieve higher Pixel output bandwidth.
A small additional production of \acp{IM} was planned for the \ac{LHC} \ac{LS2} to provide the required connectivity in Run~3.
 
\begin{figure}[htbp]
\begin{center}
\includegraphics[width=0.85\textwidth]{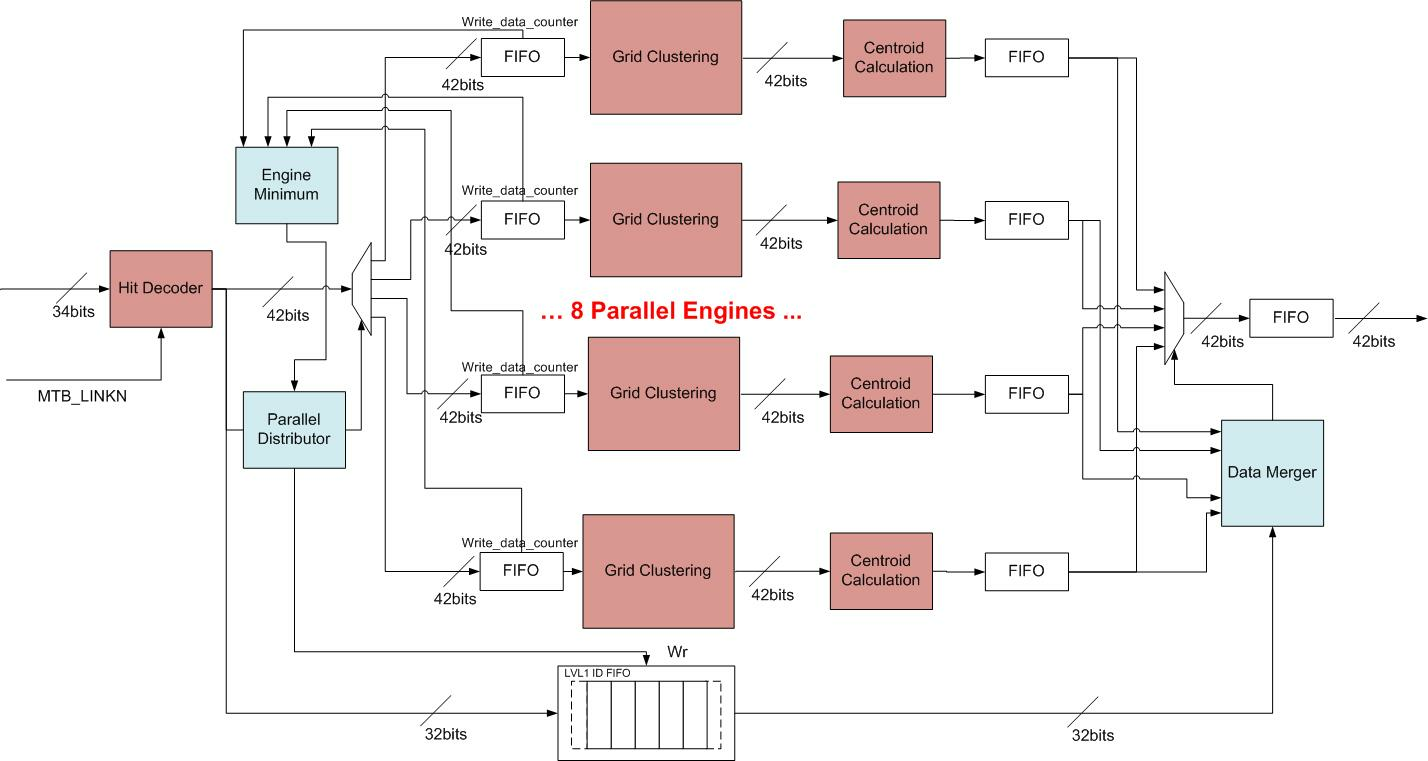}\\
\end{center}
\caption{Diagram showing the organization into parallel engines for the Pixel clustering logic.
This logic is organized into three main blocks: Hit Decoder, Grid Clustering, and Centroid Calculation.
\label{fig:im-parallel8enginges}}
\end{figure}

The Pixel clustering logic is organized into three main blocks, as shown in Figure~\ref{fig:im-parallel8enginges}: Hit Decoder, Grid Clustering and Centroid Calculation.
The Hit Decoder block performs the decoding of incoming data.
As a part of decoding, it duplicates hits from the ganged region (merged pixels near the edges) of the Pixel module to avoid an efficiency loss in that area.
It also rearranges data that arrive from the 16 front-end~ASICs of a single Pixel module so that the data are approximately sorted in increasing column order for the grid of pixels in the module.
The Pixel module is read out in pairs of columns.
For the eight front-end ASICs on one side of the Pixel module, the columns are read out in increasing order, while for the other side the readout is by decreasing order.
The described firmware block reorders the column pairs to have the same order.
As described in Ref.~\cite{TNS-Clustering}, this is required by the next block, Grid Clustering, described below.
Finally, the Centroid Calculation block receives Pixel hits grouped into clusters, calculates the cluster centroid and outputs a single word with the centroid coordinates local to the module and the cluster size projected in two dimensions.
This word is the cluster position that is used by the rest of the FTK system for processing.
 
The \ac{IM} input logic also allows selective processing of events, by stripping out cluster words, based on the L1 trigger that fired the event and a versatile prescale function.
These features are useful for controlling which trigger streams are processed and to control the processing load in the FTK system.
For example, it could be used to restrict FTK processing to events where full tracking is required, e.g.\ those with jets and \ac{MET}.
The prescale has the functionality to send every $n^{\text{th}}$ event full or empty, which is useful for debugging problems that only occur in successive events.

The Grid Clustering is the most resource-intensive block of the firmware.
It processes incoming data to find hits that are contiguous along horizontal, vertical or diagonal lines.
This operation is performed using a grid that represents an $8\times21$ portion of the Pixel detector.
The grid is aligned to the first hit received, called the seed, and loaded with all other hits present.
This grid, replicated in the FPGA, uses local logic to select hit pixels that are contiguous to the seed, or contiguous to selected pixels.
In this way, all pixels of the cluster containing the seed are selected.
The readout logic then extracts all selected pixels.
The output is the sequence of all hits of the clusters, with the last hit identified by an end-of-cluster flag.
Next, iterative processing is done to cluster the data present on the grid but not belonging to the seeded cluster.
 
The logic is organized into parallel cores as described in Figure~\ref{fig:im-parallel8enginges}.
The baseline Pixel clustering firmware instantiates eight parallel engines.
Data received from the Hit Decoder block is organized into a sequence of detector modules for each event.
For load balancing, the Parallel Distributor assigns data for an entire module to the engine with the smallest load as informed by the Engine Minimum block.
Each of the eight parallel engines has enough buffering to compensate for fluctuations in the processing time without causing a significant number of control signals to be sent upstream.
Each of these eight parallel engines works in a \ac{FIFO} mode, thus preserving the order in which data blocks flow.
A parallel \ac{FIFO} stores the event identifier which is sent for each data block from the Pixel and SCT detectors.
This event identifier is used by the Data Merger at the right side of the diagram to collect the correct data, for a given input S-LINK, for each event from the engines before sending the built event to the \ac{DF}.
The output data are organized into module packets with each packet corresponding to the data from a single \ac{pixel} or \ac{SCT} module.
For most of the tests reported in this paper, the output packet size was limited to 32 words including one header and one trailer word.
The words are 32 bits long.
In the body of the packet, $\eta$ and $\phi$ hit-coordinate pairs correspond to one 32-bit word.
The \ac{IM} works independently on data from each S-LINK.
Synchronization of data from different S-LINK inputs is performed in the \ac{DF}.

The firmware used for tests reported here in the Spartan-6 FPGA to handle SCT--Pixel processing with two parallel clustering cores utilized approximately 56\% of the logic and 71\% of the memory.
The equivalent firmware on the Artix-7 FPGA with eight parallel clustering cores utilizes approximately 71\% of the logic and 31\% of the memory.
 

\subsection{\acl{DF}}
\label{sec:df}
The \ac{DF} receives packets of hits from the \acp{IM} and other \acp{DF} and routes them to downstream boards, sorted by towers, or to other \acp{DF}.
Its main task is to assign all module data to the appropriate towers, duplicating data where necessary.
The \ac{DF} is based on the Pulsar~\cite{Olsen_2014,Okumura_2015} ATCA board with full mesh connectivity via the backplane to up to ten other  modules in the ATCA shelf over dual duplex multi-gigabit-transceiver connections.
In addition to the connections to the  ATCA backplane,  the \ac{DF} has connections to an \ac{RTM} that hosts ten \ac{QSFP} ports that provide up to 40 duplex connections.
The  \ac{DF} hosts four mezzanine cards which can transmit data to the DF via \ac{LVDS} lines.
The switching functionality in the DF is performed using a Xilinx Virtex-7 690T \ac{FPGA} \cite{Xilinx:7series}.
In the DF application, the Pulsar board uses an \ac{IPMC} for some module control functions.

\begin{figure}[htbp]
\begin{center}
\includegraphics[width=0.85\textwidth]{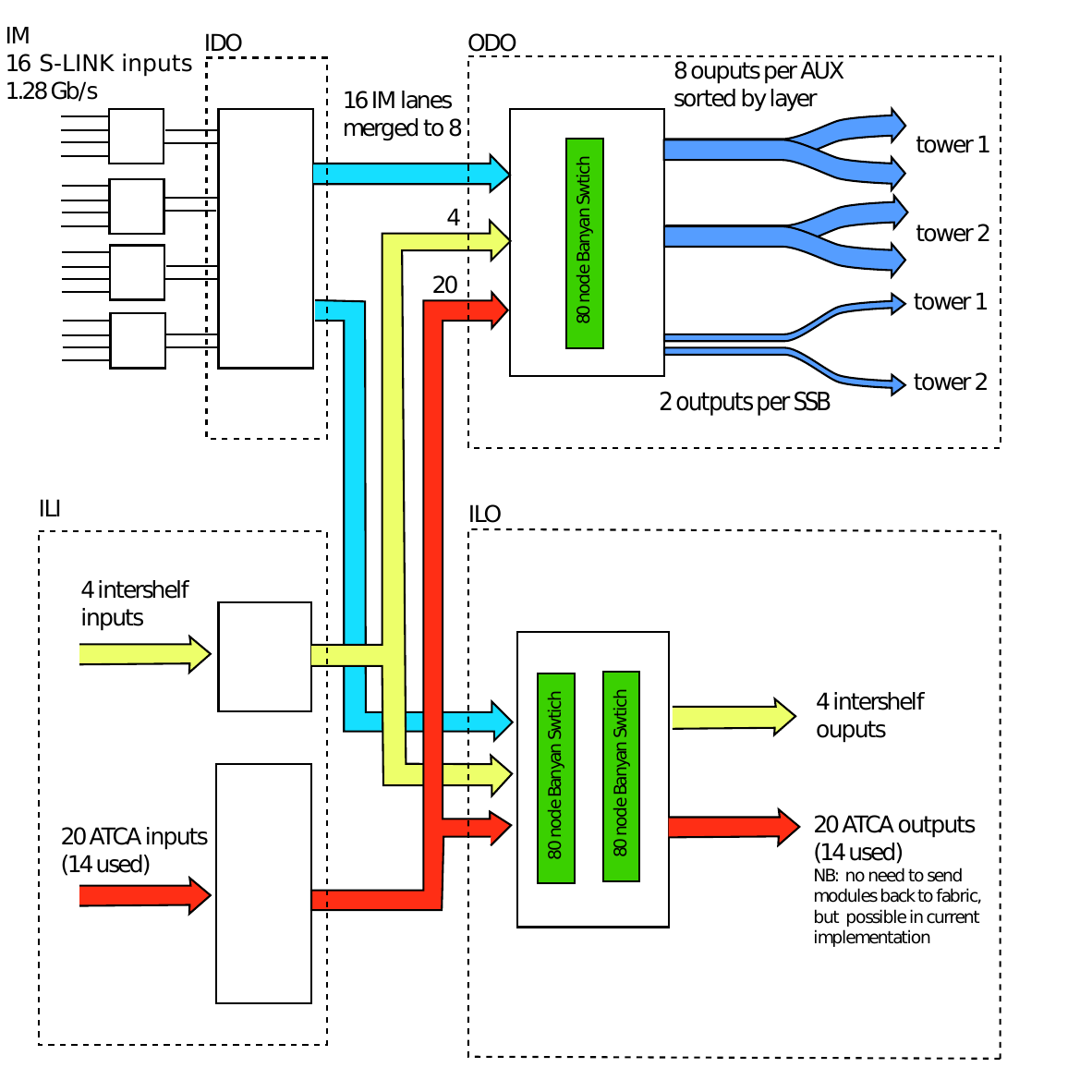}\\
\end{center}
\caption{Diagram of dataflow in the \acl{DF} firmware design.  Clustered hit data from the \aclp{IM} are sent to downstream boards via the \acf{ODO} block or are shared with other \aclp{DF} via the \acf{ILO} block. \label{fig:df-df}}
\end{figure}
 
Four \ac{IM} cards described in the previous section are installed in the four mezzanine slots on the \ac{DF}.
Each \ac{IM} transmits two lanes of hit data over six-bit DDR links operating with a clock speed of 200\,MHz.
In each clock cycle the \ac{DF} receives eight data bits, three control bits and one parity bit giving a data transfer rate for each lane of 1.6\,Gb/s.

As mentioned in the previous section, module packets from the \ac{pixel} and \ac{SCT} detectors consist of a header, a trailer, and coordinate hits in 32-bit words.
The DF switches and duplicates these module packets and does not alter their content.
As can be seen in Figure~\ref{fig:df-df}, data from the \ac{IM} are received in the \ac{DF} firmware module called the \ac{IDO}.
The \ac{IDO} detects L1 ID mismatches among the 16 \ac{IM} inputs and discards inputs with corrupted L1 IDs.
 
If a module packet  has a destination in the tower hosted by the current \ac{DF} board it is copied into one of eight lanes which are transmitted to the \ac{ODO} switch via 32-bit-wide \ac{FIFO} buffers.
If a module packet has a destination hosted by another \ac{DF} board,  it is copied into one of a separate set of eight lanes which are transmitted to the \ac{ILO}.
It should be noted that the \ac{IDO} will send some module packets via both connections.
 
Module data packets from other DF boards are received using
8b10b encoding by the \ac{ILI}.
Module data packets from other \ac{DF} modules in the same shelf are received over the ATCA backplane.
Module data packets from other ATCA shelves are received via optical fibres.
The data speed for the payload on a single fibre or backplane connection  is approximately  5\,Gb/s for both backplane and fibre transmission.
For both the backplane and fibre connections, there are pairs of duplex fibres, giving a total bandwidth between DF boards, in each direction, of  approximately
10\,Gb/s.
 
\begin{figure}[htbp]
\begin{center}
\includegraphics[width=0.85\textwidth]{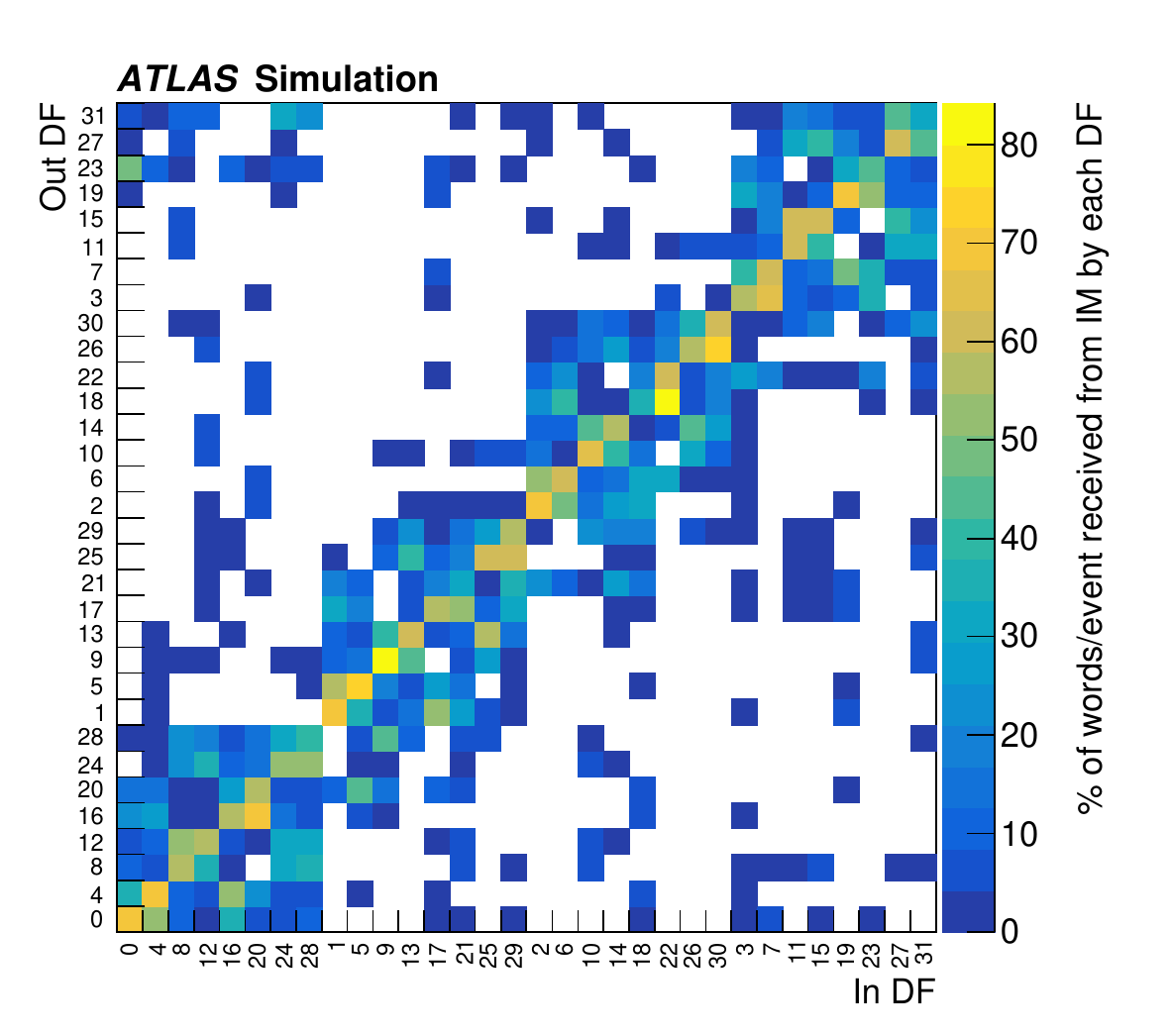}\\
\end{center}
\caption{Fraction of input \acl{IM} module (hosted by \aclp{DF} labelled `In DF') data shared with other \acl{DF} boards (labelled `Out DF') for distribution to \aclp{AUX} and \aclp{SSB}.
Data whose destination is within the same shelf are shared over the ACTA backplane.
The board numbers in the figure are arranged in blocks of eight corresponding to one ATCA shelf each.
Sharing of data with other \acl{DF} shelves requires at least one hop on an optical fibre. \label{fig:df-calc}}
\end{figure}

Figure~\ref{fig:df-calc} shows the fraction of data from the \acp{IM} of one `In DF' that is sent to various \acp{DF} in the system.
Typically, 60\% or more of the input data from the IM modules is used by the DF that hosts the IM modules.
The rate of data shared with other \acp{DF} corresponds to the off-diagonal elements in the figure.
The amount of data which must be shared with other ATCA shelves is much smaller than the amount of data which remains in a given ATCA crate.
Given the constraints inherent in the Pixel and SCT readout, the cabling was optimized to reduce data sharing between the \ac{DF} boards and between ATCA shelves.
A worst-case scenario of $t\bar{t}$ Monte Carlo simulation with at least one leptonic-decay and an average pile-up of 60 was used to generate the plot in Figure\,\ref{fig:df-calc}.

The \ac{ILO} receives module data packets from the  \ac{IM} cards via the \ac{IDO} and also receives data from other \ac{DF} modules for transmission over the backplane or fibres connecting the \ac{DF}s to different shelves via the \ac{ILI}.
This allows any given module data packet to be routed to any of the 32 \acp{DF} in the system.
The routing tables and layout of the system were configured to minimize the number of instances in which
a module packet requires several hops over fibres and backplanes to reach its final destination.  The minimal extra bandwidth needed by these rare extra hops is not reflected in Figure~\ref{fig:df-calc}.
 
The \ac{ODO} receives module data packets from the \ac{IDO} and the \ac{ILI}.
This block is responsible for sending data to the downstream boards, out of the \ac{DF} system.
Module data packets needed by the \ac{AUX} and \ac{SSB} are sorted into two towers of eight and four layers each, respectively.
Module data packets needed by the \ac{SSB} are transmitted directly to the \ac{SSB}.
If a given module data packet is needed by more than one tower it will be duplicated by the switch in the \ac{IDO}.
 
\begin{figure}[htbp]
\begin{center}
\includegraphics[width=0.50\textwidth]{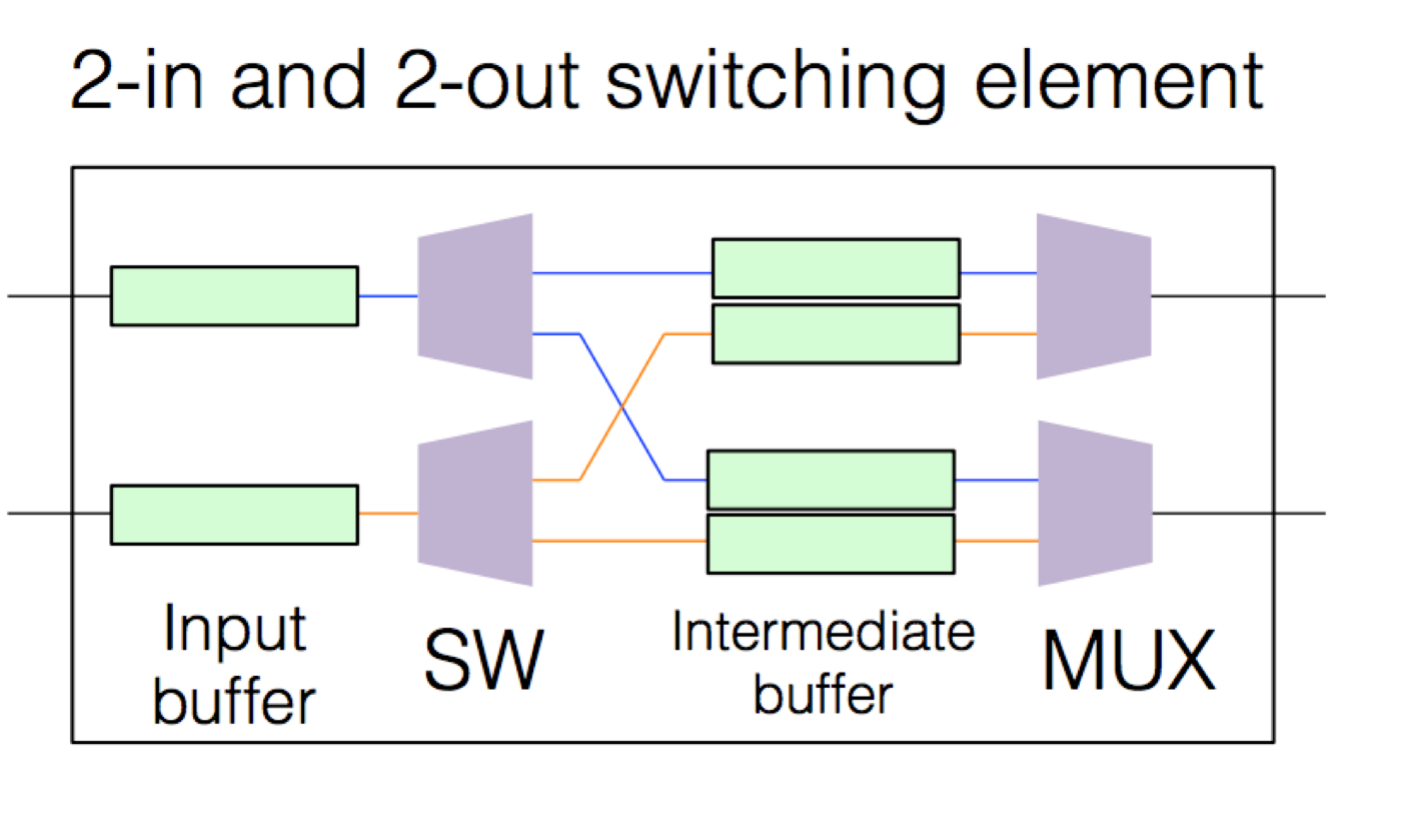}
\end{center}
\caption{ A single node in the Banyan switch showing input buffers and intermediate buffers.  The buffers are 32 bits wide.   \label{fig:df-banyan}}
\end{figure}

The data switching in both the \ac{ODO}  and \ac{ILO}  is performed using a Banyan switching network  which consists of $16 \times 5$ nodes allowing each input module data packet to be switched to one or more of the 32 output nodes.     Each node (see Figure~\ref{fig:df-banyan}) consists of two buffered  inputs and intermediate memory large enough to store the largest module data packets produced by the IM.  The nodes can switch packets to the output nodes according to a 32-bit address mask and duplicate the packet as required by the address mask.
The Banyan switching network was simulated using a \ac{UVM} simulation and it was demonstrated that the network could handle the expected Run 3 FTK load at the full 100\,kHz input rate.
 
The firmware used in the Virtex-7 690T FPGA on the Pulsar board for the tests reported here uses approximately 78\% of the logic lookup tables, 49\% of the flip-flops and 70\% of the block RAM.


\subsection{\acl{AUX}}
\label{sec:AUX}
The AUX transports hit data to the \ac{AMB} and evaluates the returned track candidates, before sending candidates which pass the selection requirement to the \ac{SSB}.
Each of the 128 \acp{PU} is composed of an AUX and AMB pair.
The AUX  is a 9U VME \ac{RTM} that is 280~mm deep.
It communicates via a P3 connector with the AMB and with the \ac{DF} and \ac{SSB} via QSFP connections.
 
\begin{figure}[htbp]
\begin{center}
\subfloat[]{\includegraphics[width=0.48\textwidth]{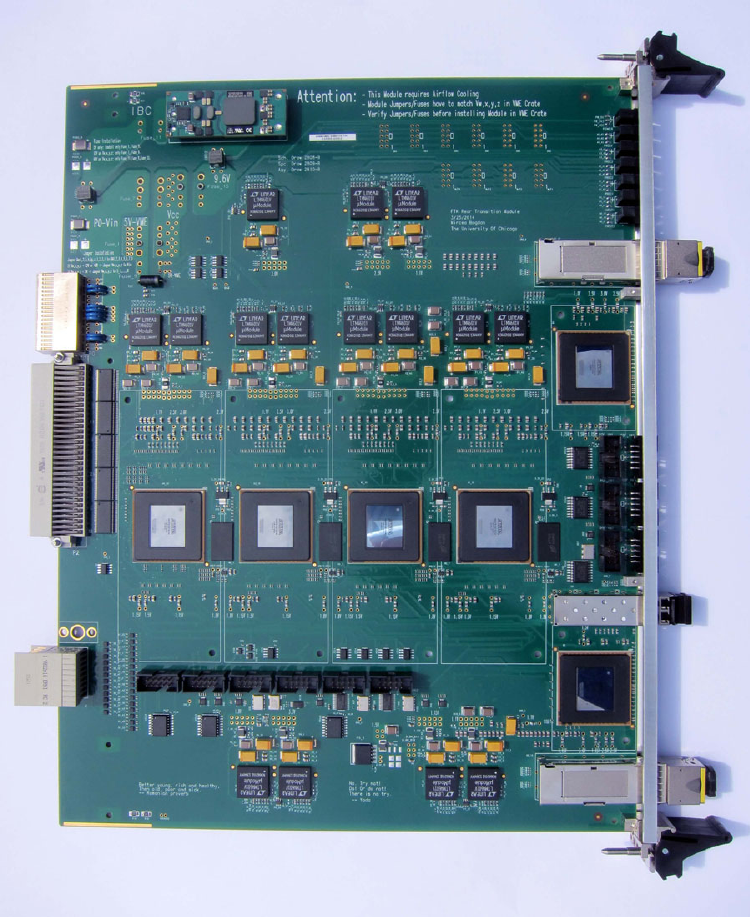}\label{fig:AUX:a}}
\hfill
\subfloat[]{\includegraphics[width=0.48\textwidth]{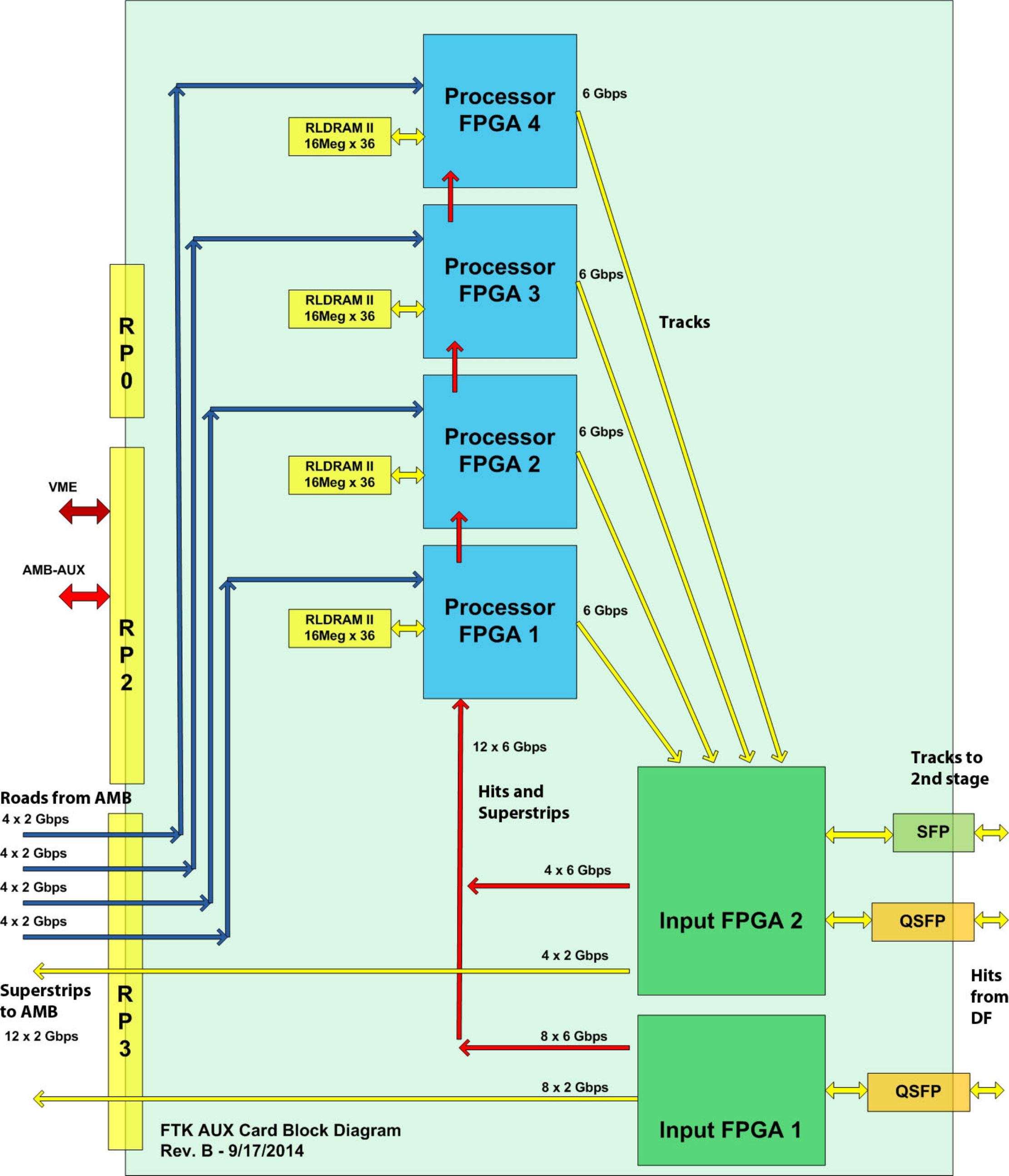}\label{fig:AUX:b}}
\end{center}
\caption{\label{fig:AUX} The (a) physical \acl{AUX} and (b) dataflow through the \acl{AUX}, starting from the lower-right.}
\end{figure}

The AUX contains six Altera Aria-V GX FPGAs \cite{AlteraV}, two FB7H4 for input/output and four FB5H4 processing chips.
Each processing chip is connected to one of the \ac{AMB} \acp{LAMB}.
The AUX board is shown in Figure~\ref{fig:AUX:a} and the dataflow is sketched in Figure~\ref{fig:AUX:b}.
Clusters from eight silicon detector layers are sent from DF boards to the AUX via a \ac{QSFP} connection; the Input-1 FPGA receives data from three \ac{pixel} layers and one \ac{SCT} layer, while the Input-2 FPGA receives data from the other four \ac{SCT} layers.
Coarse resolution hits, called \acfp{SS}, are sent to the AMB for pattern recognition over $12\times2$~Gb/s lines through the P3 connector.
Both the \acp{SS} and full resolution hits are sent to the processor chips.
The processors also receive from the AMB the addresses of roads (matched SS patterns)
that have hits on at least seven of the silicon detector layers over $16\times2$~Gb/s lines back through the P3 connector.
For each of those roads, the \ac{DO} portion of a processor FPGA retrieves all of the full resolution hits in the road.
The \ac{TF} portion then fits each track candidate (each combination of one hit per layer) and calculates the $\chi^2$.
Those tracks passing a $\chi^2$ cut are sent to the Input-2 FPGA where the four processor streams are synchronized and duplicate-track removal is performed before the remaining tracks are sent to an \ac{SSB} via S-LINK over a \ac{SFP} connection.
All of the data processing is completely pipelined using local control signals for dataflow control.
 
\begin{figure}[htbp]
\begin{center}
\includegraphics[width=\textwidth]{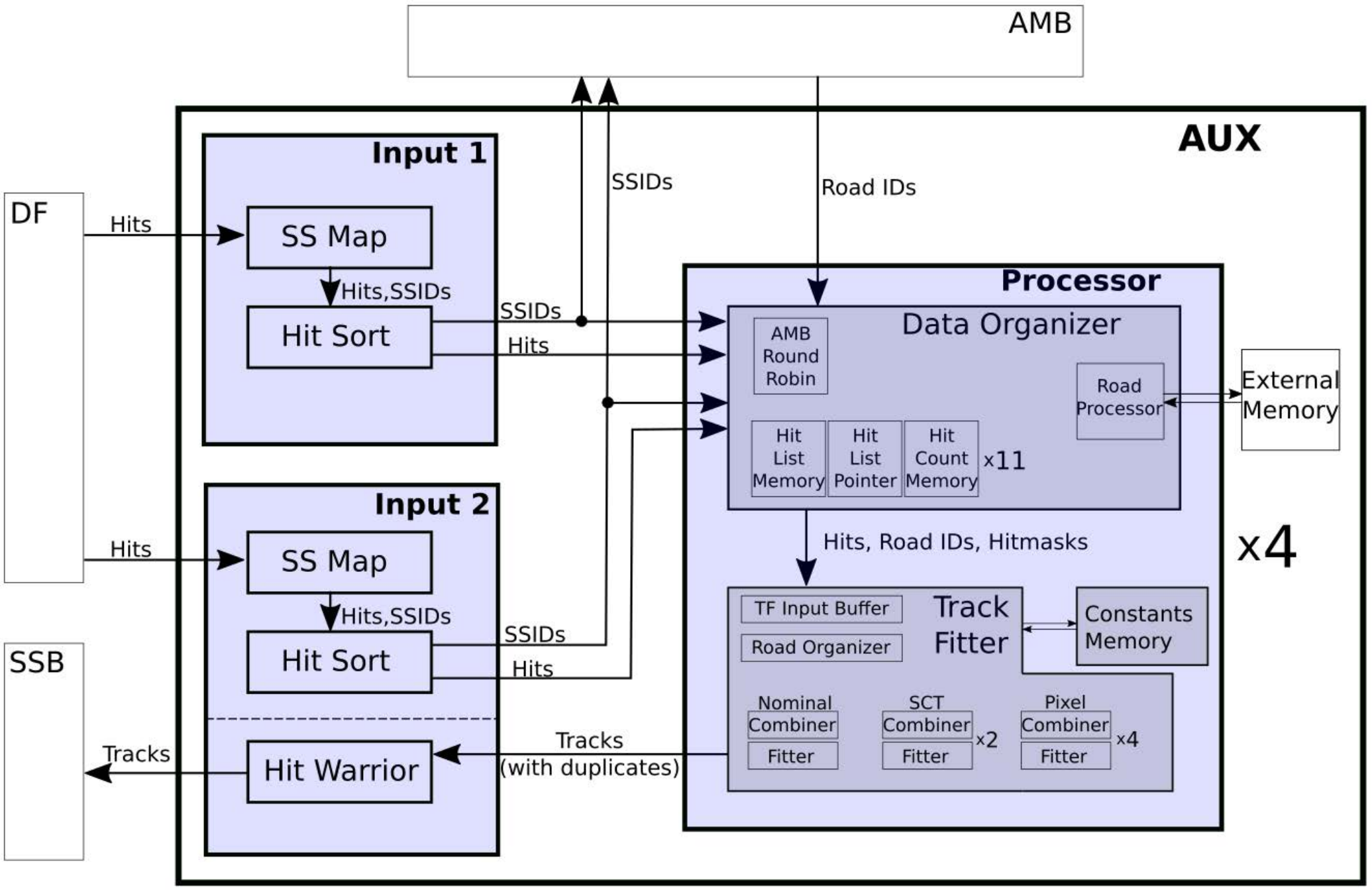}
\end{center}
\caption{\label{fig:AUXfirmware} Diagram of the \acl{AUX} firmware design.}
\end{figure}

An overview of the AUX firmware is shown in Figure~\ref{fig:AUXfirmware}.
Hits from the DF enter the SS Map in which the 32-bit hit word is converted to a 16-bit \ac{SS} using lookup tables.
The Hit Sort module uses a base-4 radix sort to guarantee that all hits in the same \ac{SS} are sent out sequentially.
Each \ac{SS} is sent to the AMB while the \ac{SS} plus full resolution hit (a combined 48-bit word) is sent to all four processor chips.
 
There are two major functions in a Processor FPGA, the \ac{DO} and the \ac{TF}.
The DO receives roads from one of the AMB \acp{LAMB} on four serial links and uses a round robin to process them with minimum latency.
The hits from the eight detector layers are processed in parallel.
The DO is a database built on the fly for rapid retrieval of all hits in a road.
Three sets of memory are used:  the \ac{HLM} stores each hit sequentially as it arrives; the Hit List Pointer contains a pointer to the first HLM location for each \ac{SS}; and the Hit Count Memory contains the number of hits in the \ac{SS}.
These databases are replicated for each of the 11 coordinates used in Eq.~\ref{8-layer-chi}, as shown in Fig.~\ref{fig:AUXfirmware}.
There are two copies of the DO which act as a ping-pong buffer.
At any instant one is in Write mode, filling the memories with the data for event $N$, while the other is in Read mode, retrieving hits in roads for event $N-1$.
For each road, the data sent from the DO to the TF consist of the road number, all of the full resolution hits in the road, and a hit mask noting which layers have hits.
 
The TF block fits each track candidate using a linear approximation and prestored constants, as described in Section~\ref{sec:ftk_system_overview} and Eq.~(\ref{8-layer-chi}), which take up about 25~Mb per \ac{AUX}.
In the AUX, only the $\chi^2$ is calculated.
For each road received from the AMB, the Road Organizer stores the road and hits and initiates the fetching of the fit constants from memory.
The data and constants are then sent to the appropriate fitter where each combination of one hit per layer is fit and its $\chi^2$ tested.
The tracks that pass the test are sent to the \ac{HW} in the Input-2 FPGA, which does duplicate-track removal.
If two tracks share more than a programmable number of hits, the one with the lower $\chi^2$ is retained.
 
There are three types of fitters: the SCT Majority Fitter, the Pixel Majority Fitter, and the Nominal Fitter.
The Majority Fitters are used if one hit is missing from their respective detectors to improve the track-finding efficiency.
For the Nominal Fitter, in parallel with the full fit, eight Recovery Fits are done in which one of the hits is dropped from each layer.
If the full fit fails the $\chi^2$ cut but a recovery fit passes the cut, that fit is retained because it is assumed that a good track had a missing hit, but a random hit landed in the dropped layer.
Based on analysis of simulated data, the optimal use of FPGA resources are one, two, and four implemented Nominal, SCT Majority, and Pixel Majority fitters respectively.
These fitters operate in parallel.

The firmware in the AUX FPGAs used for the tests here, including monitoring and error handling, came close to saturating the available resources.
For the processor chips, $>70\%$ of the logic and $75\%$ of the memory are used.
For the input FPGAs, about $40\%$ of the logic and $>80\%$ of the memory are utilized.

\subsection{\acl{AMB}}
\label{sec:AMB}
\begin{figure}[htbp]
\begin{center}
\includegraphics[width=0.75\textwidth]{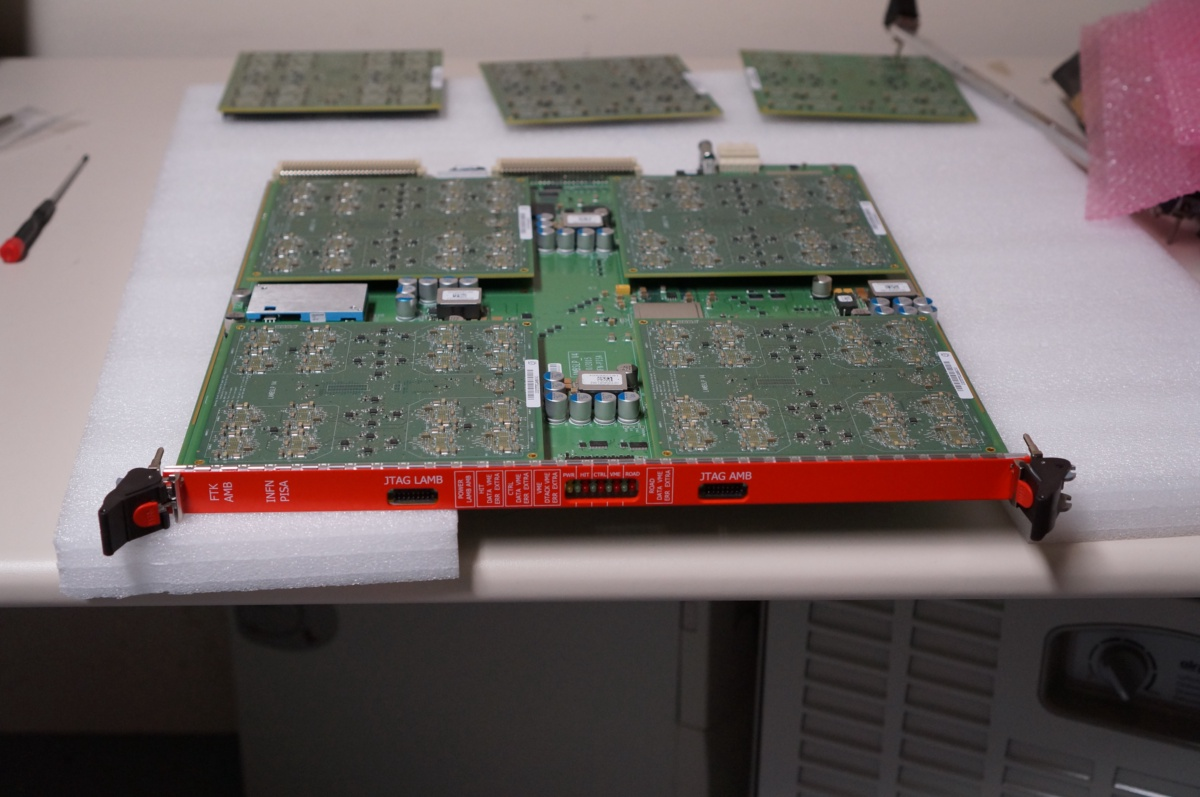}
\end{center}
\caption{\label{fig:FTH_Hardware:AMB:AMB} Photo of an Associative Memory Board with four LAMBs mounted. Additional LAMBs can be seen in the background.}
\end{figure}

The \ac{AMB}~\cite{Sotiropoulou:2295706}, shown in Figure~\ref{fig:FTH_Hardware:AMB:AMB} is the part of the \ac{PU} that is
devoted to pattern matching between the \acp{SS} sent by the \ac{AUX}
and the patterns simulated for each 8-layer road.
This functionality is implemented through a
custom ASIC: the \ac{AM}06 described in Section~\ref{sec:AM}.
 
An \ac{AMB} consists of a 9U-VME mainboard and four identical mezzanines, the \acp{LAMB}.
Four \acp{FPGA} are installed on the mainboard:
two Xilinx Spartan-6 (45T and 16)~\cite{Xilinx:Spartan6}, which control the VME interface and the state of the board respectively;
and two Xilinx Artix-7 200T~\cite{Xilinx:7series}, which control the input and output logic.
Each \ac{LAMB} is equipped with 16 \ac{AM} chips and a Xilinx Spartan-6 FPGA,
dedicated to the configuration and monitoring of the mezzanine.
 
\begin{figure}[htbp]
\begin{center}
\includegraphics[width=0.75\textwidth]{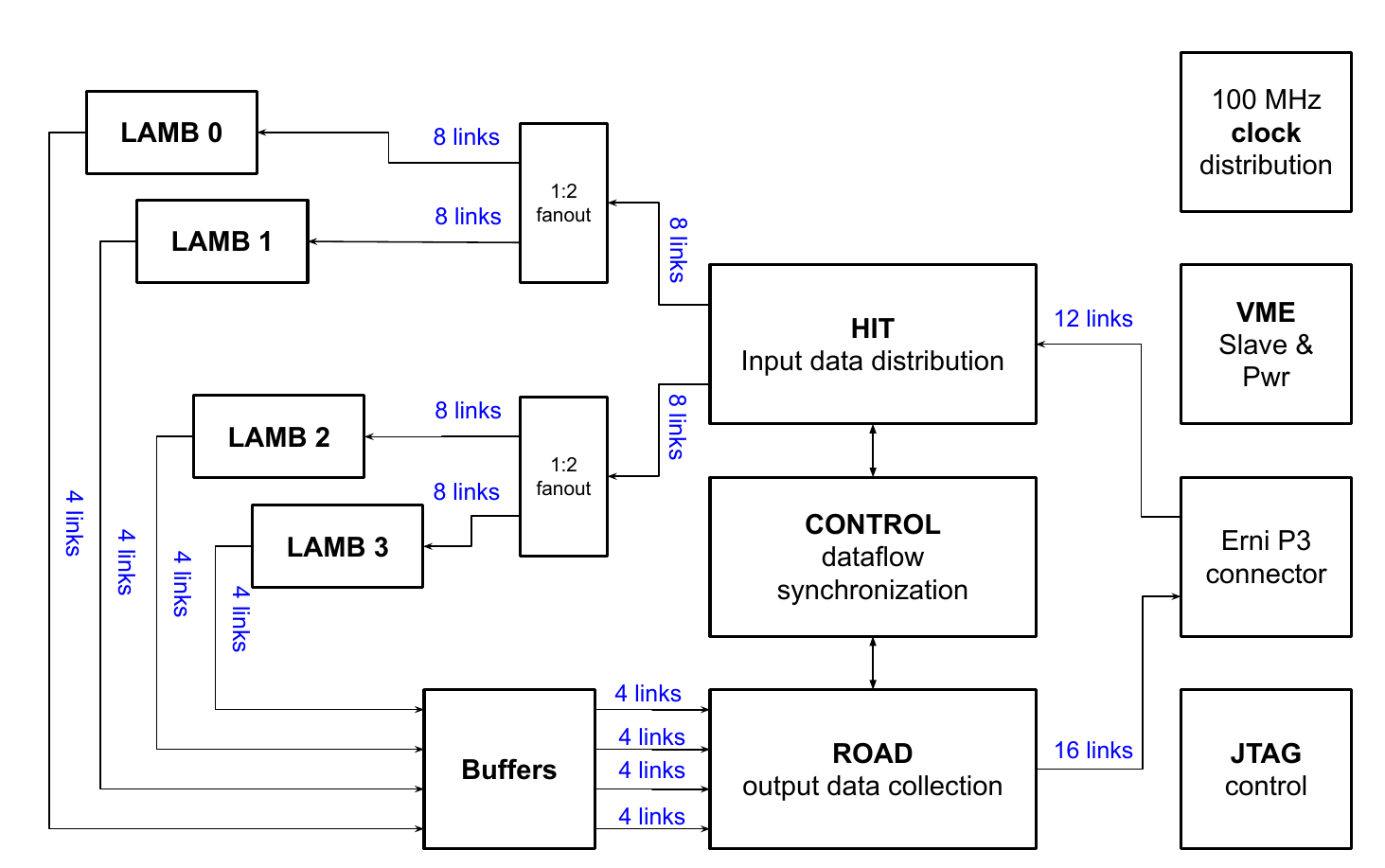}
\end{center}
\caption{\label{fig:FTH_Hardware:AMB:dataflow} Block diagram of the Associative Memory Board (AMB) structure. The black arrows show the dataflow in the AMB.}
\end{figure}

The \ac{AUX} and \ac{AMB} are connected through the VME P3 connector,
as shown in Figure~\ref{fig:FTH_Hardware:AMB:dataflow}.
High-speed serial links provide 12~Gb/s as input from the \ac{AUX} to the \ac{AMB} and 16~Gb/s as output from the \ac{AMB} to the \ac{AUX}.
The \acp{SS} provided by the \ac{AUX} are received by the Xilinx Artix-7 FPGA handling the input, called Hit.
This FPGA sends a copy of the \acp{SS} to each of the four \ac{LAMB} mezzanines installed on the \ac{AMB}.
Two sets of eight 1-to-2 fan-out buffers (one per input link) distribute the same data to all the \ac{LAMB} mezzanines.
The Hit FPGA buffers the data and synchronizes the inputs by sending a new event only when all the mezzanines have finished processing the previous one.
 
On the \ac{LAMB}, the \acp{SS} are distributed to the 
\ac{AM} chips.
A chain of 1-to-4 fan-out buffers in cascade allows each \ac{AM} chip to get a copy of the \acp{SS}.
All of the selected roads are sent to the Xilinx Artix-7 FPGA handling the output, called Road.
This FPGA collects all the roads belonging to the same event before sending them to the \ac{AUX} for the 8-layer fits described in the previous section.
A dedicated Spartan-6 FPGA, named Control, controls the dataflow on the board and synchronizes the Hit FPGA and Road FPGA.
 
A fourth Spartan-6 FPGA, named VME, handles the communication on the VME bus and takes care of the board configuration and monitoring.
 
An \ac{AMB} is able to load $\sim 8 \cdot 10^6$ independent patterns
and to perform pattern comparisons at a rate of $\sim 12$~petabytes/s,
with a peak power consumption below 250~W (100~W of which is from the AMB).
The entire \ac{FTK} system contains 128 \ac{AMB}, and it includes more than one billion (1\,098\,907\,648) track patterns.
A collection of patterns is referred to as a pattern bank.
 
During the tests performed in 2018, the \ac{AMB} demonstrated the ability
to process events at a rate  greater than $100$~kHz.
Dedicated high-power tests~\cite{Sottocornola:2741522} confirmed the designed power consumption of the \ac{AMB}, and
the ability of the custom cooling system, described in Section~\ref{sec:input},
to keep the boards within temperature bounds while in the intended Run~3 configuration.
 
The firmware used in tests reported here for the Hit, Control, and \ac{LAMB} \acp{FPGA} uses up to 50\%, 60\%, and 45\% of the logic resources respectively.
The firmware for the Road \ac{FPGA} uses approximately 30\% and 50\% of the logic and memory resources respectively.


\subsection{\acl{AM}}
\label{sec:AM}
\begin{figure}[htbp]
\begin{center}
\includegraphics[width=0.5\linewidth]{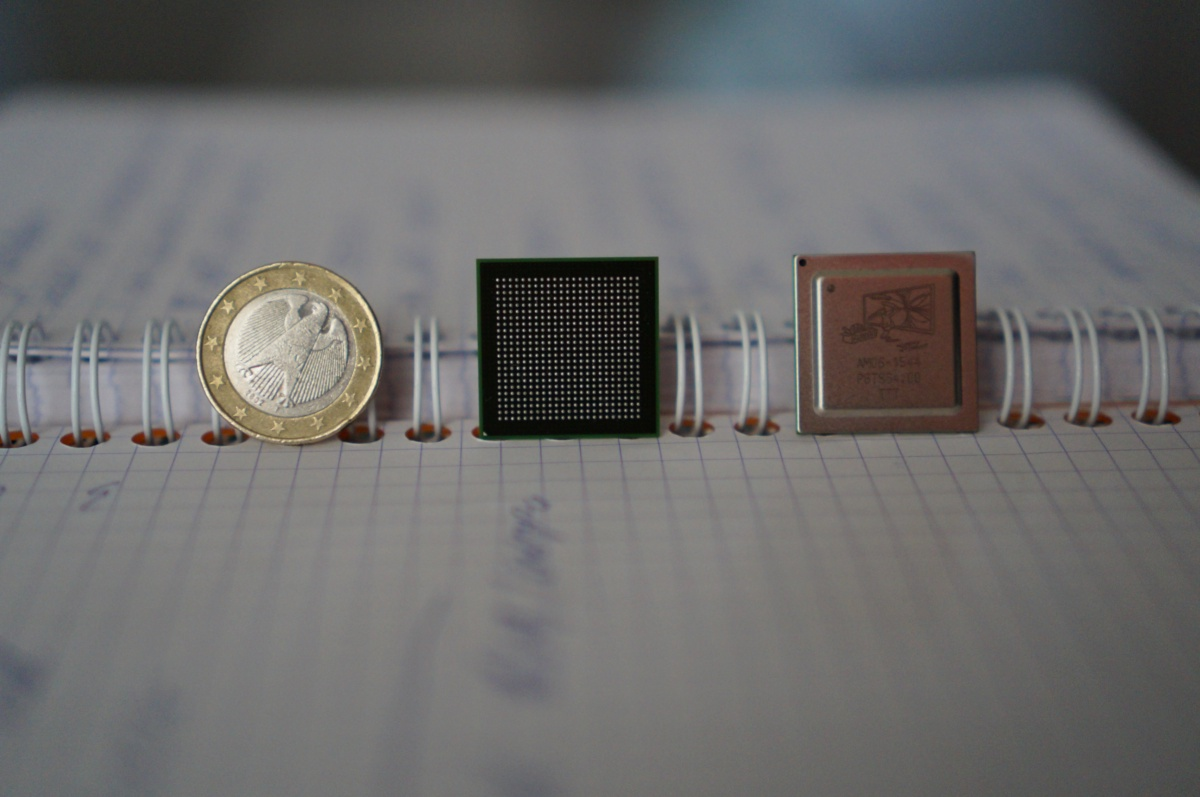}
\caption{ Photo of the bottom (centre) and top (right) of the AM06. On the left a 1-euro coin is shown for size comparison. }
\label{fig:FTK_Hardware:AM:AM06}
\end{center}
\end{figure}

The \ac{AM} ASIC is a highly parallel processor for pattern matching, hosted on mezzanines on an \ac{AMB}, which feeds in \aclp{SS} and reads out roads.
The \ac{AM} ASIC used by the \ac{FTK} system, the AM06~\cite{Annovi2228284} shown in Figure~\ref{fig:FTK_Hardware:AM:AM06}, is designed in 65~nm CMOS technology.
The design of the \ac{AM} chip combines full-custom content-addressable memory arrays, standard logic cells and SerDes IP blocks at 2~Gb/s for input/output communication.
The \ac{AM} chip is able to perform pattern matching in parallel on all the 131\,072 patterns stored on its $168~\text{mm}^2$ die.
The stored patterns are composed of a set of eight \acp{SS}, one for each detector layer used by the FTK. 
The 16 bits of each \ac{SS} are stored on 18 XORAM~\cite{Frontini2012ANX} cells forming an \ac{AM} cell.
The XORAM  cell is composed of a conventional six-transistor SRAM cell, which stores the data, merged with an XOR gate executing the matching with the input at a clock rate of 100~MHz.
Two XORAM cells can be paired to allow the presence of ternary bits in the pattern to be stored. 
When two XORAM cells are used to store the value of a single bit, the bit-value can assume three values ``0'' (or `00'), ``1'' (or `11'), and ``$X$'' (either `01' or `10').
While for the ``0'' and ``1'' values the pattern behaves as if no ternary bit is present, the ``$X$'' value can be used as a \ac{DC} value.
The \ac{DC} bits are bits in the pattern that are always considered matched, regardless of the value of the corresponding bits in the input word.
Effectively, these \ac{DC} bits increase the width of the pattern by a factor of $2^{N_{X}}$, where $N_{X}$ is the number of \ac{DC} bits used in a pattern.
This provides variable resolution patterns, allowing some patterns to be wider than others.
Figure~\ref{fig:ftk_tracking} uses a dotted line as a visual representation of this on the second layer.
While the configuration used by the FTK uses a 16-bit SS with up to two DC bits, the AM chip can be configured to accommodate from two to nine DC bits for each pattern.
The XORAM cell is optimized to achieve a power consumption of 1~fJ/bit-comparison, allowing 64 AM06 to be installed on a single VME board.
 
\begin{figure}[htbp]
\begin{center}
\includegraphics[width=0.5\linewidth]{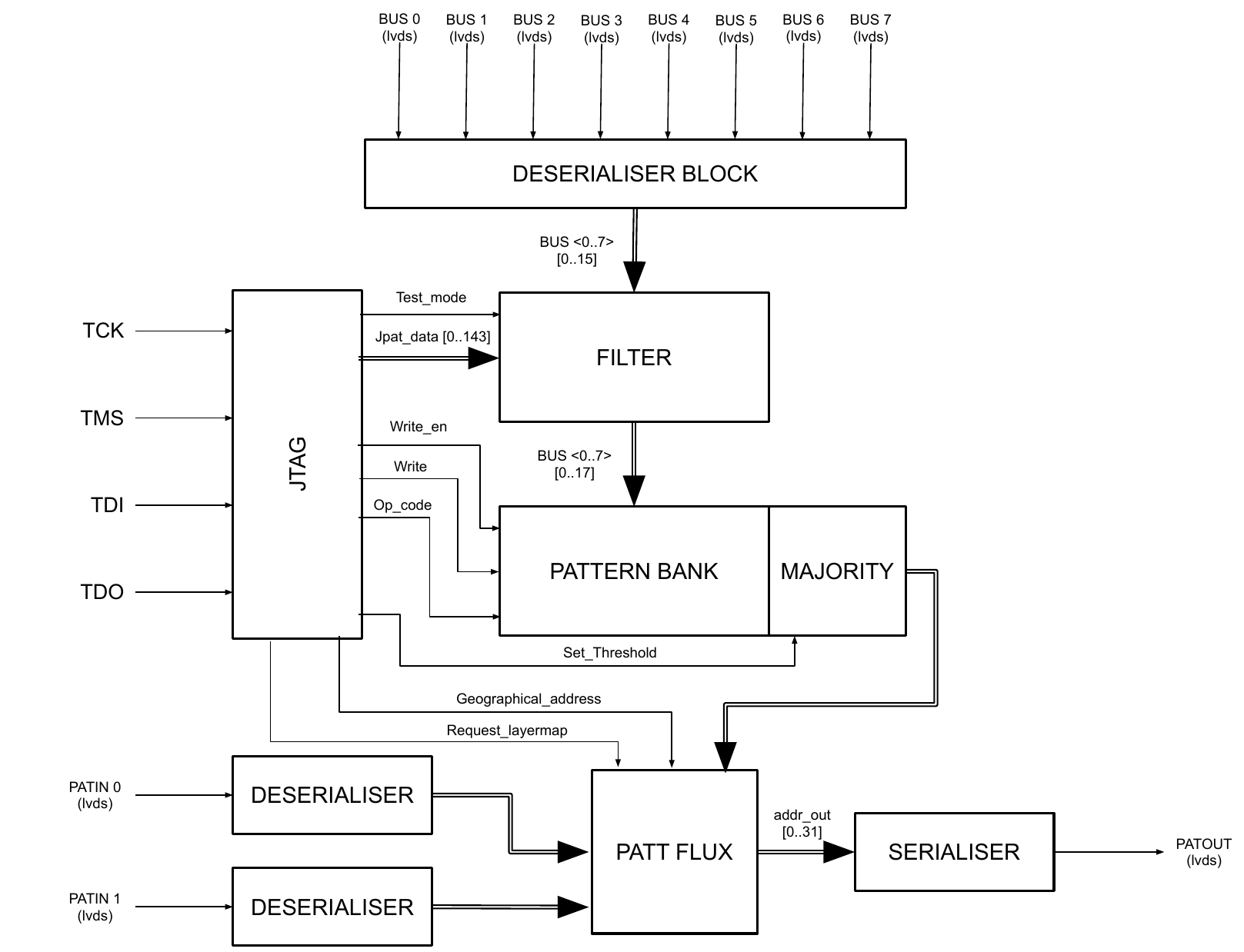}
\caption{ Diagram of the dataflow through the AM06 chip. }
\label{fig:FTK_Hardware:AM:schema_AM06}
\end{center}
\end{figure}

As shown in Figure~\ref{fig:FTK_Hardware:AM:schema_AM06}, \acp{SS} are received by the AM06 asynchronously on the eight input serial-buses.
The \acp{SS} are deserialized and transmitted in parallel to the FILTER module that takes care of correctly encoding ternary bits by mapping the 16-bit input words to 18-bit words.
This module can also receive data through the JTAG connection in order to write the pattern data and to test the stored pattern.
The $8\times18$-bit data are propagated to all of the memory locations in the chip where they are compared to the stored pattern.
The match status for each of the eight layers in a pattern is held until all the data belonging to the same event are received.
At the end of the event, a majority block counts the number of matched layers for each pattern.
If the number of matched layers is greater than a selectable threshold, the pattern identifier is sent to the PATT FLUX module.
This module adds a chip geographical address to the internal pattern address and handles the chip output.
Patterns received from previous chips in the daisy chain are deserialized and sent to the PATT FLUX module, which buffers the data and ensures they are transmitted before the given chip's data.
The end of an event resets all of the \ac{AM} cells in the ASIC in one clock cycle so that the \ac{AM} chip can process data belonging to a new event while sending out the matched patterns.
 
The \ac{AM} chips can be connected in a daisy chain to maximize the number of stored patterns on each \ac{AMB} while maintaining a simple routing at the board level.
Two dedicated inputs are present on each AM06 to receive data from previous \ac{AM} chips in the chain.
The pattern identifiers collected on the two inputs are copied to the \ac{AM} chip output.
 
To allow the full system of 128 \acp{PU} to be commissioned, more than 9000 AM06 were produced in two production rounds.
A first batch of 2159 preproduction chips, including slow corner and fast corner chips,\footnote{The fast corner and slow corner chips are obtained with a  variation of fabrication parameters that are optimized for set-up and hold time respectively. This has the effect of producing chips that on average will work at slightly faster or slower clock frequencies.} was delivered in 2014 and tested in-house, achieving an estimated yield of $83\%$ for the slow corner and $89\%$ for the fast corner.
The production batch composed of 8126 ASICs was delivered in 2016 and tested by an external company.
The average yield for the production batch was $83\%$. 
\ac{AM} chips from this batch were split into two groups.
Normal chips are required to pass all functional tests at nominal power consumption and clock speed.
High-performance chips are those which pass all functional tests with a $10\%$ increase in clock frequency and $5\%$ decrease in voltage.
High-performance chips account for roughly $65\%$ of the produced \ac{AM} chips, and normal-performance chips account for roughly $15\%$ of the produced \ac{AM} chips.
A correlation of the yield with the mechanical wear of the \ac{AM} chip socket
was observed, with more chips ending up in the lower performance group.
Only high-performance chips were expected to be mounted on the production boards.
Retesting of chips to recover more of them was planned after having understood the effect of the aging of the test-stand.


\subsection{\acl{SSB}}
\label{sec:SSB}
\begin{figure}[htbp]
\centering
\includegraphics[width=0.8\linewidth]{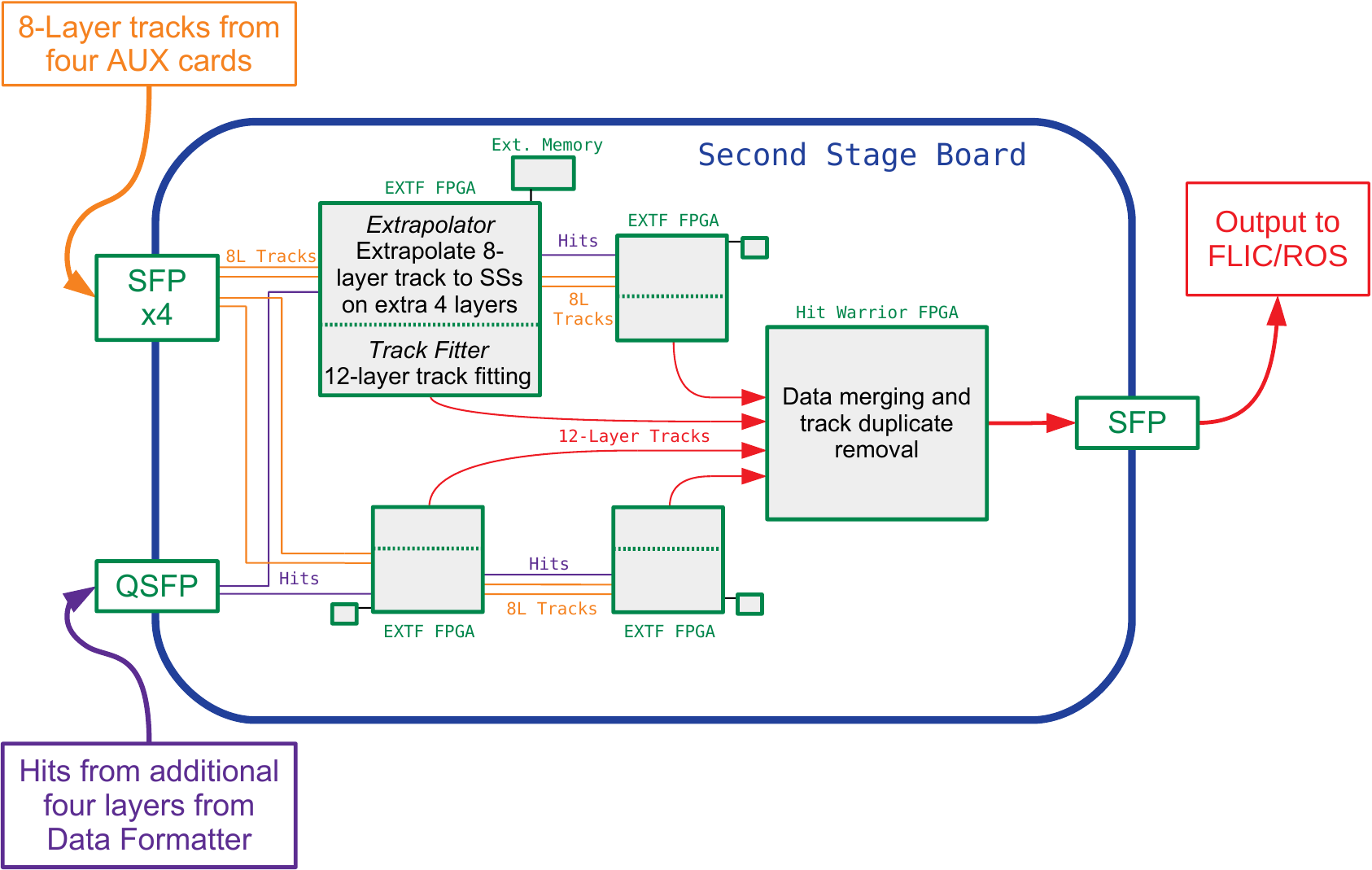}
\caption{Diagram of the dataflow through the \acl{SSB}.  }
\label{fig:ssb_func}
\end{figure}

\begin{figure}[htbp]
\centering
\includegraphics[width=0.8\linewidth]{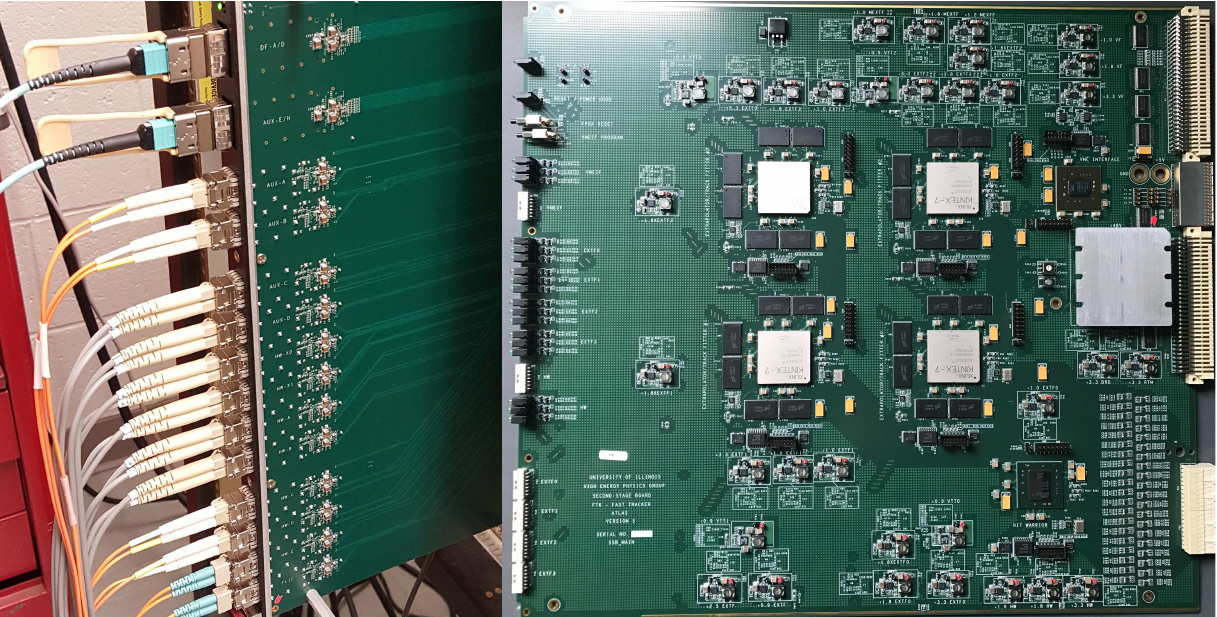}
\caption{The physical RTM (left) and MAIN \aclp{SSB} (right).}
\label{fig:ssb_RTM_MAIN}
\end{figure}

The \ac{SSB} receives 8-layer tracks from the \acp{AUX} and clustered hits from the \ac{DF} on the four layers that are unused in the pattern matching.
It extrapolates these tracks into the four layers to search for hits, performs 12-layer track fits and removes duplicate tracks before outputting the final FTK tracks.
A full track packet consists of its helix parameters, $\chi^2$ value, and associated cluster coordinates.
A schematic description of this functionality and the dataflow through the \ac{SSB} is shown in Figure~\ref{fig:ssb_func}.
The second-stage processing hardware consists of a VME main board and \ac{RTM}, shown in Figure~\ref{fig:ssb_RTM_MAIN}.
 
The processing of hits and tracks is performed on the \ac{SSB} main board (SSB MAIN), which hosts six Xilinx Kintex-7 FPGAs \cite{Xilinx:7series} (four 410T-FBG900 and two 410T-FBG676).
Four of the FPGAs, \acp{EXTF}, perform the extrapolation and 12-layer track fitting.
One FPGA, the \acf{HW}, merges the streams, performs duplicate-track removal, and sends final tracks to the \ac{FLIC} or \ac{ROS}.
The remaining FPGA provides the VME interface.
The high-speed serial interfaces to the \ac{DF}, \ac{AUX} cards, and \ac{FLIC} are provided by the \ac{SSB} \ac{RTM}, which also has serial interfaces to other SSBs intended for duplicate-track removal.
Data arrive at the \ac{SSB} from a \ac{DF} via a \ac{QSFP} connection and from \acp{AUX} via \ac{SFP} connections.
Two of the \ac{QSFP} channels and two \ac{SFP} connections are routed to each of the two `primary' \acp{EXTF}, those that are directly connected to the input data streams.
 
Each \ac{EXTF} FPGA has three main functions.
First, the \ac{EXTF} synchronizes the data streams from up to two \acp{AUX} and one \ac{DF} covering one \ac{FTK} tower.
Second, 8-layer track information is used to extrapolate the track into \acp{SS} on the additional \ac{IBL} and three \ac{SCT} layers.
The \acp{SS} of the most likely hit positions are then retrieved.
Last, the clustered hits in the extrapolated \acp{SS} are retrieved and combined with the 8-layer track hits to perform full 12-layer track fits, as described in Section~\ref{sec:ftk_system_overview} and Eq.~(\ref{lin-track-fit}).
 
\begin{figure}[htbp]
\centering
\includegraphics[width=\linewidth]{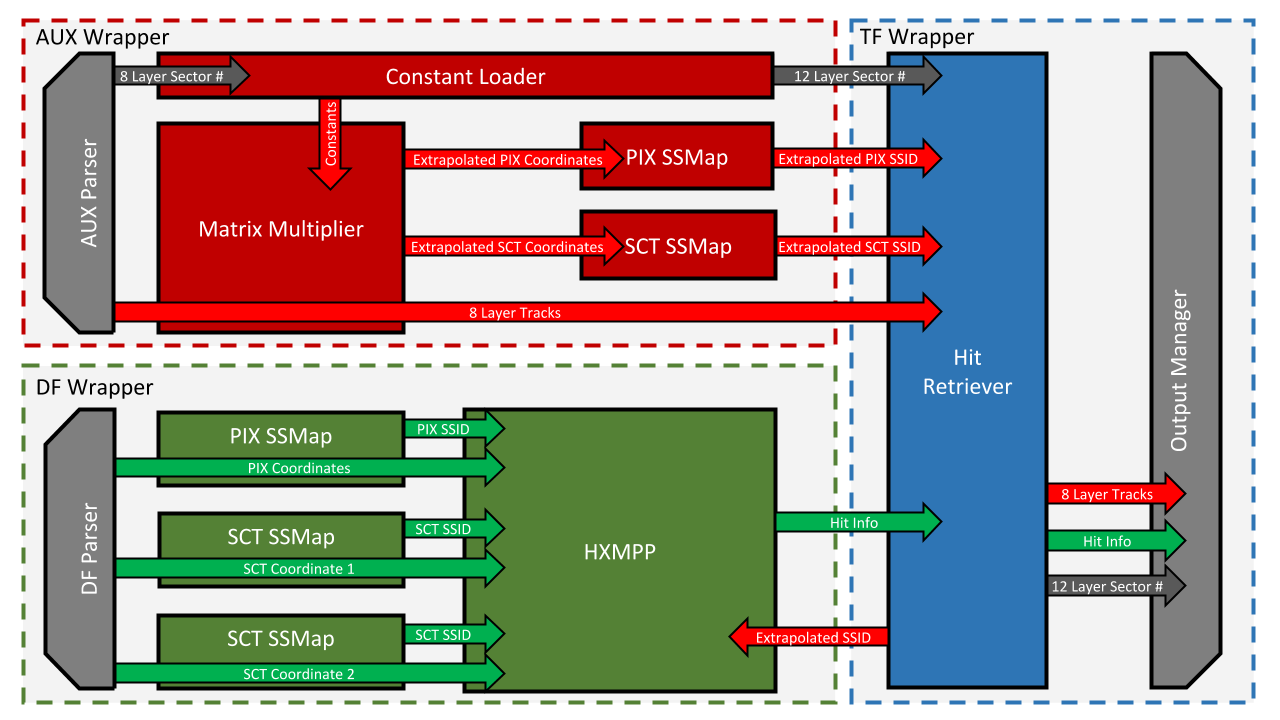}
\includegraphics[width=\linewidth]{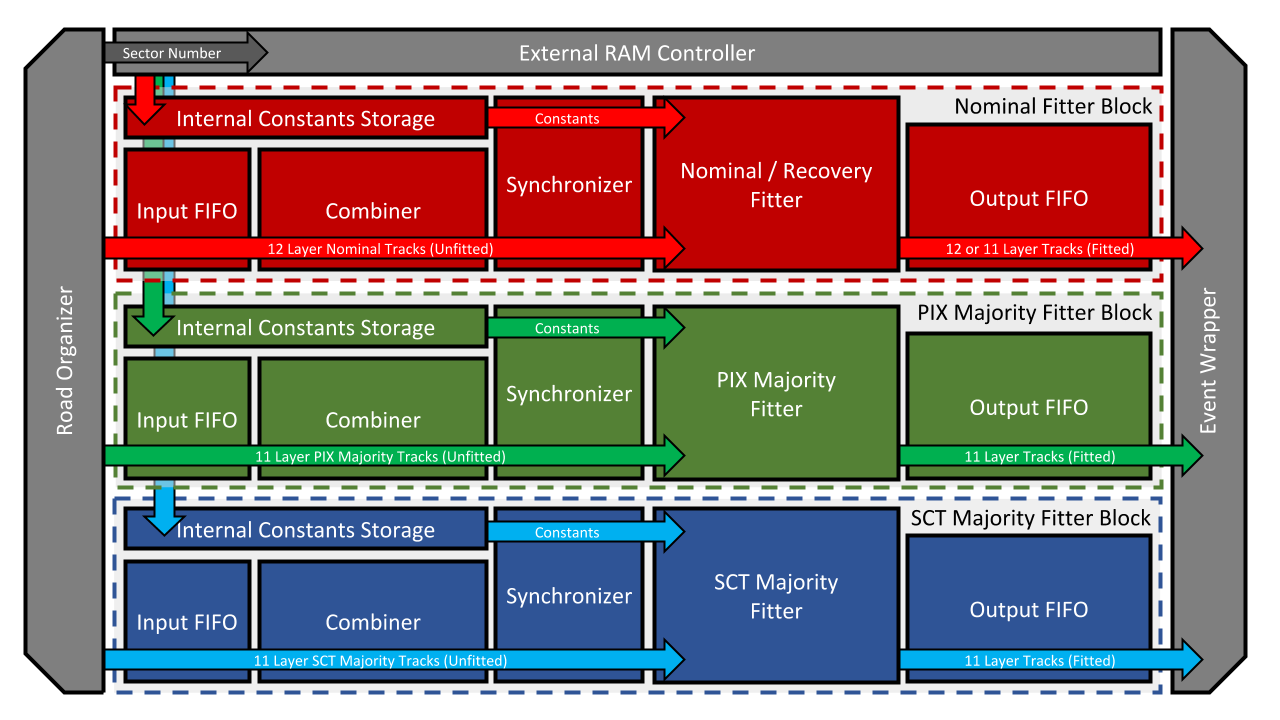}
\caption{Block diagrams of the firmware in the \acf{EXTF} FPGA on the \acl{SSB} split into the Extrapolator (top) and Track Fitter (bottom) firmware design.  The AUX and DF wrappers receive the 8-layer track and additional hits respectively.  The Extrapolator's output manager feeds into the Track Fitter's Road Organizer, and the Track Fitter sends tracks to the Hit Warrior via the Event Wrapper.}
\label{fig:ssb_extf}
\end{figure}

Tracks are required to pass a $\chi^2$ selection before being sent to the \ac{HW} FPGA.
Up to one hit in the additional four layers is allowed to be missing or dropped in a recovery fit as in Section~\ref{sec:AUX}.
When using all four \acp{EXTF} on the board, two of the \acp{EXTF} each receive their input data via pass-through from one of the primary \acp{EXTF}.
Configuration and monitoring data are read in and out through the VME interface.
The firmware design is illustrated in Figure~\ref{fig:ssb_extf}.
Up to 80\% of the logic resources and 90\% of the memory of the FPGA are used in the design.

Constants for extrapolation and track fitting are stored on external RLDRAM, accessed with a 400~MHz DDR clock.
Two chips per EXTF, with memory totalling 576~Mb, are available for the extrapolation constants, and four chips per EXTF, totalling 1152~Mb, are available for the track-fitting constants.
The constants used during comissioning were at most 70 and 550~Mb in size for the Extrapolator and Track Fitter respectively.
 
\begin{figure}[htbp]
\centering
\includegraphics[width=0.8\linewidth]{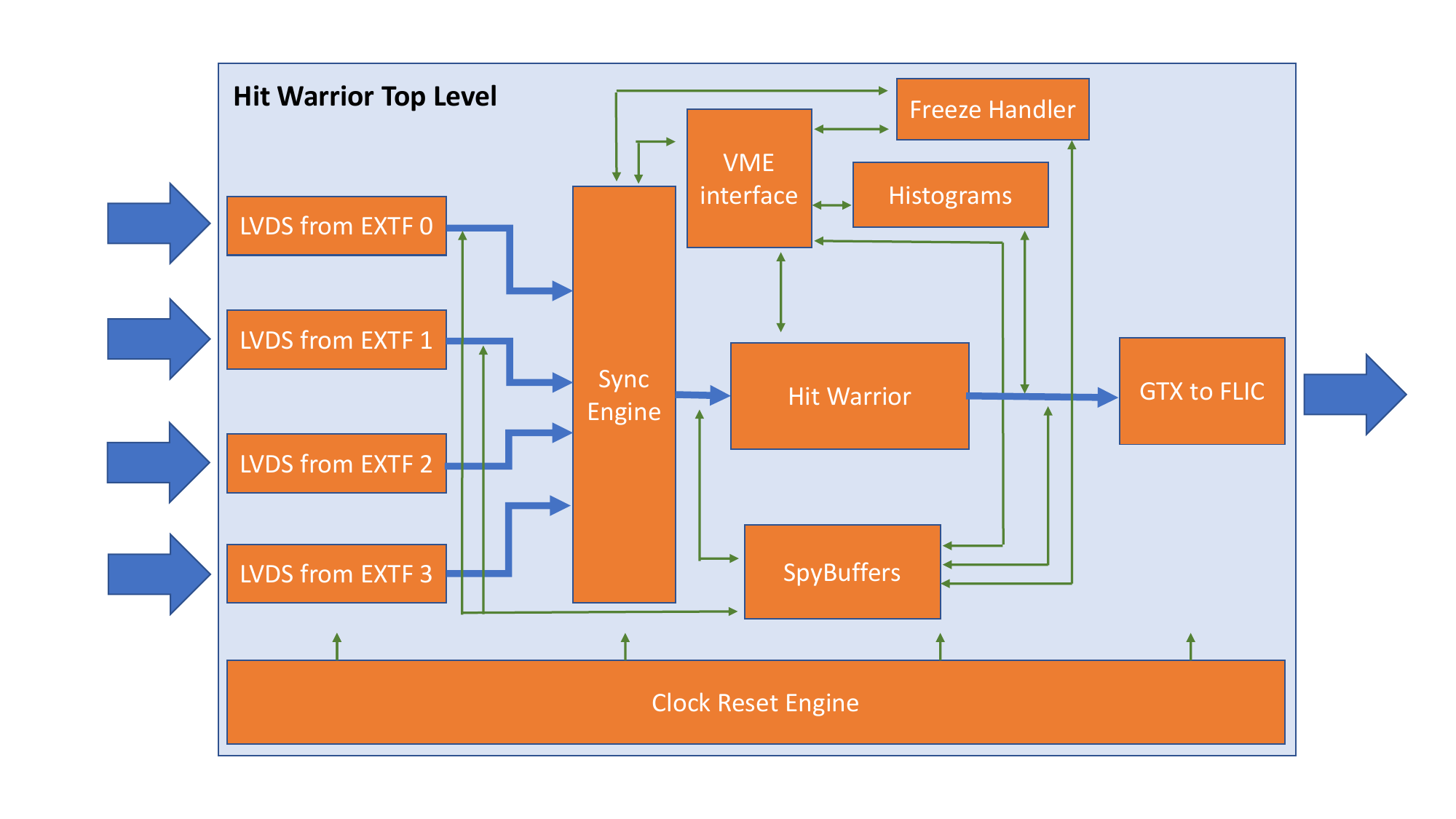}
\caption{Diagram of the Hit Warrior firmware design for the \acl{SSB}.}
\label{fig:ssb_hw}
\end{figure}

The \ac{HW} FPGA has two primary functions on the SSB. First, the \ac{HW} combines the event data streams from up to four EXTF FPGAs.
Second, it searches for and removes any duplicate tracks from the combined data stream and outputs the filtered event data stream to either the ROS or FLIC.
In addition to its primary functions, the \ac{HW} also provides monitoring and histogramming of output track parameters.
As in the \ac{EXTF} FPGA, these monitoring data are read out over the VME interface.
A top-level block diagram for the firmware is shown in Figure~\ref{fig:ssb_hw}.

Data arrive from each EXTF FPGA over a parallel \ac{LVDS} interface operating at 180~MHz.
The input logic extracts the event packets and presents them to a multiplexer called the Sync Engine.
The Sync Engine combines the event packets pertaining to the same event identifier.
The \ac{HW} track filter algorithm is then applied.
This algorithm writes every incoming track data to its own block of memory with corresponding comparison logic, and a FIFO, not shown in Figure~\ref{fig:ssb_hw}.
When each track arrives, it is simultaneously compared with all previously received tracks for that event.
Thus by the time the last track is received, all track comparisons have been performed with minimal additional latency.
Any detected match is marked.
Tracks are considered to be duplicated if the tracks come from the same sector and they have the same hit coordinates for at least eight layers.
If a duplicate is found, the track with more hits is kept.
If both tracks contain the same number of hits, the track with a smaller $\chi^2$ value is kept.
 
When the \ac{HW} algorithm FIFO is read out, any duplicate-track data are removed such that the data leaving the \ac{HW} only contains unique tracks.
These output tracks are formatted to be sent to either the \ac{ROS} or \ac{FLIC}.
The final version of the \ac{HW} firmware can handle an average of 141 tracks per SSB at an event rate of 100~kHz.
After duplicate-track removal, the \ac{HW} is able to maintain an average output event size of 28 tracks per SSB at 100~kHz.
Up to 20\% of the logic resources and memory are used in the FPGA design.
 
The full FTK system requires 32 SSB MAIN and 32 SSB RTM boards.
By the beginning of spring 2019, the required production boards were fully tested for integration at CERN and by the beginning of summer 2019, three additional spare SSB MAIN boards and five additional spare SSB RTM boards were at CERN.
 

\subsection{\acl{FLIC}}
\label{sec:flic}
The \ac{FLIC} is a custom ATCA card that interfaces the upstream \ac{FTK} components with the \ac{HLT}.
The \ac{FLIC} boards receive tracks via the eight 2~Gb/s optical links on the front panel, which are connected to the \acp{SSB}.
The key functionality of the \ac{FLIC} is to find and insert the global module IDs into the output by using lookup tables.
These are used by the HLT to obtain global coordinates from the local \ac{FTK} coordinates.
In addition, the output packet is formatted to the standard ATLAS event format before it is sent to the HLT farm.
This final packet includes an error word to inform the HLT of possible problems with the contents of the output packet.
Besides data processing, the \ac{FLIC} sends data over the backplane to multiple processor blades mounted in the same ATCA shelf in order to perform data quality monitoring.
 
\begin{figure}[htbp]
\centering
\includegraphics[width=0.53\linewidth]{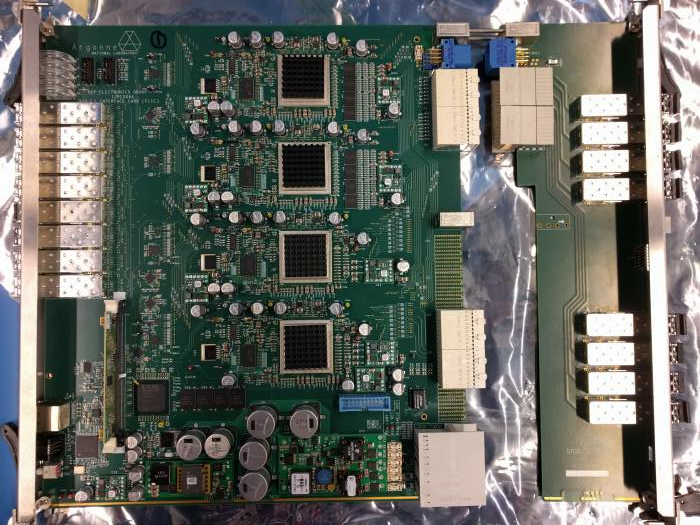}
\caption{The \acl{FLIC} (left) and its RTM (right).}
\label{fig:flic_card}
\end{figure}
 
\begin{figure}[htbp]
\centering
\includegraphics[width=0.7\linewidth]{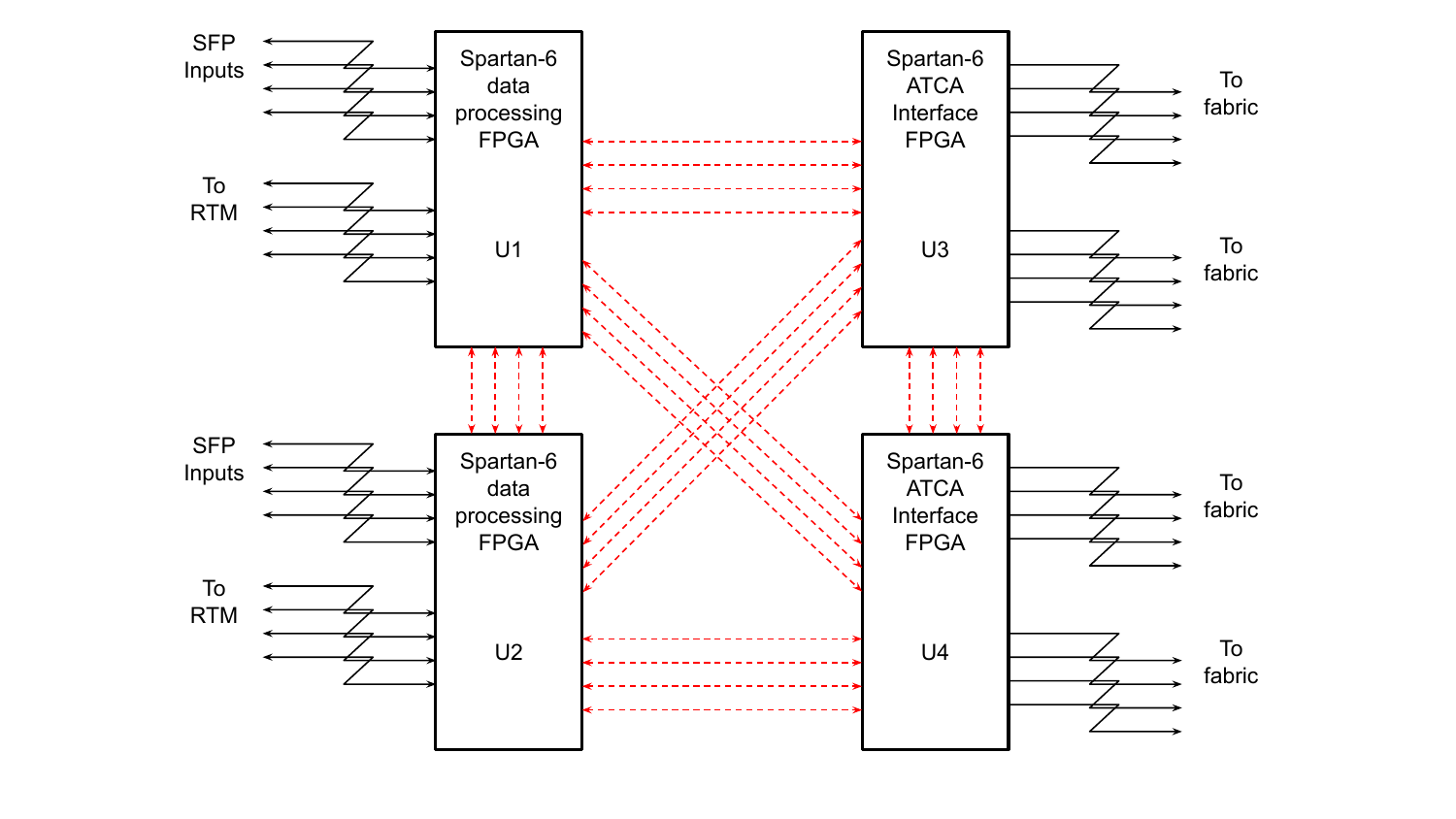}
\caption{Diagram of the connections including the internal mesh connecting each \acl{FLIC} FPGA to the others with four high-speed GTX lines. }
\label{fig:flic_internal_mesh}
\end{figure}

There are two \ac{FLIC} boards in the FTK system, hosted in the hub slots of the ATCA shelf, and two processor blades hosted in the payload slots.
Event packets can be sent to any of the processor blades via 10~Gb/s Ethernet.
The \ac{FLIC} board consists of the input card and the \ac{RTM}, shown in Figure~\ref{fig:flic_card}.
The input card has two areas, the data processing area and the management area.
Four Xilinx Spartan-6 75T-FGG676 FPGAs~\cite{Xilinx:Spartan6} are installed in the data processing area.
Two of them handle the data processing and the other two communicate with the processor blades over the ATCA backplane.
The first two are referred to as the processing FPGAs while the other two are referred to as the interface FPGAs.
The four FPGAs are connected to each other via a full internal mesh of high-speed \ac{LVDS} lines with a designed bandwidth of 4~Gb/s, shown in Figure~\ref{fig:flic_internal_mesh}.
 
The management area provides slow control and monitoring of the \ac{FLIC} board.
One small Spartan-3 S400-FG320 FPGA \cite{Xilinx:Spartan3}, one Microchip \ac{PIC}, four Flash RAMs, and one \ac{IPMC} card are located in the management area.
The Spartan-3 FPGA provides level translation and bus control logic to interface the microprocessor and Flash RAMs to the Spartan-6 FPGAs.
The firmware and lookup tables are stored in the Flash RAM.
Commands are sent through the front-panel Ethernet port to the \ac{PIC} to perform register and data accesses to all other devices by controlling the small FPGA.
All monitoring information can be transferred through the \ac{IPMC} to the shelf manager.
 
\begin{figure}[htbp]
\centering
\includegraphics[width=0.8\linewidth]{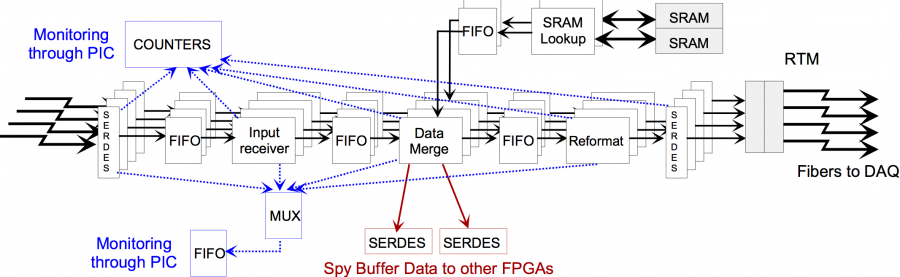}
\caption{Diagram of the data processing pipeline in the \acl{FLIC} processing FPGA.  The input lines (left) receive data from the \aclp{SSB}.}
\label{fig:flic_data_processing}
\end{figure}

Data processing in each processing FPGA has four pipelines.
Each pipeline receives data from one \ac{SFP} connection on the front panel and processes the data.
The core data processing functionality, ATLAS global module ID lookup, is performed with the help of 32~Mb of fast SRAM per input link.
Each track header has the address for the lookup embedded so that the lookup can be initialized immediately after the track starts being processed.
The lookup and track processing are done in parallel to ensure the module IDs are ready to be inserted when the track processing finishes.
The merge state machine combines the module IDs and the remaining event packet, formatted in the ATLAS event format.
The packets are then sent out from the \ac{RTM} to the HLT over eight fibres using 2~Gb/s S-LINKs.
The data processing pipeline is summarized in Figure~\ref{fig:flic_data_processing}.
For stand-alone diagnostic purposes and baseline performance tests, a second version of the firmware that emulates the output of the upstream FTK system can be loaded to either or both of the processing FPGAs.
FPGAs with the nominal firmware loaded can receive emulated event packets to perform a loop-back test.
 
The \ac{FLIC} also implements higher-level monitoring functionality where the two interface FPGAs copy selected fragments from the processing FPGAs via the
internal mesh.
Those fragments are then assembled by using the event identifier and sent to the processor blades using Ethernet.
Each pipeline only receives information from a part of the FTK system, while the assembled fragments contain information for the whole system, allowing advanced event-level monitoring.

\section{Online software, control and  monitoring}
\label{sec:online}
The \ac{FTK} online software integrates the \ac{FTK} system into the ATLAS trigger and data-taking infrastructure.
It provides features to configure, run and monitor the \ac{FTK} boards from the ATLAS \ac{RC} infrastructure, and handle the internal commands and synchronization operations.
 
\subsection{Run Control integration}
 
The ATLAS \ac{RC} is based on a \ac{FSM} guiding the entire ATLAS detector through  standby, configuration, and running states.
This general behaviour is inherited by the \ac{FTK} online software.
 
The \ac{FTK} software required a dedicated redesign and extension of the ATLAS \ac{FSM} in order to handle the special cases of the \ac{FTK} system.
For example, it takes about 50~minutes to configure and load the pattern banks into the 16 PUs in one VME crate, which exceeds the time allotted to configure the ATLAS detector.
The long configuration time is driven by the need to transfer about 500~MB/board with limited bandwidth and CPU resources.
This step was optimized to maximize the usage of the resources on the single-board computer in the VME crate and throughput over the VME backplane.
To accommodate this, two configuration transitions were implemented.
During the LHC inter-fill, pattern banks and fit constants are loaded onto the boards.
During data-taking, only pattern-bank and fit-constant checksums are computed for validation.
If a checksum validation failure occurs, the \ac{FTK} software prevents the \ac{FSM} transition from completing successfully.
 
The complexity of operations performed by the \ac{FTK} boards for different transitions required the development of a multi-threaded approach in order to fit within the ATLAS transition time windows.
This is particularly crucial for the \acp{AMB}, \acp{AUX}, and \acp{SSB} hosted in VME crates and controlled by a single-board computer where the bandwidth over the VME bus is a limiting factor.
The implemented design provides control of operations per board with options for parallelization, the number of threads, and thread queue management.
 
Another important aspect of the \ac{FTK} control software is to handle concurrent access to board resources.
In most cases, the firmware does not provide concurrent access management; therefore, access serialization was implemented at the software level by means of mutexes and system semaphores.
This approach has the cost of longer delays between operations such as state transitions and monitoring.
However, this did not impact data-taking operations.
 
The software infrastructure allows both the inclusion of the FTK in ATLAS data-taking runs while writing output to the ATLAS data stream and stand-alone parasitic runs writing to an independent data stream.
The latter was particularly useful early in the commissioning phase for performing rapid tests, without waiting for an entire LHC fill to retry a test and without negatively affecting ATLAS data-taking.
Near the end of Run~2, a vertical slice of FTK was regularly included in ATLAS runs.
Using the functionality of the \acp{IM}, the FTK system can selectively process events according to the result of the ATLAS L1 trigger.
During Run~2 a special trigger was configured to select events with muons that fell within the geometric coverage of the vertical FTK slice that was included in ATLAS data-taking.
 
\subsection{Automated procedures}
 
Two automatic procedures are required for integration into the ATLAS RC framework: Stopless-Removal and Stopless-Recovery.
The Stopless-Removal procedure removes a subsystem, or part of it, from the ATLAS FSM when it is interrupting data-taking, without changing the global FSM state.
In the \ac{FTK} software, this was implemented by means of an external application automatically triggered by missing event fragments from the \ac{FTK} subsystem.
This temporary solution proved to be particularly useful during the commissioning phase, where the software infrastructure was rapidly evolving.
 
Stopless-Recovery consists of a semi-automatic procedure to reinclude components previously excluded from the run, e.g.\ because of a Stopless-Removal.
As implemented in the \ac{FTK} system, this requires a full reconfiguration of the different boards before reinclusion of the system in the FSM and data-taking.
This reconfiguration procedure is managed by an internal ATLAS-like \ac{FSM} system.
 
\subsection{Monitoring}
 
Another important aspect of the \ac{FTK} online software is to provide the monitoring functionalities of the board status, on-going operations, and of the quality of the output data.
The first is performed by accessing specific board registers and publishing the information to the \ac{IS} and the \ac{OH}~\cite{TDAQ2016}.
Additional information is also propagated to higher-level monitoring applications, such as Grafana.
Dedicated functions are used to retrieve monitoring information from the boards.
Their parallelization can be controlled by assigning them to separate threads or by forcing serialization of the operations.
Easy access (e.g.\ via Grafana) to monitoring information and its history, ranging from basic counters to detailed firmware states, is extremely useful for commissioning, understanding system performance and tracking down bugs in the system.
 
Data quality information for each board, such as data snapshots and counters, is stored in dedicated spybuffers, which are read out periodically by online software monitoring functions.
The data are then published to the ATLAS \ac{EMON} service by means of a customized interface which provides temporary storage for multiple spybuffers from different boards that can be accessed on-demand for analysis and evaluation.
 
A freeze mechanism was implemented in order to block the overwriting of the spybuffers when an error occurs and during readout of spybuffers.
This functionality was crucial for accelerating firmware debugging and development by catching problematic events in situ.

\section{Configuration, optimization and performance}
\label{sec:simulation}
The FTK uses configuration data in the pattern matching and track fitting.
The production of this configuration and its optimization are enabled by a functional simulation of the FTK.
This simulation of the hardware provides a way to evaluate the performance and validate the output data from the hardware.
The \ac{FTK} technical design report \cite{FTKTDR} discusses the required tracking efficiencies for several physics cases.

\subsection{Simulation}
The simulation software for the FTK~\cite{Adelman:1951845} serves multiple purposes:
\begin{itemize}
\item support of hardware design decisions,
\item production and validation of the input pattern banks,
\item provision of sectors and constants needed by the hardware,
\item enabling of new trigger strategies that make use of the FTK,
\item and validation of the output of the FTK hardware, both online and offline.
\end{itemize}
It must simulate the full FTK chain, including clustering, data organization and distribution, pattern matching, track fitting in both the first and second stages, and any final FLIC processing.
The output of the FTK simulation, e.g.\ FTK tracks at either the first or second stages, can be stored in simple ROOT~\cite{BRUN199781} files as well as ATLAS-specific file formats.
The simulation is integrated into the ATLAS Athena software environment~\cite{athena}, and as such, the option to store FTK tracks inside of simulated events as if they were part of the real hardware output was also developed. 
This integration allows it to run in the ATLAS offline validation framework.
It also allows its use in reprocessing data to study trigger algorithms for use in the \ac{HLT}.
The FTK simulation is fully integrated and included by default in such reprocessings.
 
A big challenge for the FTK simulation is that it is slow to run on commodity hardware such as CPUs that
lack the parallelism found in the \ac{AM} chips for pattern matching and in the FPGAs used to compute linear fits.
Moreover, the databases of pattern banks require very large amounts of run-time memory
which can only be loaded slowly from the compressed on-disk storage format, making it unfeasible to process all FTK towers for a given event at the same time.
Typically, the constants and patterns required to simulate a single tower take up roughly 500~MB of memory.
 
To get around the challenges above, the FTK simulation uses a multi-step approach.
It starts by processing events, before any reconstruction, from the detector to create an intermediate output containing the inputs to the 64 towers of the FTK.
Next the full pattern-matching and simulation chain is run as 64 separate jobs.
The clustering algorithm can be run in either the first or second step.
The separate jobs for each tower are run serially, but configured and controlled by a single command, and their outputs are merged.
In this last step, the simulation of the duplicate-track removal in the hardware can also be applied.
 
The simulated output of the FTK system can also be merged with the original detector-level input file.
This creates a combined file including tracks as they might be found by the FTK.
This last step is very useful for studying how the FTK can be used as part of the trigger system.
The same simulation environment that is used for this full simulation of the FTK system is used to develop FTK pattern banks, as described below.

It is crucial to ensure that the pattern-matching and track-finding codes are robust and efficient, as these two steps are the slowest parts of the FTK simulation.
The linear fits and matrix algebra of the track-fitting code use the highly optimized Eigen libraries \cite{eigenweb}.
The number of patterns matched typically rises linearly with detector occupancy; however, the number of fits rises much faster than this, as it depends on the number of combinations across multiple layers and the number of hits in each \ac{SS}.
The CPU time is dominated by pattern matching at lower pile-up and by track finding at higher pile-up.
The exact balance depends on the pile-up value for which the pattern banks were optimized.
 
In order to simulate a large number of events for trigger studies, a `fast' simulation to emulate the full FTK performance was also developed.
The fast simulation starts with fully simulated tracks as found by \ac{HLT}-based tracking algorithms, which are always more efficient with respect to offline tracking than the \ac{FTK}.
It works by smearing the track parameters $d_0$, $z_0$, $q/p_\text{T}$, $\eta$ and $\phi$ to account for the loss of resolution due to using linear approximations and by randomly rejecting tracks using parameterizations from the full simulation to account for any FTK inefficiencies relative to the \ac{HLT} tracking.
In principle, similar fast simulations seeded by offline tracks or a Monte Carlo simulation record could be developed.
The limitation of the fast simulation is that it does not account for the true rate of fake tracks, which typically account for 1--2\% of reconstructed FTK tracks.
The full simulation described above, rather than this fast simulation, is used for the comparison in Section~\ref{sec:result}.
 
\subsection{Sectors and constants}
The sectors and constants needed by both the FTK hardware and simulation are derived from a full ATLAS simulation of single muons traversing the inner detector.
Muons are chosen because they do not scatter and interact as much as electrons and hadrons do.
Typically, ten muons per event were used, drawn from uniform distributions in the five track parameters.
The same ranges as for the pattern banks, described below, are used, except for
a slightly wider $\eta$ range in order to not lose any efficiency at the edge of the inner detector.
Around one billion muons in total are needed to produce sectors and constants with the highest track finding efficiencies.
 
When producing the fit constants, the FTK simulation provides an option to measure $d_0$ relative to either the beamspot or the origin.
Nominally, the beamspot option more closely reflects the running conditions, but it also requires the HLT to know which beamspot was used during the single-muon detector simulation.
Since the beamspot may shift, it was decided to always compute $d_0$ relative to the ATLAS coordinate system's origin and then update $d_0$ in accord with the beamspot in software.
During Run 2, the LHC beamspot was around 1~mm from the ATLAS origin in the transverse plane.
The best guess of the beamspot position is used when generating the patterns, described below.
A change in the beamspot position by 1~mm relative to the one used to produce the pattern banks results in a reduction of the tracking efficiency of up to 5\%.
Changes in the beamspot position during Run 2 were typically much smaller than this.
Over the course of a given year, the position moved by less than about 0.2~mm in the $x$ or $y$ coordinate.
 
The sectors and constants produced can be validated before beginning the time-consuming pattern-bank production by running the FTK simulation in `sector as pattern' mode.
In this configuration, the nominal pattern matching is not performed, and instead, all combinations of hits that fall within a sector are deemed to have fired a pattern.
This can be done for sectors in the first stage only, or sectors in the full 12-layers used by the \ac{SSB}.
This mode can only be used on samples of single particles without pile-up, otherwise the combinatorics of hits within a `sector as pattern' lead to too many track fits for the simulation to run.

A separate set of sectors and constants is produced to be used in the hardware.
This is needed because the nominal ATLAS simulation does not fully reproduce small detector misalignments.
Differences in the detector alignment between data and simulation would lead to large losses in efficiency and misreconstructed track parameters.
The ATLAS `data overlay' infrastructure~\cite{Haas_2017}, used to add minimum-bias events from real ATLAS zero-bias data to simulated hard-scatter events, is used to simulate single muons in the full ATLAS detector environment observed in data.

\subsection{Pattern-bank production}
\label{sec:pattern_prod}
FTK pattern-bank production begins with the generation of approximately $10^{11}$ pattern candidates (about $10^9$ per tower). 
Each pattern candidate corresponds to eight coarse resolution hit positions on different detector layers which are traversed by a track.
The tracks used to produce the candidate patterns are drawn from uniform distributions in the five track parameters: $d_0 \in [-2.2, 2.2]~\si{\milli\meter}$, $z_0 \in [-120, 120]~\si{\milli\meter}$, $q/p_\text{T} \in [-0.8, 0.8]~\si{\giga\electronvolt}^{-1}$, $|\eta|<3$, and $\phi \in [-\pi, \pi]$.
Due to the large number of required pattern candidates, determining the corresponding hit positions using the full ATLAS tracker simulation is not feasible.
Instead, the previously determined sectors and fit constants are used to estimate the hit positions of each track.
 
It is possible that a single track is compatible with multiple valid 8-layer patterns if it traverses a region of overlapping silicon modules.
In this case, the pattern corresponding to the sector with the most tracks in the sectors-and-constants production step is selected.
 
The multiplicity of tracks producing the same pattern is referred to as the pattern's coverage.
Patterns with a high coverage are the most important for finding tracks.
The full set of unique pattern candidates, ordered by region, coverage, and sector, is called the \ac{TSP} bank.
 
The standard 128-PU FTK configuration allows $2^{24}$ (16.8 million) patterns per tower to be stored in the AM chips.
The use of \ac{DC} bits (see Section~\ref{sec:AM}) increases the effective number of patterns stored in the available hardware.
Including a DC bit within a pattern results in that pattern being valid for two hit numbers in the respective layer as that bit is always considered a match.
Setting a total number of $N_{X}$ \ac{DC} bits across all eight layers increases the effective number of patterns stored in the AM's space for a single pattern to $2^{N_X}$.
For example if a single \ac{DC} bit is used on two layers of a single pattern, then it is effectively covering four patterns.
 
With a larger effective number of patterns stored in the AM chips via the use of DC bits, the FTK's track-finding efficiency will increase.
However, patterns with many DC bits will suffer from poorer spatial resolution and a higher rate of combinatorial background (fake tracks).
Therefore, the maximum number of DC bits per pattern, summed across all layers, is limited to a fixed number ($N_X\leq N_{X,\text{max}}$).
This maximum number is adjusted to ensure that the dataflow is kept within the hardware limits.
The maximum number of DC bits per pattern is defined separately for the barrel ($N_{X,\text{max}}^\text{B}$) and endcap ($N_{X,\text{max}}^\text{EC}$) regions, and both are discussed in the following sections.
The barrel region consists of the two innermost towers in $\eta$, and the remaining outer towers are in the endcap region.
 
Patterns are selected for storage on the AM chip in order of decreasing coverage.
Where possible, patterns are merged with previously stored patterns using DC bits.
This process continues until all AM addresses are filled and there is no room to include further pattern candidates.
The full set of $2^{24}$ patterns including DC bits, determined for each of the 64 regions, is termed the AM pattern bank.
 
\begin{figure}[htbp]
\centering
\includegraphics[scale=0.4]{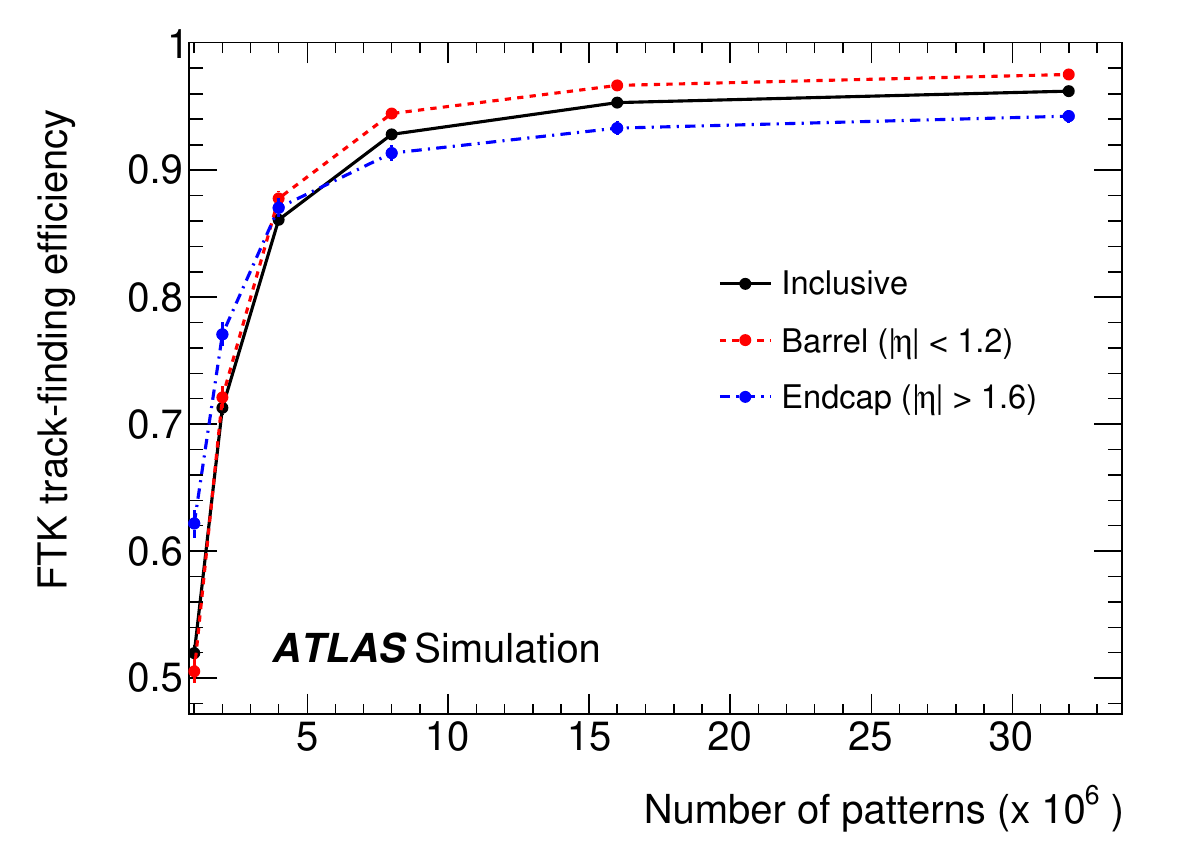}
\caption{Track-finding efficiency, relative to the Monte Carlo `truth record', of the \acl{AM} pattern bank as a function of the number of patterns in the bank.
Tracks are taken from a single-muon Monte Carlo simulation sample and are required to have $|z_0|<110$~mm and |$d_0|<2.2$~mm.
The efficiency is calculated separately for all tracks with $|\eta|<2.5$ (black), for tracks in the barrel region with $|\eta|<1.2$ (red), and for tracks in the endcap region with $1.6<|\eta|<2.5$ (blue).
The barrel/endcap transition region $1.2<|\eta|<1.6$ is excluded from the latter two as it has a low efficiency before mitigation is applied.
}
\label{fig::effVspatt}
\end{figure}

The efficiency of an AM pattern bank, as a function of the number of patterns in the bank, is shown in Figure~\ref{fig::effVspatt}.
The inclusive curve does not always lie between the others because of the inclusion of the barrel/endcap transition region $1.2<|\eta|<1.6$, which has a low efficiency.
Mitigation of this is discussed in the next section.
With 16.8 million patterns per tower in the 128-PU FTK configuration, the track-finding efficiency is on the plateau of the efficiency curve.
 
\subsection{Pattern-bank optimization}
 
\begin{figure}[htbp]
\centering
\includegraphics[scale=0.4]{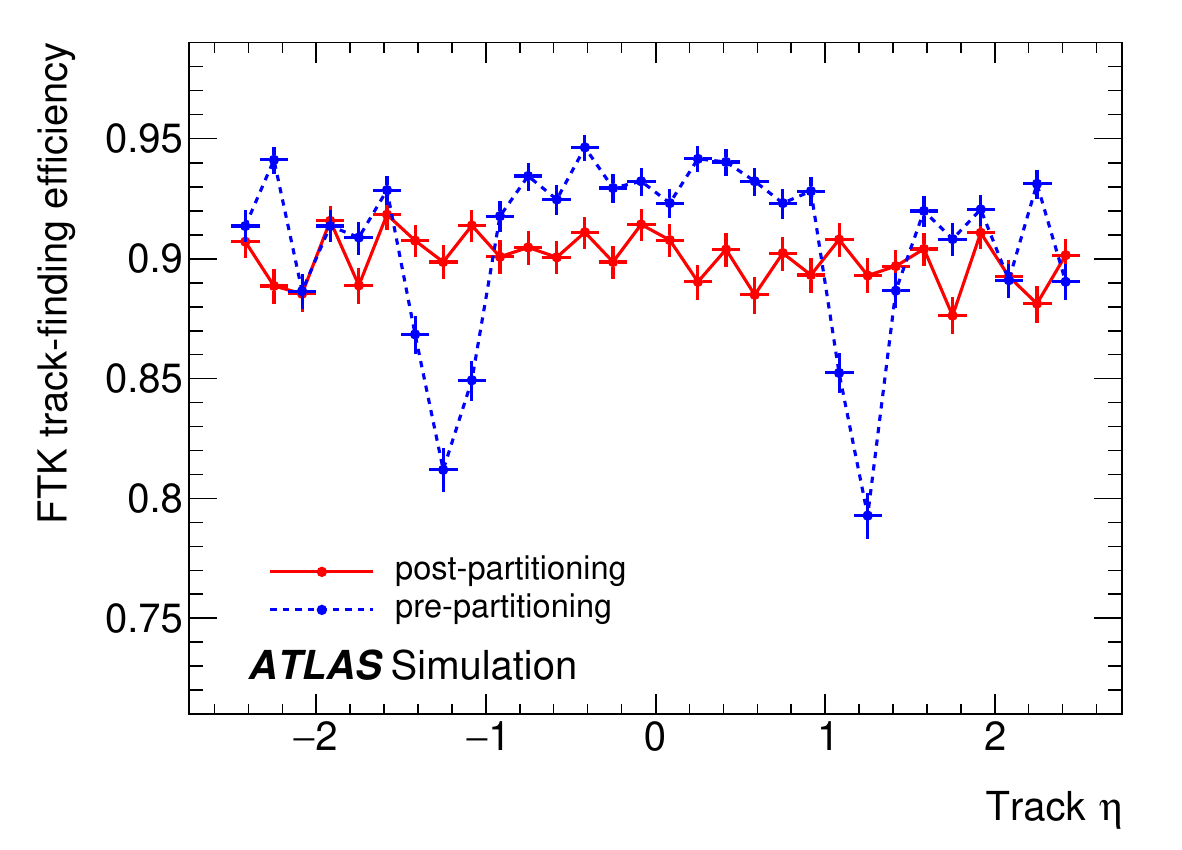}
\caption{FTK track-finding efficiency, relative to the Monte Carlo `truth record', as a function of the track $\eta$, before and after applying partitioning. Tracks are taken from a single-muon Monte Carlo sample and are required to have $|z_0|<110$~mm and |$d_0|<2.2$~mm. The pattern bank is created for a 64-Processing-Unit configuration, with $N_{X,\text{max}}^\text{B}=9$ and $N_{X,\text{max}}^\text{EC}=6$.}
\label{fig::trackparams_part}
\end{figure}

The track-finding efficiency of an AM bank as a function of track $\eta$ is shown in blue in Figure~\ref{fig::trackparams_part}.
In this example, a pattern bank designed for the 64-PU configuration with $N_{X,\text{max}}^\text{B}=9$ and $N_{X,\text{max}}^\text{EC}=6$ is used.
Two drops in efficiency are observed at $\eta\approx \pm1.2$, which is related to the change in geometry from the barrel ($|\eta|< 1.2$) to endcap disk ($1.6<|\eta|< 2.5$) detector layers in these regions.
To make the efficiency more uniform as a function of $\eta$, \ac{TSP} patterns are reassigned to AM addresses in bins of $\eta$.
A larger budget of AM addresses is assigned to $\eta$ bins with low efficiency.
To keep the overall number of AM addresses constant, the budget of AM addresses in $\eta$ bins with a high efficiency is lowered accordingly.
This iterative process, called partitioning, is repeated until the track-finding efficiency is roughly constant as a function of $\eta$.
The track-finding efficiency of the AM bank after partitioning is shown in red in Figure~\ref{fig::trackparams_part}.
The remaining variation in the efficiency is likely due to a binning-like effect of the procedure where the sectors are assigned to one $\eta$ bin, while in reality they have a shape in $\eta$.
The relative change in efficiency with more patterns is roughly similar to that shown in Figure~\ref{fig::effVspatt}.
Thus, in order to improve the efficiency from roughly 80\% to 90\%, the number of patterns needs to be increased by a factor of 3--4.

\subsection{Production of a pattern bank for commissioning of the 128-PU system}
In 2018, the FTK was commissioned with a pattern bank compatible with a 64-PU system, with $N_{X,\text{max}}^\text{B}=9$ and $N_{X,\text{max}}^\text{EC}=6$.
The full FTK system is designed to have 128-PUs, allowing the storage of twice as many patterns.
To facilitate the scaling up of the FTK system from a 64-PU to a 128-PU system, a new pattern bank was created.
With more patterns in the pattern bank, a higher track-finding efficiency is expected.
However, more patterns also results in a higher dataflow rate.
In order to keep the dataflow within the limits of the hardware, the maximum number of DC bits was adjusted.
 
Two 128-PU pattern banks were created with different DC bit configurations.
The configuration with fewer DC bits ($N_{X,\text{max}}^\text{B}=7$ and $N_{X,\text{max}}^\text{EC}=4$) has narrower patterns and therefore a lower dataflow rate, while the configuration with more DC bits ($N_{X,\text{max}}^\text{B}=9$ and $N_{X,\text{max}}^\text{EC}=6$) has wider patterns and a higher dataflow rate.
The overall track-finding efficiencies of these two 128-PU banks, compared with the 64-PU bank used previously for commissioning, are listed in Table~\ref{tab::128pu}.
For commissioning, a choice was made to deliberately push the FTK system to its performance limits.
The high-dataflow 128-PU bank, with the highest track-finding efficiency, has the highest dataflow rate.
Compared to the lower dataflow configuration, at a pile-up of 60, the high-dataflow bank results in about a factor of 2 increase in the number of roads, a factor of 1.4--1.5 increase in the number of 8-layer and 12-layer track candidates, and a factor of 1.1 increase in the number of 12-layer track candidates after overlap removal.
To run this configuration at 100~kHz would likely have required reducing the number of track candidates by tuning of the number of fits performed.

\begin{figure}[htbp]
\centering
\subfloat[]{\includegraphics[scale=0.4]{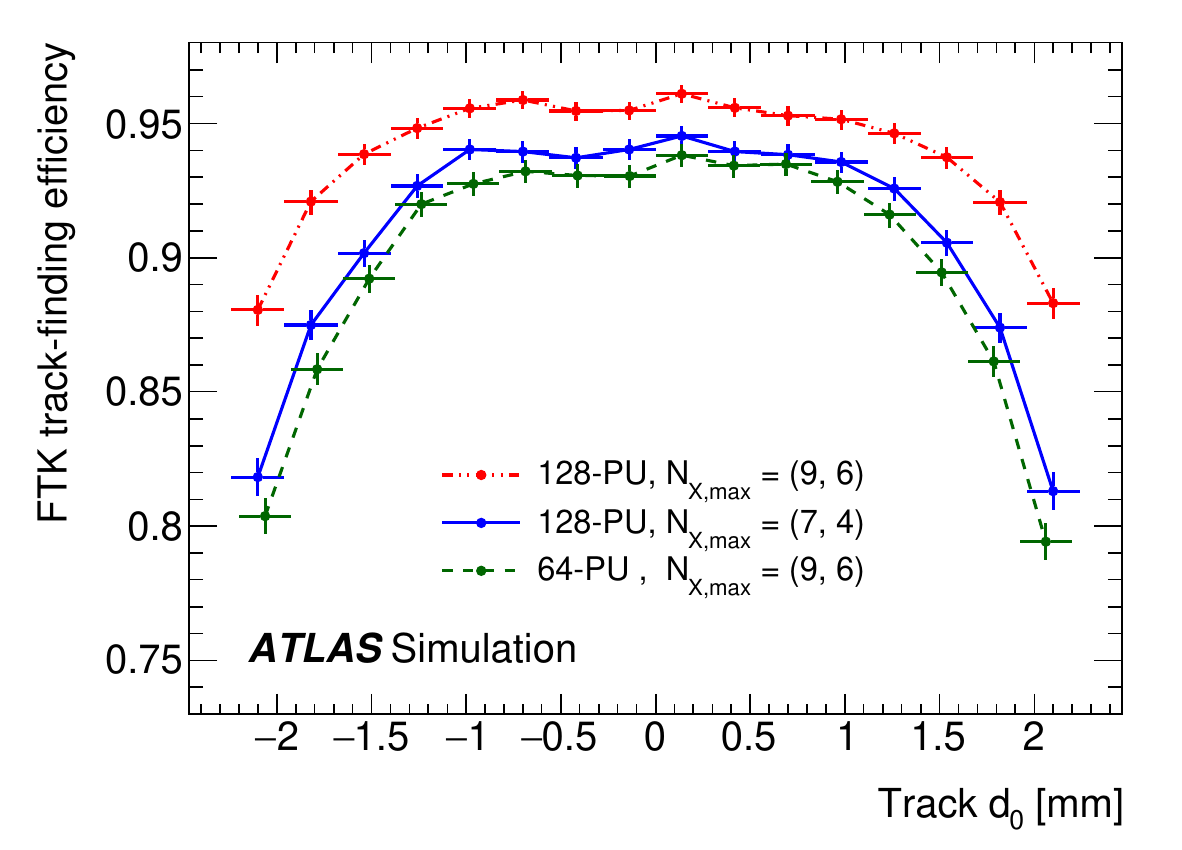}}
\subfloat[]{\includegraphics[scale=0.4]{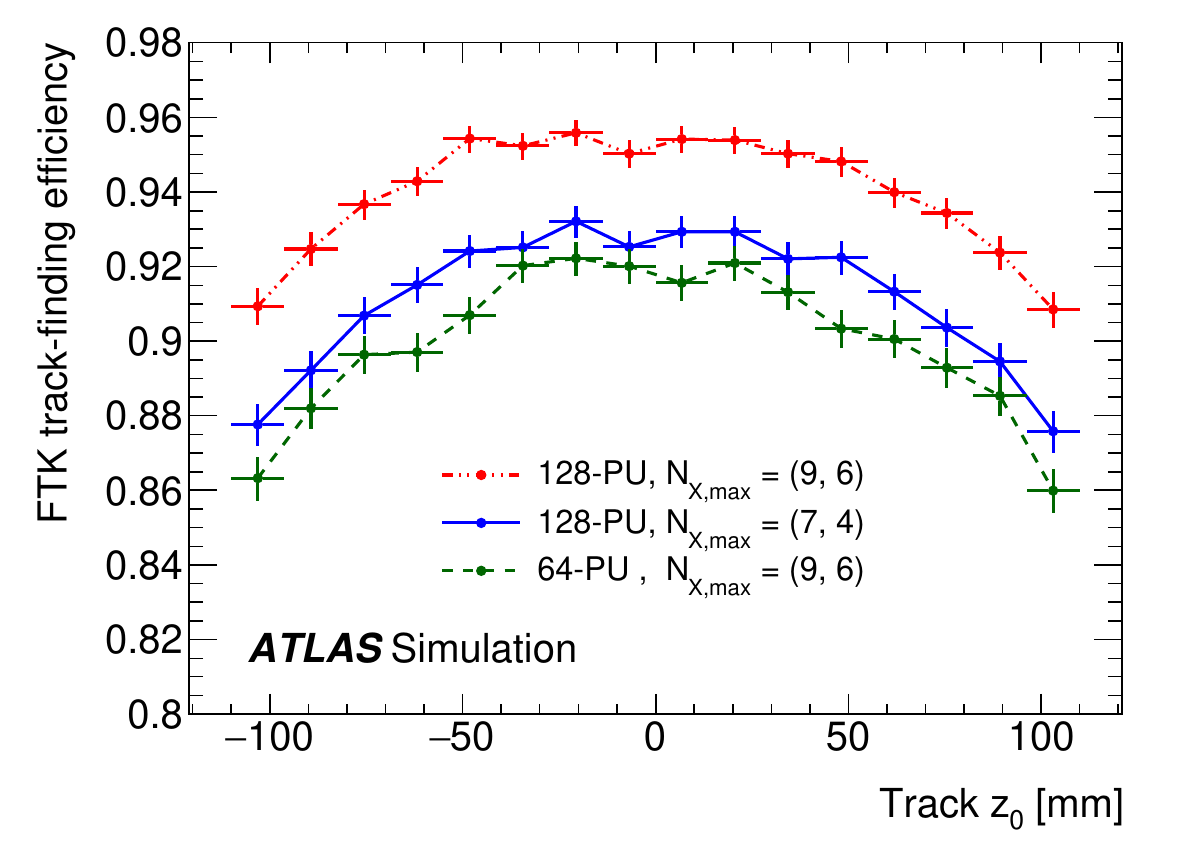}}
\caption{FTK track-finding efficiency with respect to the Monte Carlo `truth record' as a function of the (a) $d_0$ and (b) $z_0$ of tracks in a single-muon Monte Carlo sample satisfying $|\eta|<2.5$,  $|z_0|<110$~mm and |$d_0|<2.2$~mm. The \acl{DC} bit configuration, $N_{X,\text{max}}$~=~$(N_{X,\text{max}}^\text{B},N_{X,\text{max}}^\text{EC})$, and the number of processing units for each pattern bank are specified in the legend.}
\label{fig::trackparams_d0z0}
\end{figure}

\begin{figure}[htbp]
\centering
\subfloat[]{\includegraphics[scale=0.4]{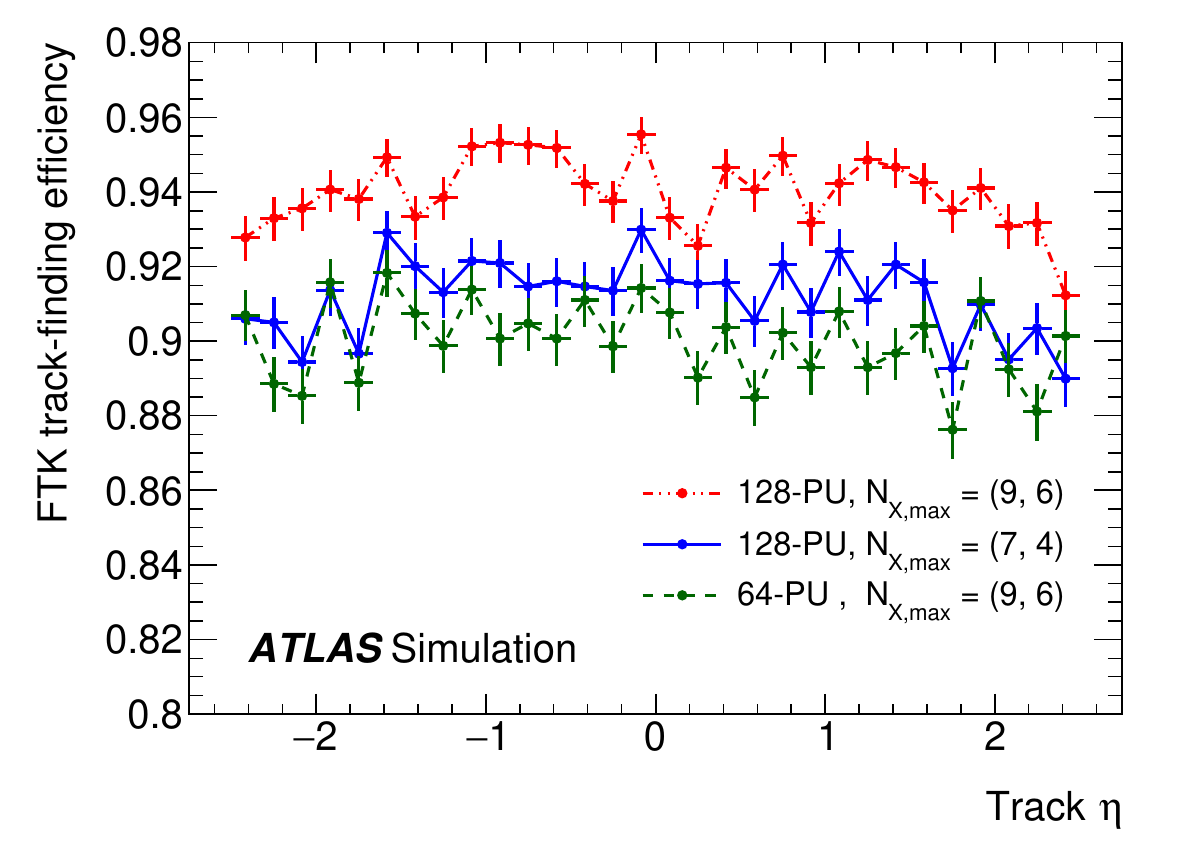}}
\subfloat[]{\includegraphics[scale=0.4]{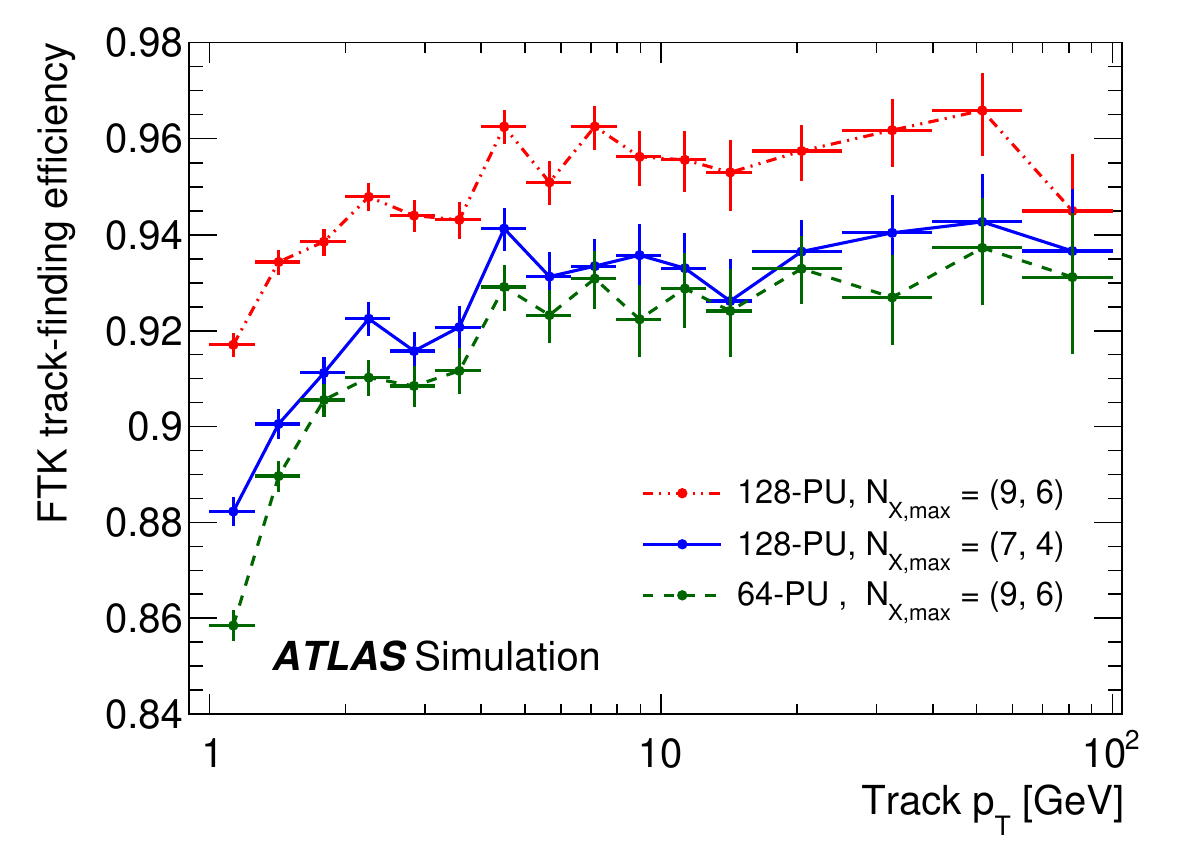}}
\caption{FTK track-finding efficiency with respect to the Monte Carlo `truth record' as a function of the (a) $\eta$ and (b) $p_\text{T}$ of tracks in a single-muon Monte Carlo sample satisfying $|\eta|<2.5$,  $|z_0|<110$~mm and |$d_0|<2.2$~mm. The \acl{DC} bit configuration, $N_{X,\text{max}}$~=~$(N_{X,\text{max}}^\text{B},N_{X,\text{max}}^\text{EC})$, and the number of processing units for each pattern bank are specified in the legend.}
\label{fig::trackparams_etapt}
\end{figure}

The track-finding efficiency of these banks, as a function of track parameter values, is shown in Figures~\ref{fig::trackparams_d0z0}~and~\ref{fig::trackparams_etapt}.
The track-finding efficiency peaks for tracks with low impact parameters ($d_0$ and $z_0$) and high $p_\text{T}$, and stays roughly constant as a function of track $|\eta|$.
The high-dataflow 128-PU bank offers the highest track-finding efficiency across the entire range of all track parameters, with the largest improvement at large $d_0$ and low $p_\text{T}$.
 
\begin{table}[htbp]
\centering
\caption{Track-finding efficiency of pattern banks with respect to `truth record' tracks in a single-muon Monte Carlo sample. Only tracks with $|\eta|<2.5$,  $|z_0|<110$~mm and |$d_0|<2.2$~mm are considered in the efficiency calculation. Each column indicates the \acl{DC} bit configuration of the bank $N_{X,\text{max}}=(N_{X,\text{max}}^\text{B},N_{X,\text{max}}^\text{EC})$.}
\vspace{0.2cm}
\begin{tabular}{c|c|c}
\hline
\multicolumn{3}{c}{Track-finding efficiency of various pattern-bank configurations} \\
\hline
64-PU, $N_{X,\text{max}}=(9, 6)$ & 128-PU, $N_{X,\text{max}}=(7, 4)$ & 128-PU, $N_{X,\text{max}}=(9, 6)$ \\
\hline
90.0\% & 91.2\% & 94.4\% \\
\hline
\end{tabular}
\label{tab::128pu}
\end{table}
 
\subsection{Extension of the pattern bank to high-$d_0$ tracks}
 
As described in Section~\ref{sec:pattern_prod}, the pattern-bank training is performed using a flat distribution of $d_0$ in order to ensure good coverage for displaced tracks associated with objects such as jets originating from $b$-hadrons.
Since the nominal banks are trained on tracks with $|d_0| < 2.2$~mm, the efficiency is approximately flat for $|d_0| < 1.2$~mm and falls off steeply outside that range, as shown in Figure~\ref{fig::trackparams_d0z0}(a).
 
In order to increase the trigger efficiency for possible new long-lived particles, an alternative pattern bank was developed.
It has 30$\%$ of the patterns dedicated to high-$d_0$ tracks, extending the coverage to $|d_0| < 10$~mm.
This high-$d_0$ pattern bank was constructed for 128-PUs, a \ac{DC} bit configuration with $N_{X,\text{max}}^\text{B}=7$ and $N_{X,\text{max}}^\text{EC}=4$ and with nominal detector geometry and conditions.
In order to compensate for the increase in the number of patterns required to cover this larger $d_0$ space, the minimum track \pt was increased to 5~\gev\ for $|d_0| > 2$~mm.
 
\begin{figure}[htbp]
\centering
\includegraphics[width=0.7\textwidth]{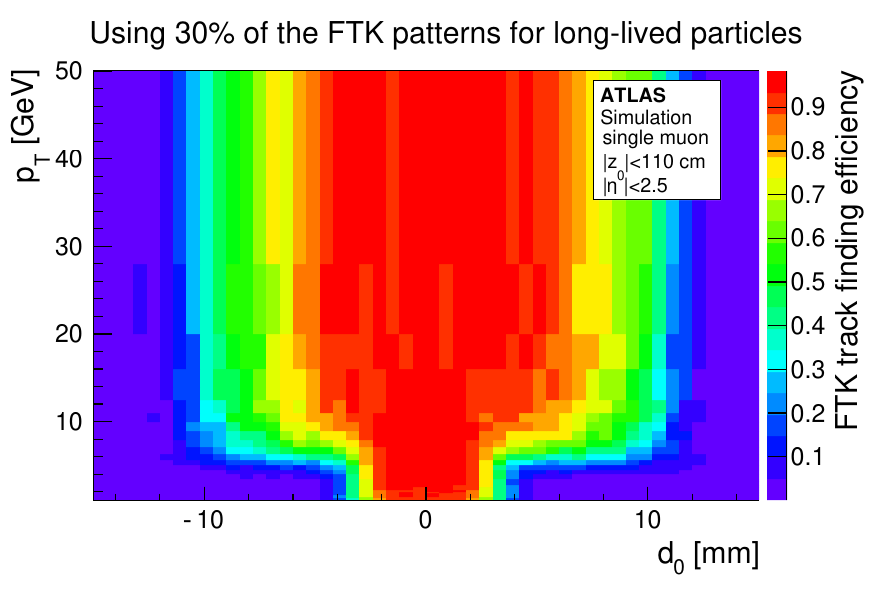}
\caption{FTK track-finding efficiency with respect to the Monte Carlo `truth record' as a function of the track \pt and $d_0$ for the extended $d_0$ pattern bank. Tracks are taken from a single-muon Monte Carlo sample satisfying $|\eta|<2.5$ and $|z_0|<110$~mm. The pattern bank is created with a 128-Processing-Unit configuration, with $N_{X,\text{max}}^\text{B}=7$ and $N_{X,\text{max}}^\text{EC}=4$.}
\label{fig::highd0_2d}
\end{figure}

The resulting track-finding efficiency as a function of both \pt and $d_0$ is shown in Figure~\ref{fig::highd0_2d}.
In the regions targeted by the training, the FTK track-finding efficiency is over 50$\%$, and for tracks with $|d_0| < 5$~mm and $\pt > 10$~\gev, the efficiency is over 90$\%$.
 
This bank was produced without reoptimizing the sectors.
Sectors correspond to a width of approximately 10~mm, so it is likely that such a reoptimization could increase the efficiency at high $d_0$, but this was not studied explicitly.
 
\begin{figure}[htbp]
\centering
\subfloat[]{\includegraphics[scale=0.4]{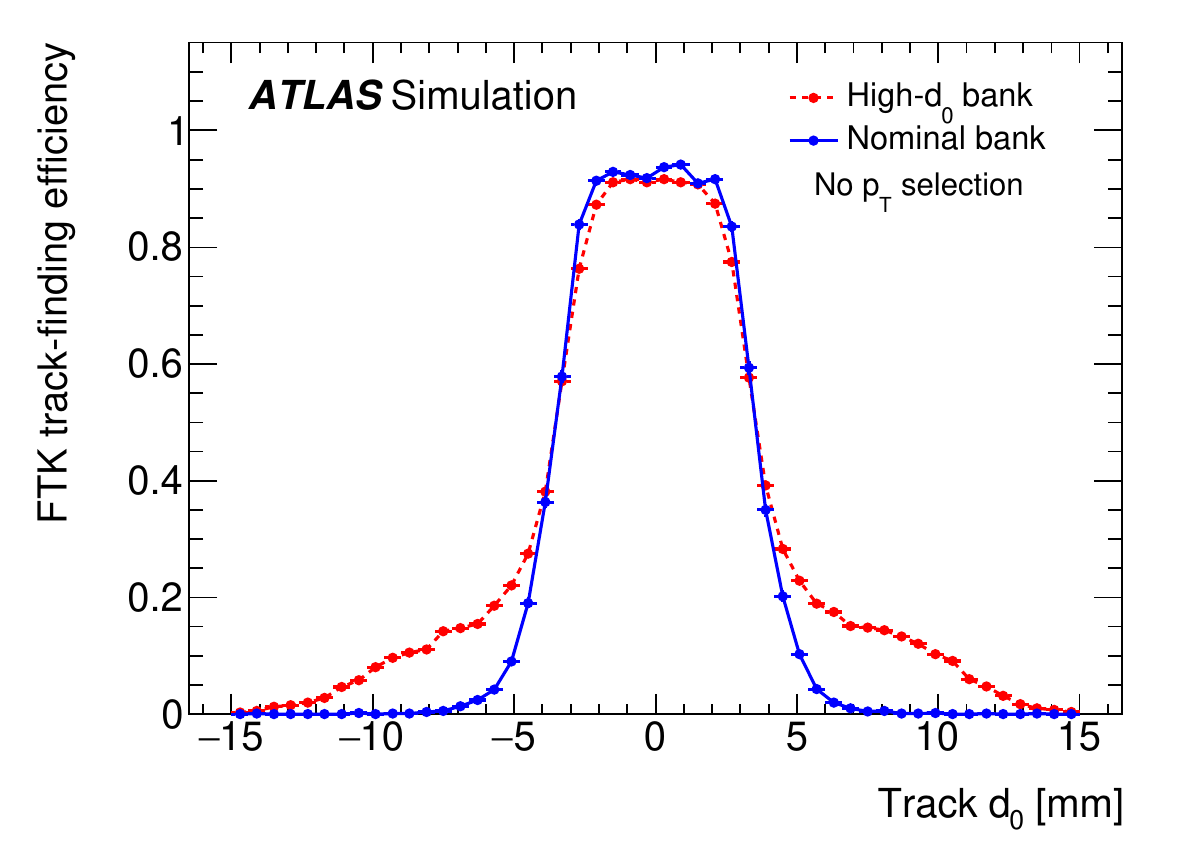}}
\subfloat[]{\includegraphics[scale=0.4]{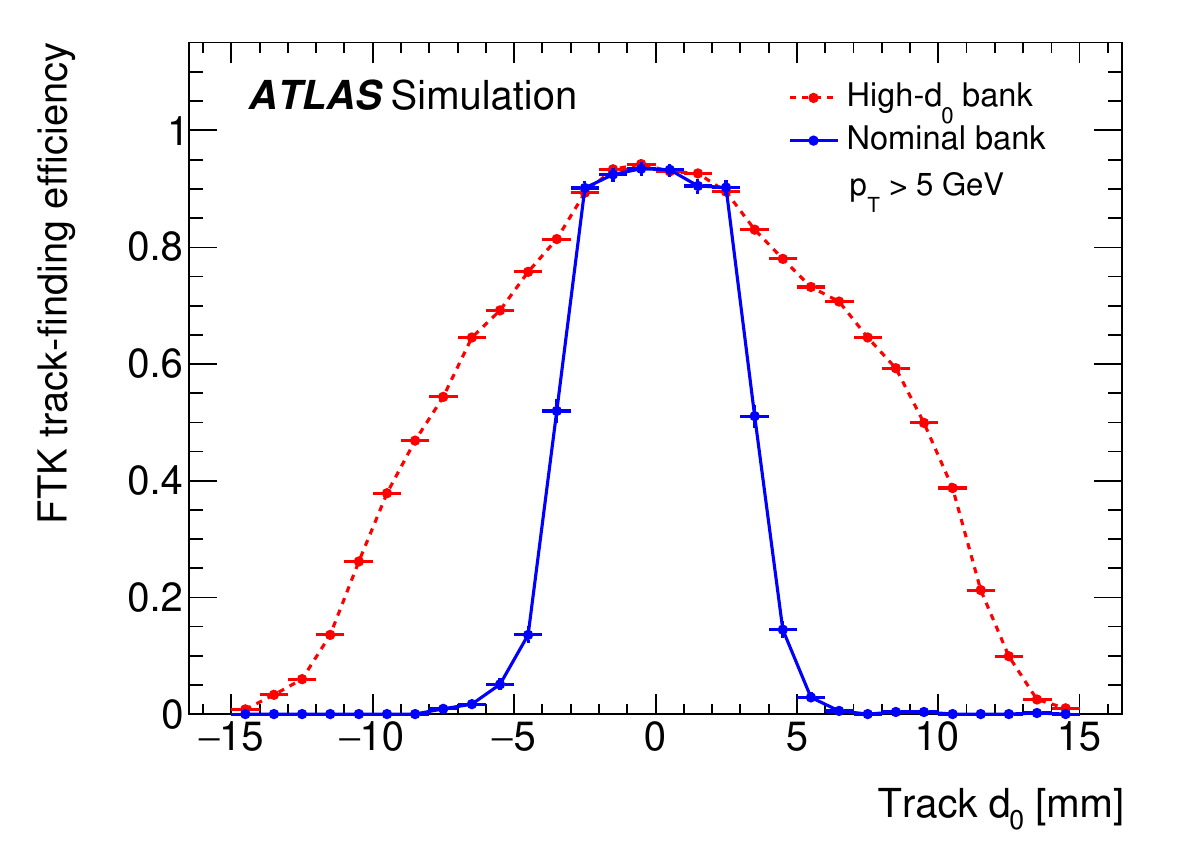}}\\
\subfloat[]{\includegraphics[scale=0.4]{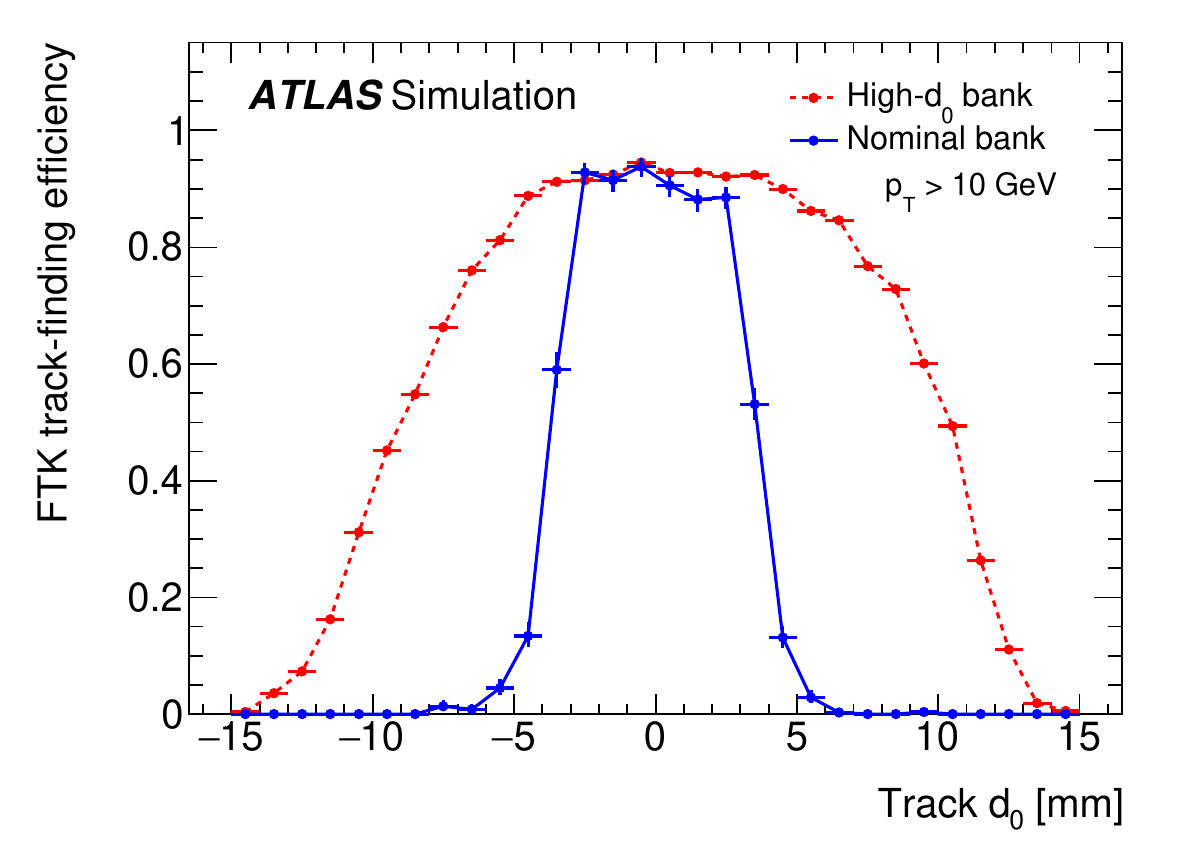}}
\caption{FTK track-finding efficiency with respect to the Monte Carlo `truth record' as a function of $d_0$ for two pattern banks, with (a) no track $\pt$ selection, (b) track $\pt>5$~GeV and (c) track $\pt>10$~GeV. Tracks are taken from a single-muon Monte Carlo sample and are required to have $|\eta|<2.5$,  $|z_0|<110$~mm. The nominal bank (blue) only includes patterns for tracks with $|d_0|<2$~mm, and the high-$d_0$ bank (red) has 30$\%$ of its patterns dedicated to tracks with $2 < |d_0| < 10$~mm and $\pt > 5$~\gev. Both banks are constructed with 128-Processing-Unit configuration, a \acl{DC} bit configuration with $N_{X,\text{max}}^\text{B}=7$ and $N_{X,\text{max}}^\text{EC}=4$, and with the nominal detector geometry and conditions.}
\label{fig::highd0_comp}
\end{figure}

The patterns dedicated to nominal tracks with the highest coverage are preferentially kept, so the 30$\%$ reduction in the number of nominal patterns does not translate into a 30$\%$ efficiency drop in the nominal low-$d_0$ region.
Figure~\ref{fig::highd0_comp} shows the FTK track-finding efficiency as a function of $d_0$ for the high-$d_0$ pattern bank, compared with a pattern bank produced with the exact same configuration but without any high-$d_0$ patterns.
The track-finding efficiency of the high-$d_0$ bank is reduced by approximately 2\% compared to the nominal bank within the nominal $|d_0|<2$~mm region. For high-$\pt$ tracks, this reduction in efficiency is even smaller.
 
\subsection{Comparison of data and simulation}
 
\label{sec:result}
In Run~2, the inputs to the \ac{FTK} vertical slice were recabled to maximize the track-finding efficiency of a particular central tower without data sharing between multiple \acp{DF}.
The hardware used corresponds to the boards needed for the `A' tower data-path in Figure~\ref{fig:overview:four-tower-system}, as described in Section~\ref{sec:hardware}.
This central tower, number 22, covers roughly $1/64$th of the ATLAS tracking volume, covering $-1.6 < \eta < 0.1$ and $1.5 < \phi < 2.1$.
The data used in the following analysis were collected using 64-PU pattern banks.
During LS2, previously recorded data from the ATLAS detector can be replayed via the \ac{QNP} system (see Section~\ref{sec:questnp}) through the \ac{FTK} vertical slice in the same way as they would have during Run~2.
 
\begin{figure}[htb]
\centering
\includegraphics[width=0.45\textwidth]{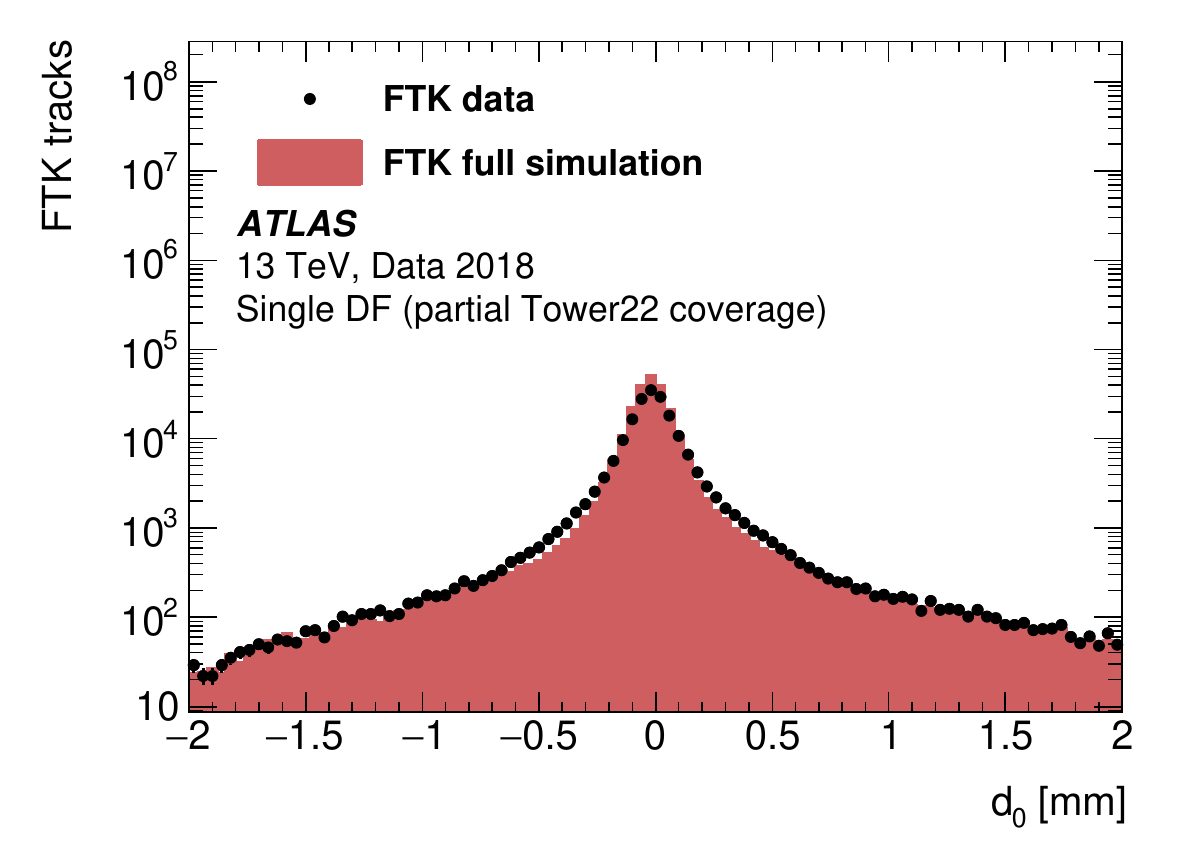}
\includegraphics[width=0.45\textwidth]{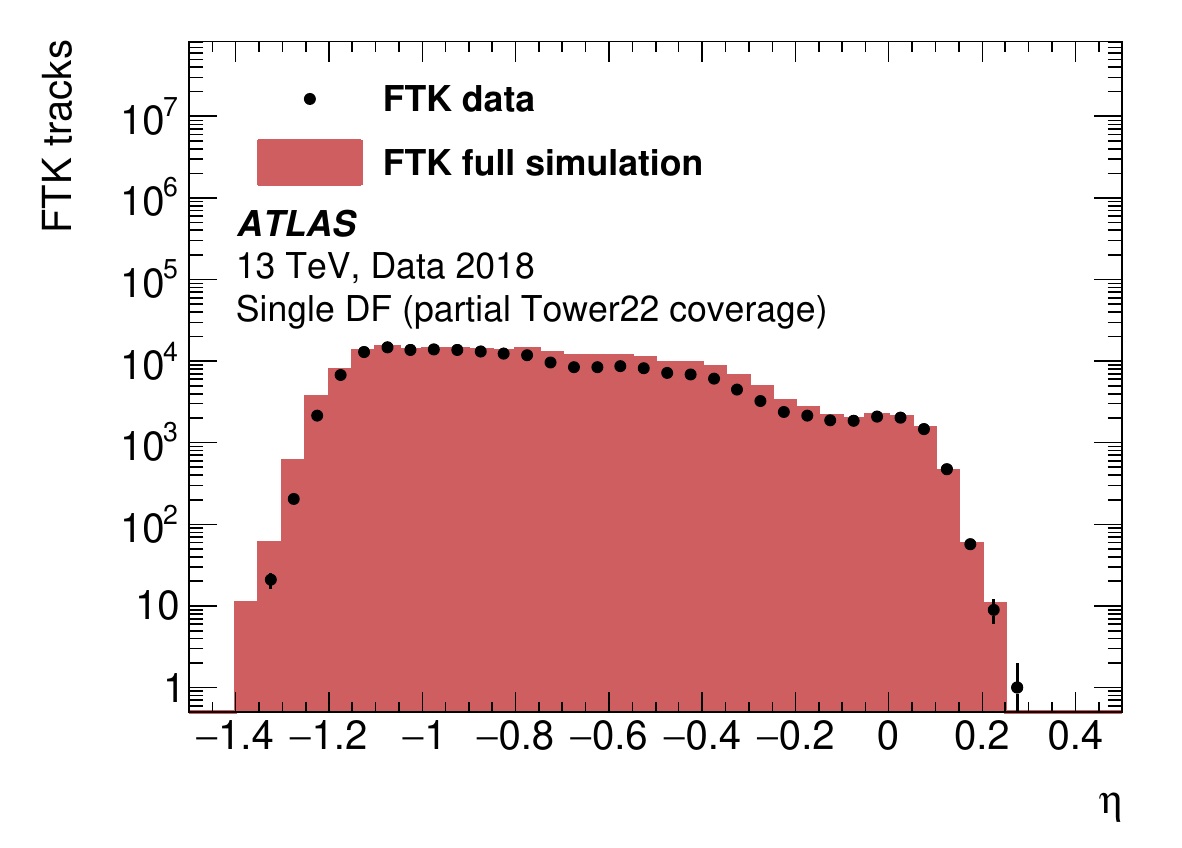}
\includegraphics[width=0.45\textwidth]{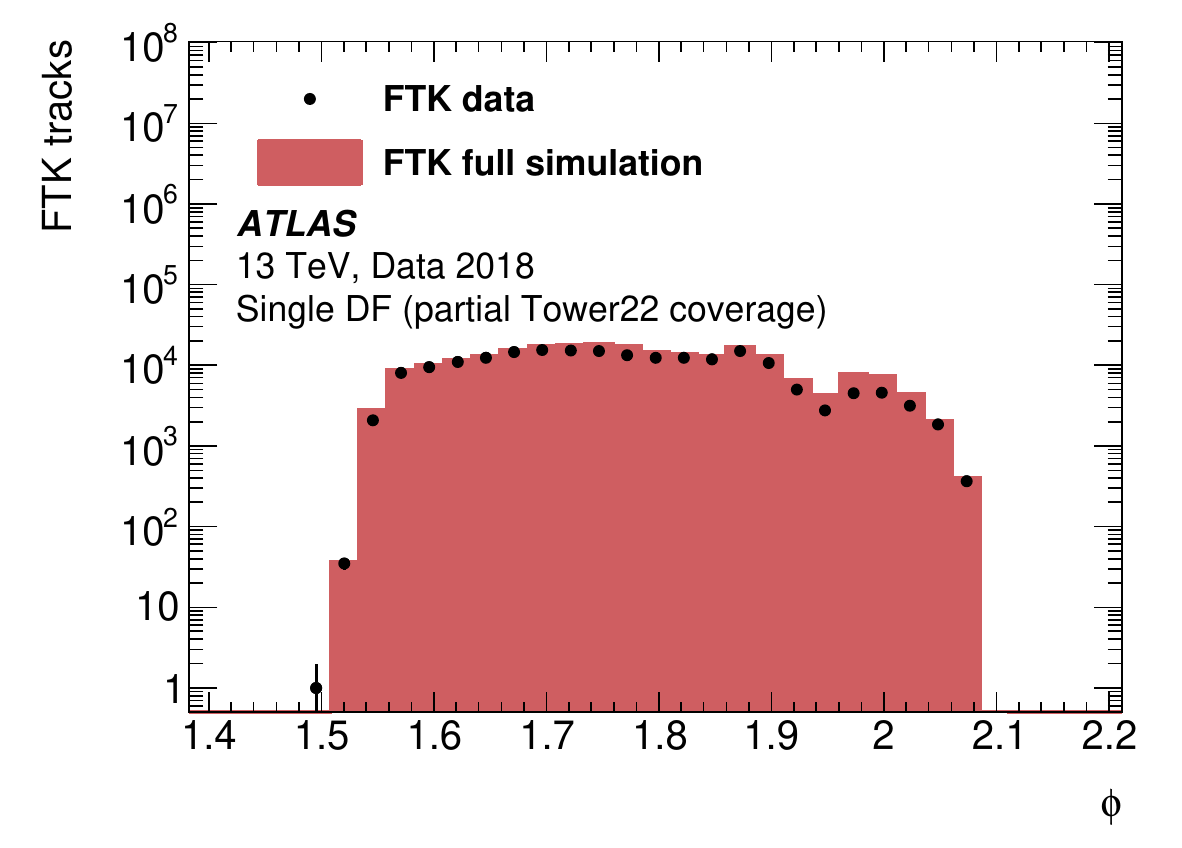}
\includegraphics[width=0.45\textwidth]{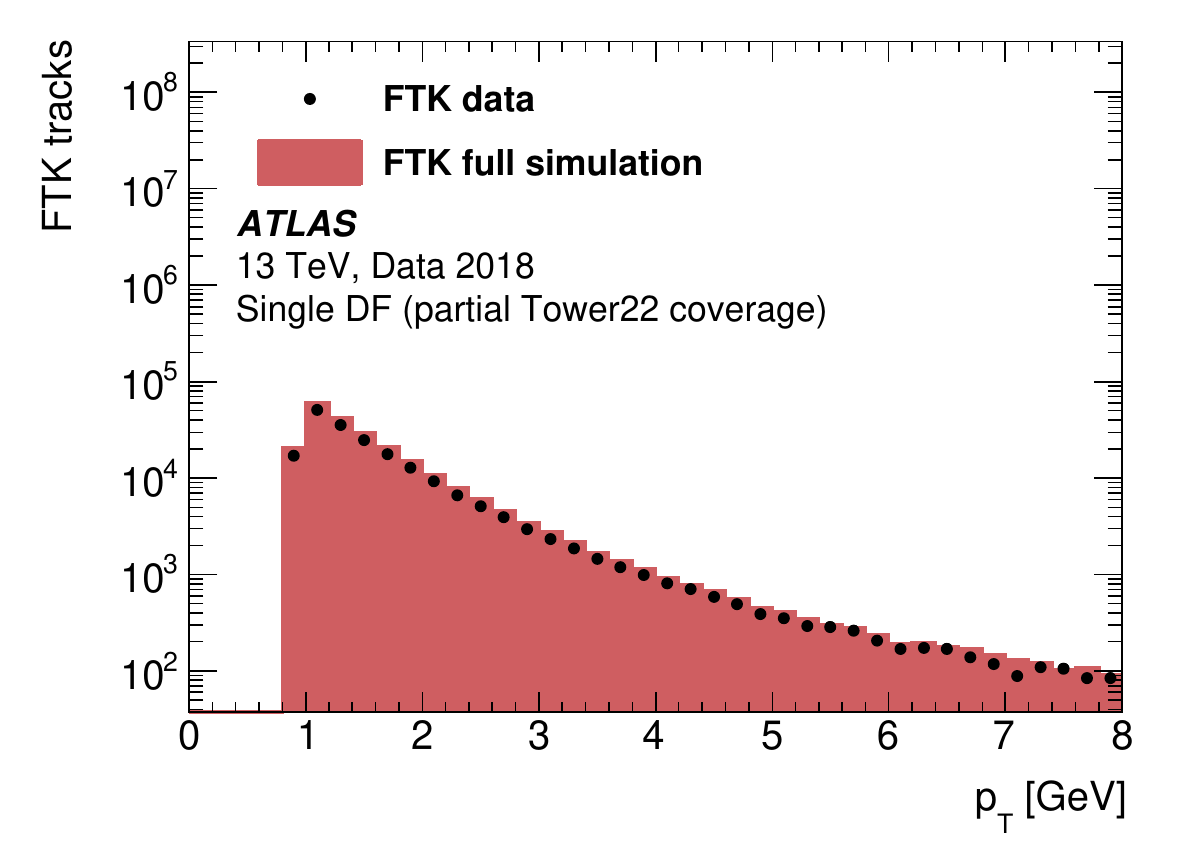}
\includegraphics[width=0.45\textwidth]{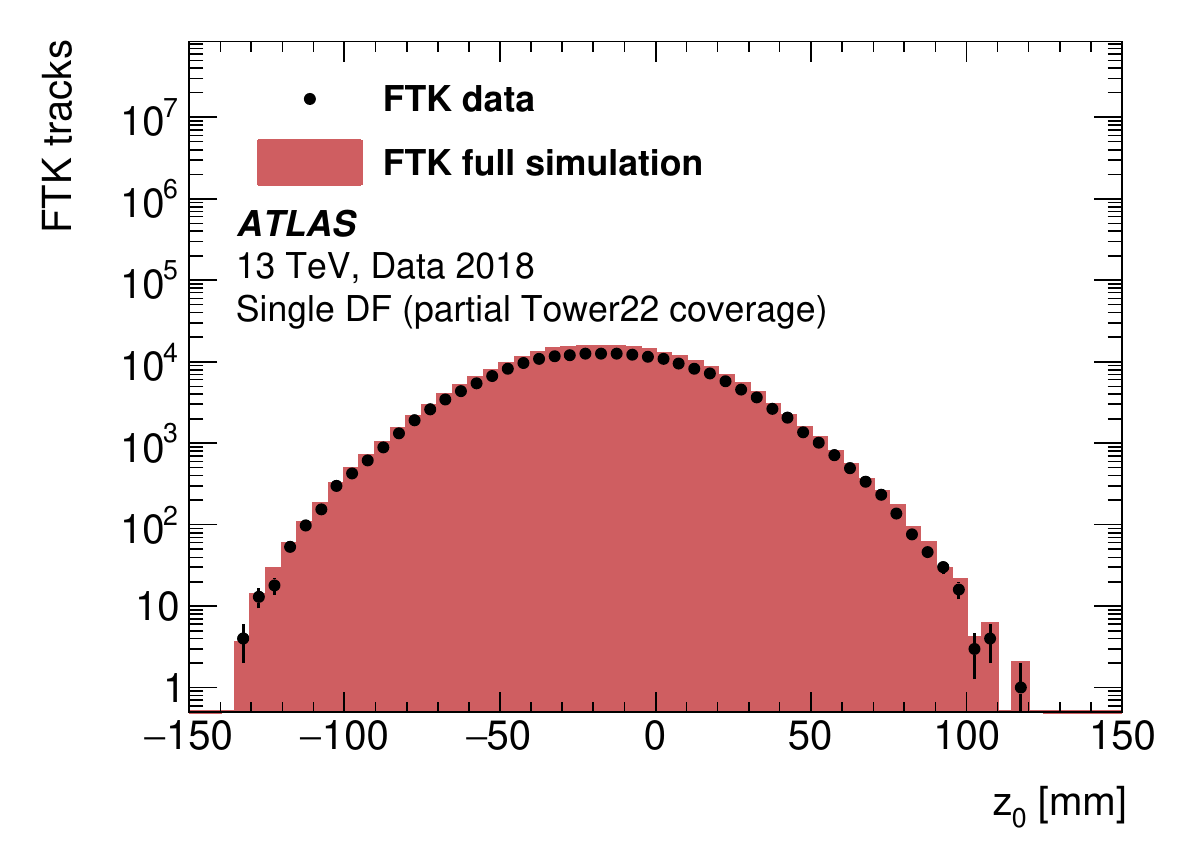}
\caption{The distributions of the five track parameters in FTK data (solid black circles) and FTK simulation (red histogram) run over the same ATLAS data.
\label{fig::kinematics}
}
\end{figure}

Events from an ATLAS run in September of 2018 with an average pile-up of 49 were input into the FTK vertical slice via the \ac{QNP} system.
A total of 192\,000 $pp$ collisions were selected for offline analysis. 
The same ATLAS data were also processed with the FTK simulation in order to compare the output of the hardware with the expectation from simulation.
The mean number of FTK tracks found by the hardware per event in the tower is 1.05, compared to an expectation of 1.25 FTK tracks per event from simulation.
The distributions of the five track parameters in FTK data vs FTK simulation are compared in Figure~\ref{fig::kinematics}.
The shapes of the distributions are well-modelled by the simulation.
The small differences seen in these distributions, and in later figures, between \ac{FTK} data and simulation are likely due to truncations in the hardware (e.g.\ the number of hits considered per \ac{SS}) that are not fully replicated in the software. 
 
\begin{figure}[htb]
\centering
\includegraphics[width=0.45\textwidth]{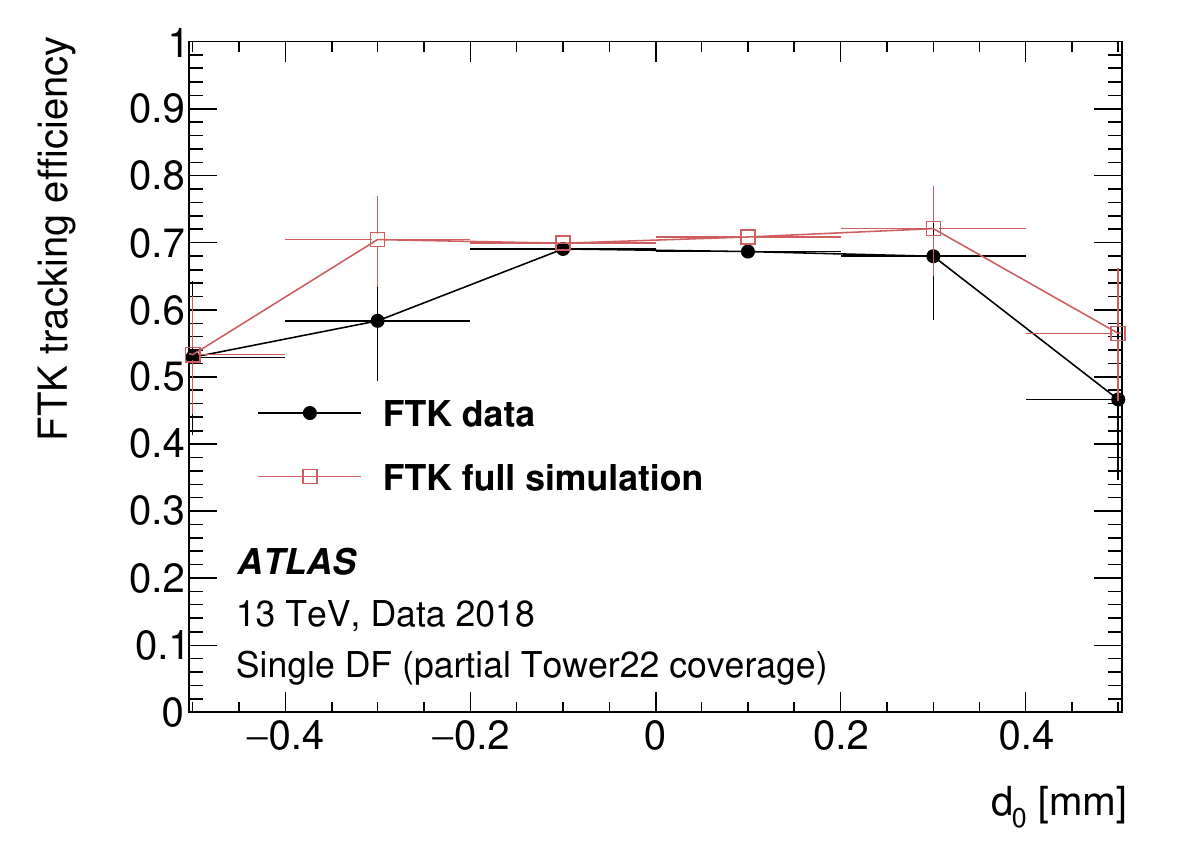}
\includegraphics[width=0.45\textwidth]{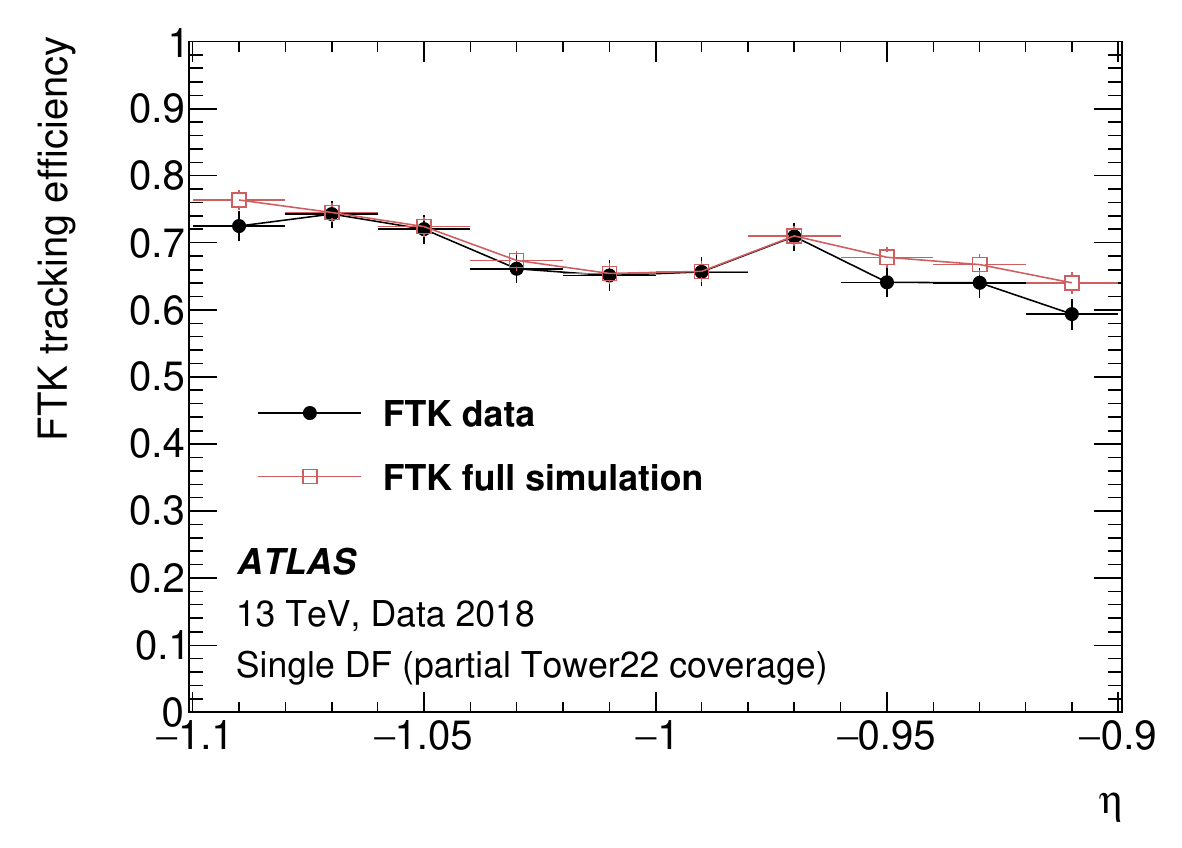}
\includegraphics[width=0.45\textwidth]{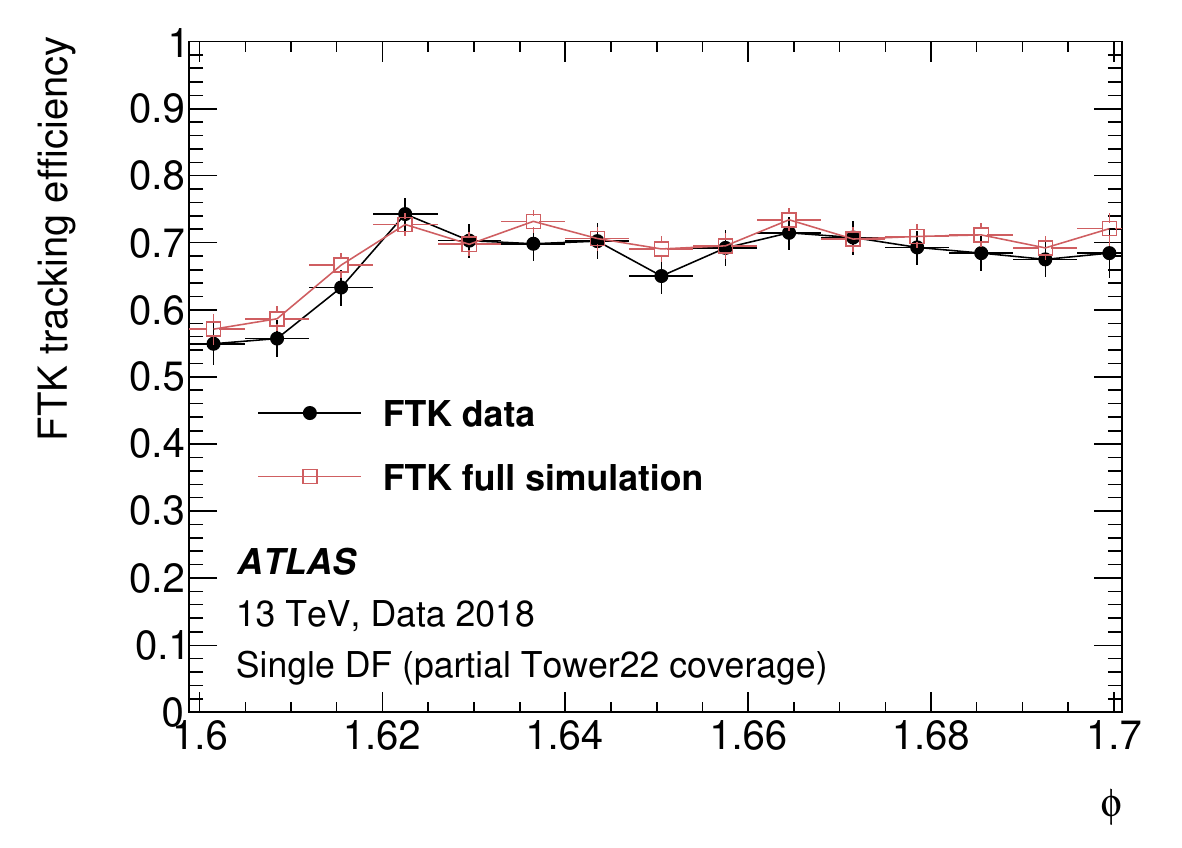}
\includegraphics[width=0.45\textwidth]{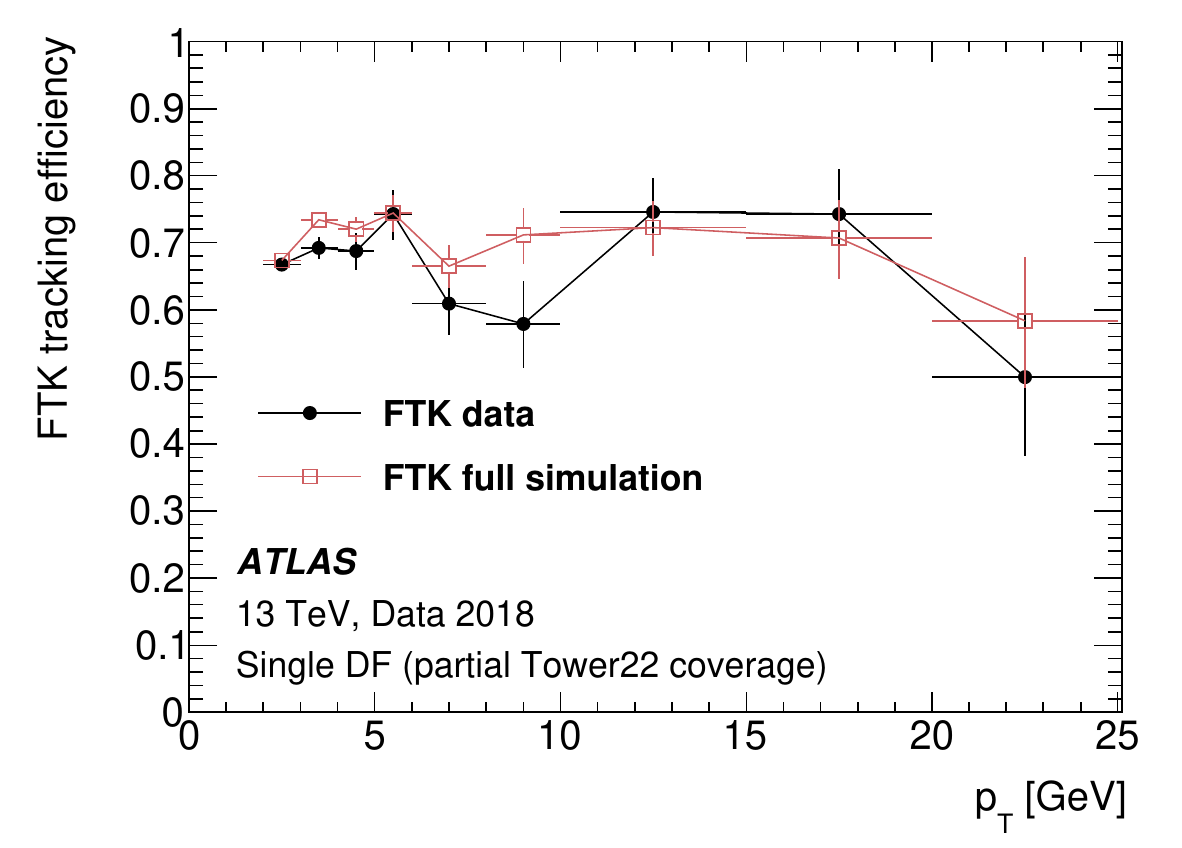}
\includegraphics[width=0.45\textwidth]{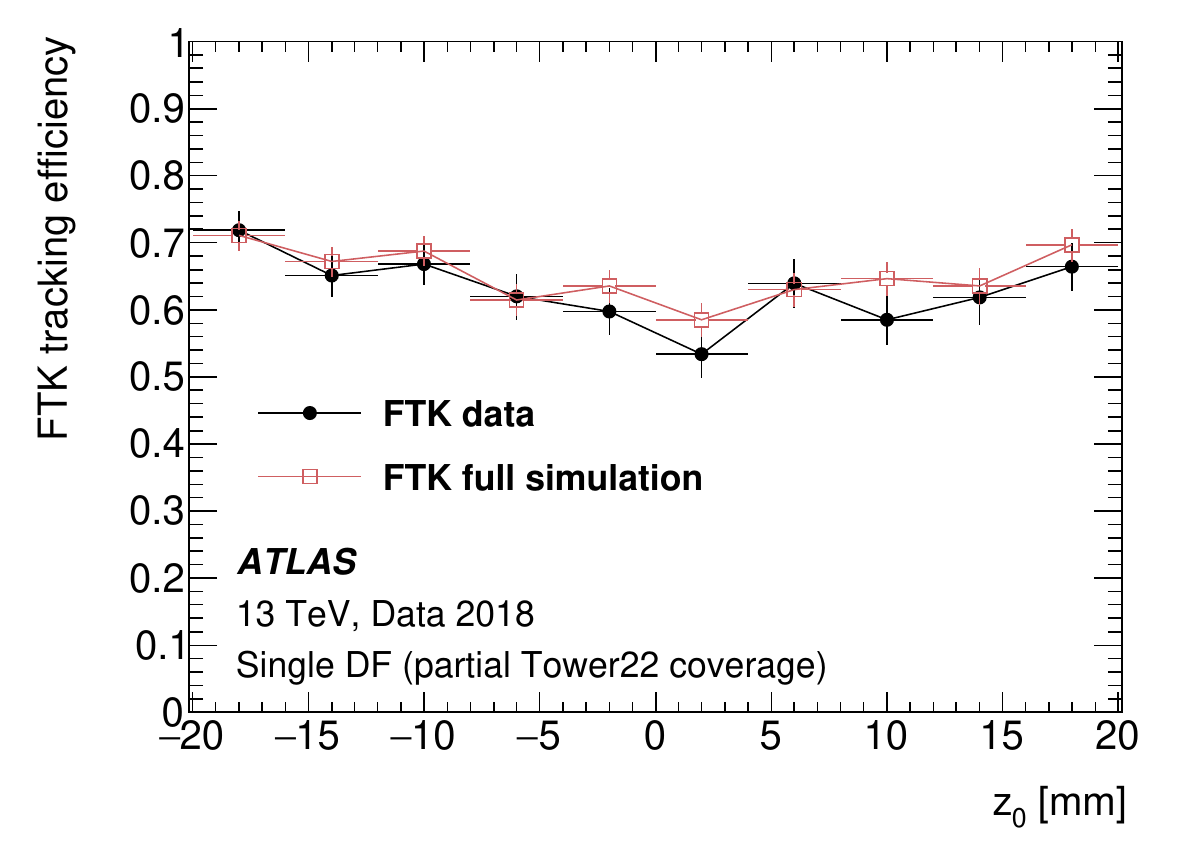}
\caption{The FTK tracking efficiency relative to offline tracks, as a function of the five track parameters,
in FTK data (solid black circles) and FTK simulation (open red squares). An offline track
is considered to be matched to FTK tracks with $\Delta R < 0.02$.
Only tracks in the core Tower 22 region defined by $-1.1 < \eta < -0.9$ and $1.6 < \phi < 1.7$ are considered here.
The error bars show the statistical uncertainties for the data and simulation, which are correlated with each other because the same input data are used.
\label{fig::sliceEfficiency}
}
\end{figure}

The FTK tracking efficiency relative to offline tracking is determined by evaluating the probability for each offline track with at least four (six) \ac{pixel} (\ac{SCT}) hits
to have a matched FTK track within an angular distance $\Delta R < 0.02$, for offline tracks with $\pt>2$~\GeV\ in the core of the Tower 22 region defined by $-1.1 < \eta < -0.9$ and $1.6 < \phi < 1.7$.
In this region, the average FTK tracking efficiency relative to offline tracks is 67\%,
compared to an expectation of 69\% from FTK simulation.
Due to the incomplete
geometric coverage and lack of data sharing of the FTK vertical slice, the FTK tracking efficiency is not expected to approach the values given in Table~\ref{tab::128pu}.
The tracking efficiency as a function of the five track parameters is presented in Figure~\ref{fig::sliceEfficiency}.
 
The fraction of FTK tracks matched to offline tracks within $\Delta R < 0.02$ is 98\% in FTK data
compared to an expectation of 99\% in FTK simulation, demonstrating that the FTK is not reconstructing large rates of spurious tracks due to random hit combinations.
 
\begin{figure}[htb]
\centering
\includegraphics[width=0.45\textwidth]{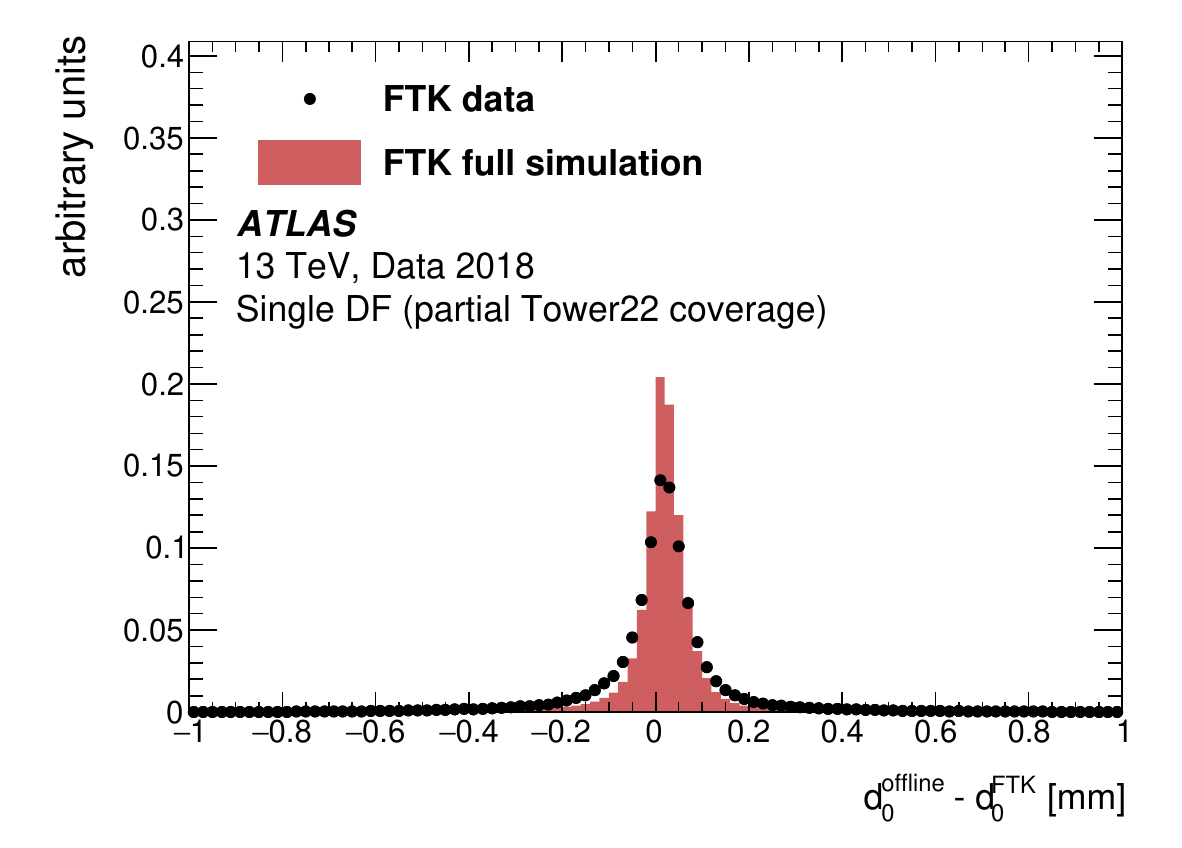}
\includegraphics[width=0.45\textwidth]{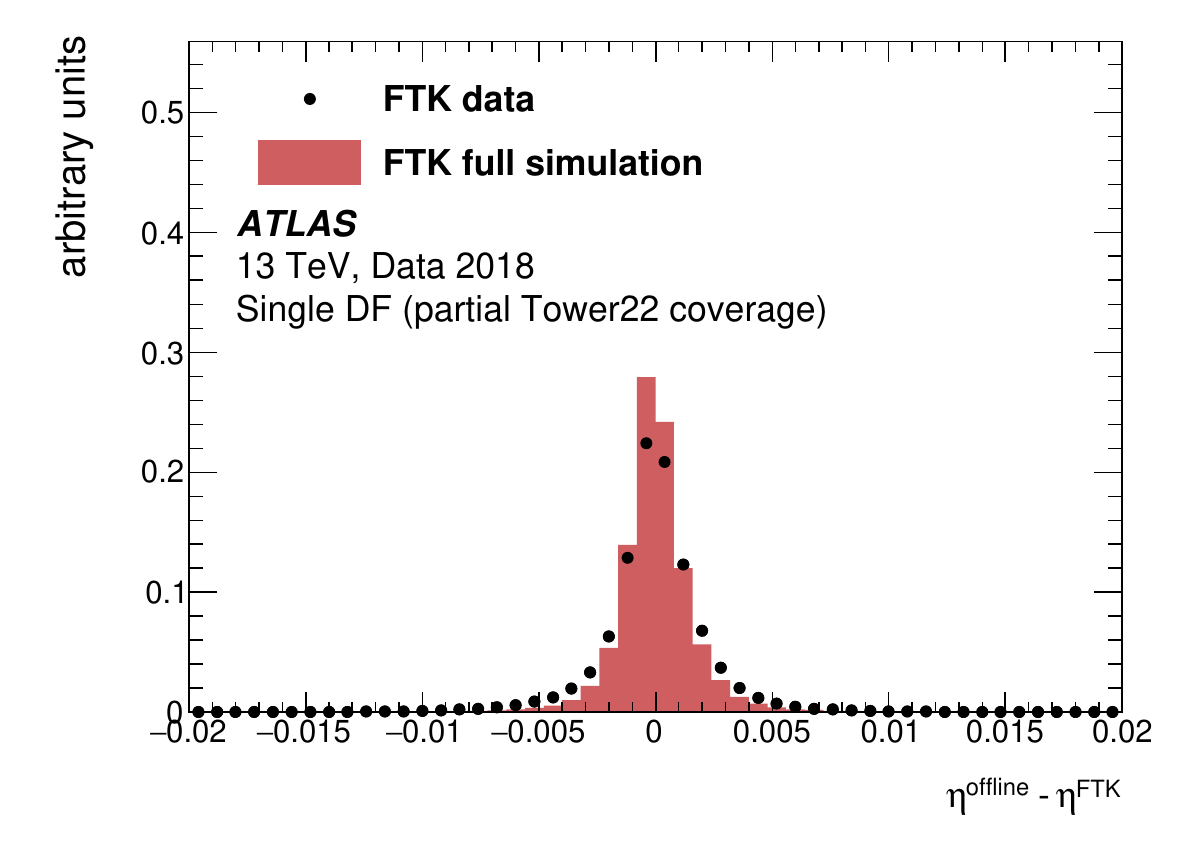}
\includegraphics[width=0.45\textwidth]{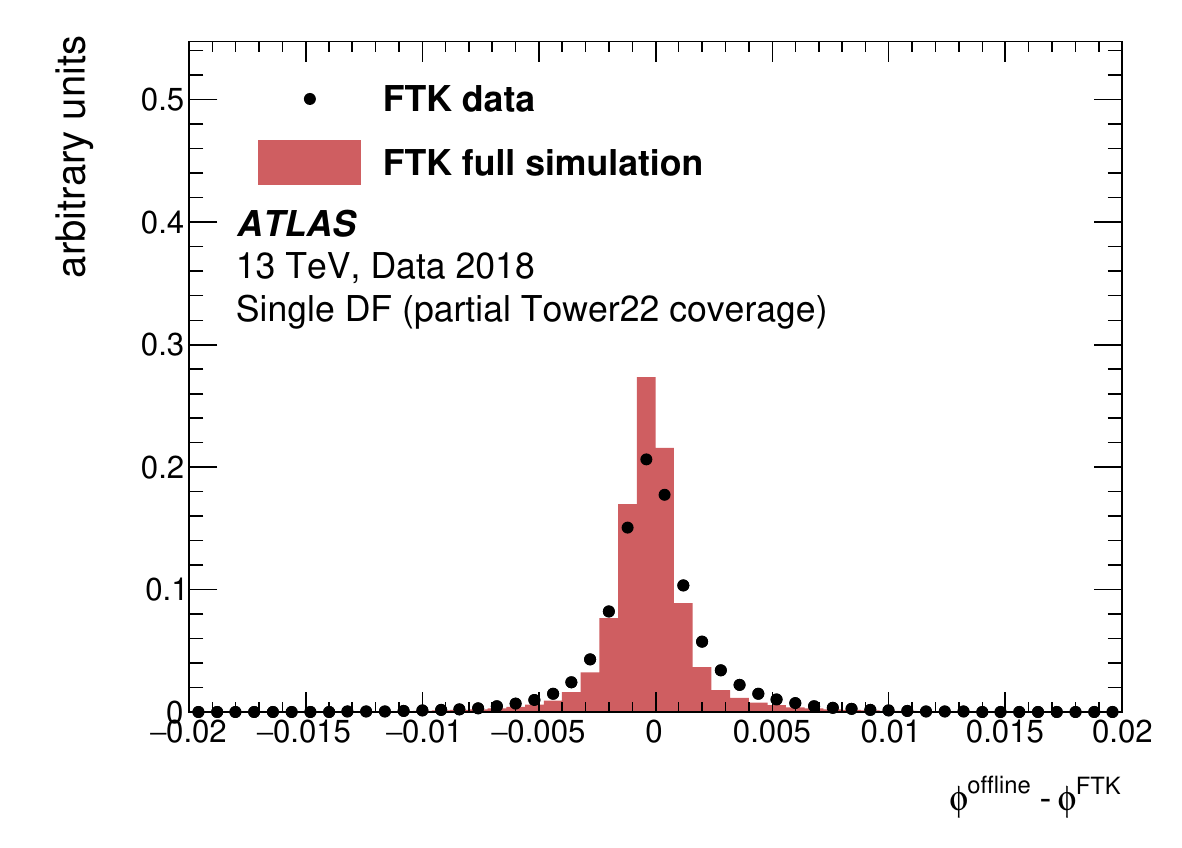}
\includegraphics[width=0.45\textwidth]{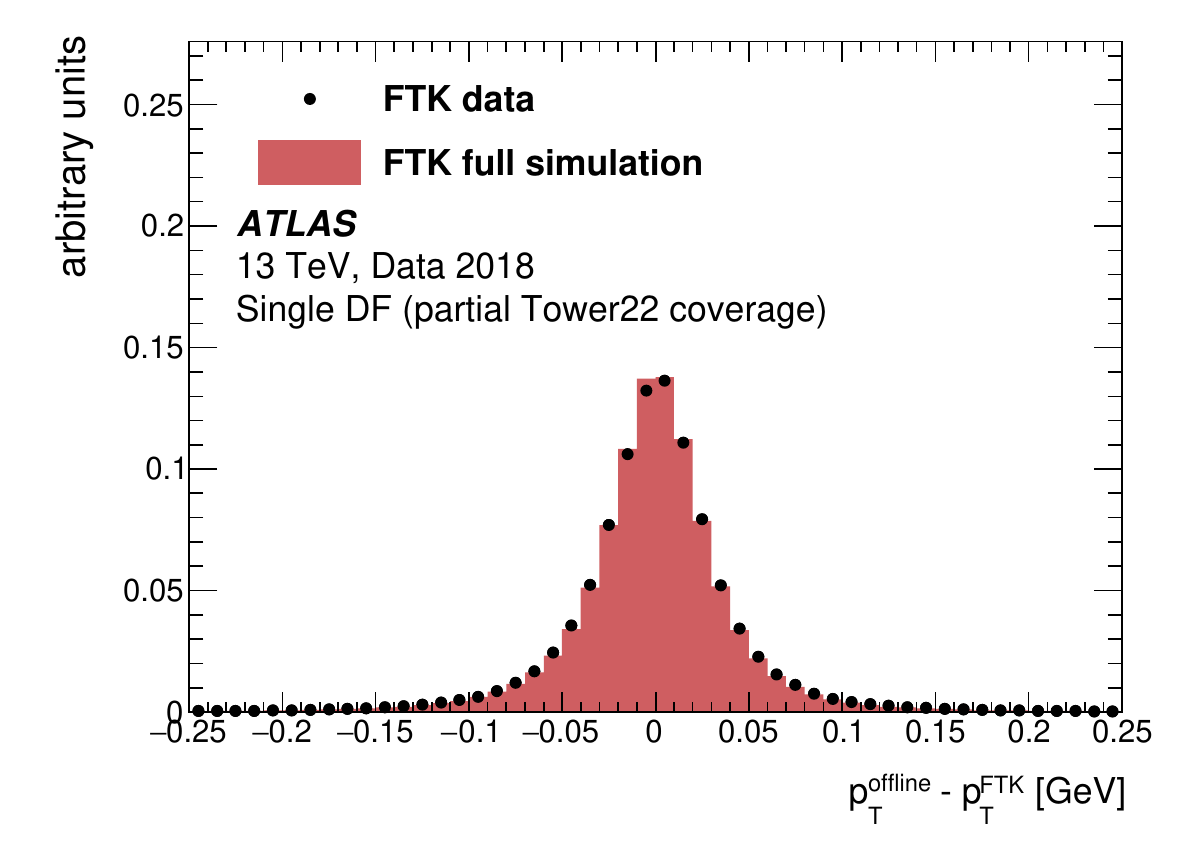}
\includegraphics[width=0.45\textwidth]{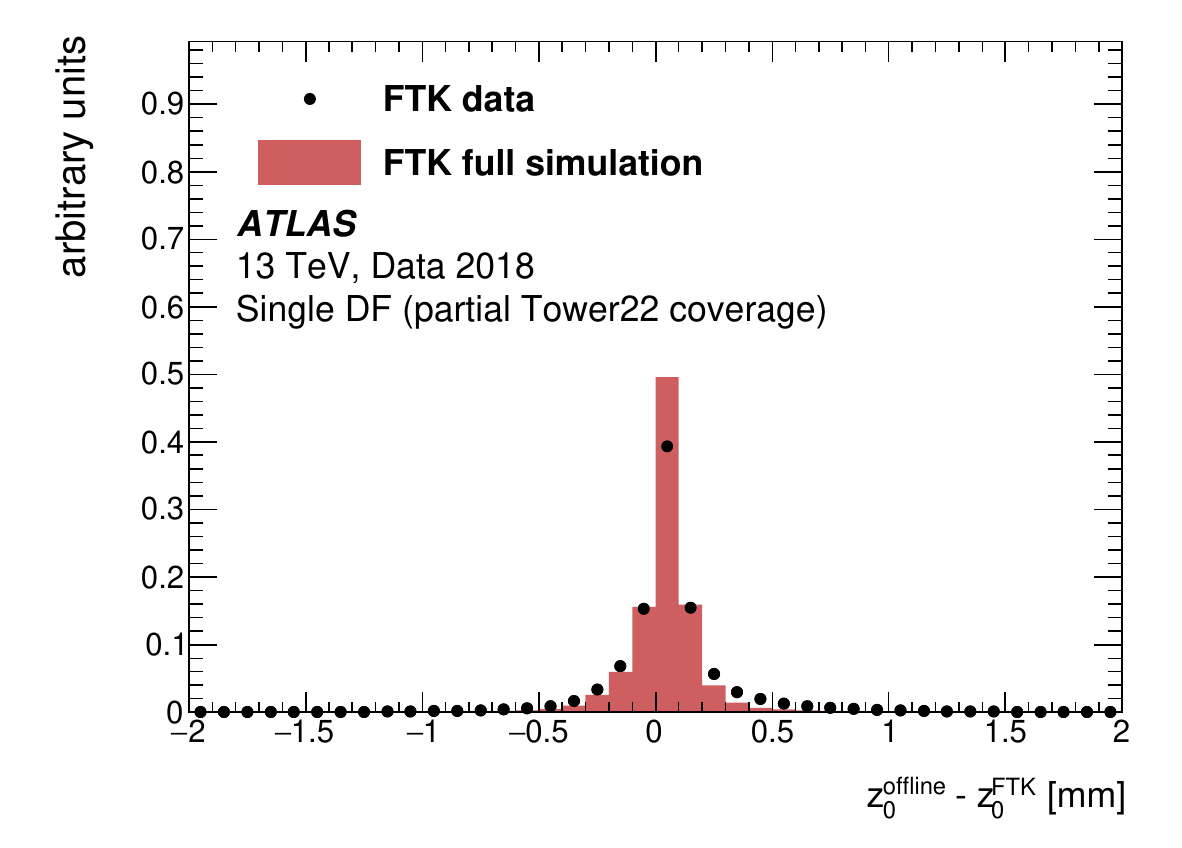}
\caption{The FTK track parameter residuals relative to offline tracks, as a function of the five track parameters,
in FTK data (black dots) and FTK simulation (red histogram). For each quantity, the difference between the
FTK and offline track parameter values is shown.
\label{fig::sliceResidual}
}
\end{figure}

The FTK track parameter residuals relative to offline tracks are shown in Figure~\ref{fig::sliceResidual}, separately for FTK vertical slice data and FTK simulation.
The $p_\text{T}$ residual in data agrees well with expectations from simulation, while the other track parameters are slightly worse in data than in simulation.

\clearpage


\section{Conclusion}
\label{sec:conclusion}
The FTK system was designed to provide tracks to the \ac{HLT} using custom hardware composed of \ac{AM} chips and \acp{FPGA} in order to improve the ATLAS HLT trigger performance.
The FTK system, the supporting software and a data playback facility are presented.
These systems were used for commissioning at the end of Run 2 data-taking and during LS2, before the project was stopped in 2019.
The pattern recognition and track-fitting performance of the FTK system has been demonstrated with a functional simulation of the system and in a vertical slice of the FTK.
The vertical slice shows that the as-built performance of the FTK system components and the associated firmware is capable of reproducing the performance expected from the simulation, with only very small differences.
Integration of the vertical slice into the ATLAS run control and regular inclusion in data-taking during 2018 demonstrate the functionality of the software, control, and monitoring.
Simulation studies of the pattern banks designed to be used in the FTK system show that a track-finding efficiency of 94.4\% for muons above 1~\GeV\ without additional pile-up could be reached for a system with the full complement of 128 PUs, as was foreseen for Run\,3 of the LHC.
For the FTK configuration with 64 PUs, the simulation shows an efficiency of 90.0\%.
Simulation studies with 30\% of the FTK patterns dedicated to high-$p_\mathrm{T}$ tracks with large impact parameters ($d_0$) show that the efficiency for tracks with $|d_0| < 10$\,mm could be significantly increased at a cost of 2\% in the overall efficiency.

\section*{Acknowledgements}
 

We thank CERN for the very successful operation of the LHC, as well as the
support staff from our institutions without whom ATLAS could not be
operated efficiently.
 
We acknowledge the support of ANPCyT, Argentina; YerPhI, Armenia; ARC, Australia; BMWFW and FWF, Austria; ANAS, Azerbaijan; SSTC, Belarus; CNPq and FAPESP, Brazil; NSERC, NRC and CFI, Canada; CERN; ANID, Chile; CAS, MOST and NSFC, China; COLCIENCIAS, Colombia; MSMT CR, MPO CR and VSC CR, Czech Republic; DNRF and DNSRC, Denmark; IN2P3-CNRS and CEA-DRF/IRFU, France; SRNSFG, Georgia; BMBF, HGF and MPG, Germany; GSRT, Greece; RGC and Hong Kong SAR, China; ISF and Benoziyo Center, Israel; INFN, Italy; MEXT and JSPS, Japan; CNRST, Morocco; NWO, Netherlands; RCN, Norway; MNiSW and NCN, Poland; FCT, Portugal; MNE/IFA, Romania; JINR; MES of Russia and NRC KI, Russian Federation; MESTD, Serbia; MSSR, Slovakia; ARRS and MIZ\v{S}, Slovenia; DST/NRF, South Africa; MICINN, Spain; SRC and Wallenberg Foundation, Sweden; SERI, SNSF and Cantons of Bern and Geneva, Switzerland; MOST, Taiwan; TAEK, Turkey; STFC, United Kingdom; DOE and NSF, United States of America. In addition, individual groups and members have received support from BCKDF, CANARIE, Compute Canada, CRC and IVADO, Canada; Beijing Municipal Science \& Technology Commission, China; COST, ERC, ERDF, Horizon 2020 and Marie Sk{\l}odowska-Curie Actions, European Union; Investissements d'Avenir Labex, Investissements d'Avenir Idex and ANR, France; DFG and AvH Foundation, Germany; Herakleitos, Thales and Aristeia programmes co-financed by EU-ESF and the Greek NSRF, Greece; BSF-NSF and GIF, Israel; La Caixa Banking Foundation, CERCA Programme Generalitat de Catalunya and PROMETEO and GenT Programmes Generalitat Valenciana, Spain; G\"{o}ran Gustafssons Stiftelse, Sweden; The Royal Society and Leverhulme Trust, United Kingdom.
 
The crucial computing support from all WLCG partners is acknowledged gratefully, in particular from CERN, the ATLAS Tier-1 facilities at TRIUMF (Canada), NDGF (Denmark, Norway, Sweden), CC-IN2P3 (France), KIT/GridKA (Germany), INFN-CNAF (Italy), NL-T1 (Netherlands), PIC (Spain), ASGC (Taiwan), RAL (UK) and BNL (USA), the Tier-2 facilities worldwide and large non-WLCG resource providers. Major contributors of computing resources are listed in Ref.~\cite{ATL-SOFT-PUB-2020-001}.
 

\printbibliography

\clearpage 
 
\begin{flushleft}
\hypersetup{urlcolor=black}
{\Large The ATLAS Collaboration}

\bigskip

\AtlasOrcid[0000-0002-6665-4934]{G.~Aad}$^\textrm{\scriptsize 101}$,    
\AtlasOrcid[0000-0002-5888-2734]{B.~Abbott}$^\textrm{\scriptsize 127}$,    
\AtlasOrcid[0000-0002-7248-3203]{D.C.~Abbott}$^\textrm{\scriptsize 102}$,    
\AtlasOrcid[0000-0002-2788-3822]{A.~Abed~Abud}$^\textrm{\scriptsize 36}$,    
\AtlasOrcid[0000-0002-1002-1652]{K.~Abeling}$^\textrm{\scriptsize 53}$,    
\AtlasOrcid[0000-0002-2987-4006]{D.K.~Abhayasinghe}$^\textrm{\scriptsize 93}$,    
\AtlasOrcid[0000-0002-8496-9294]{S.H.~Abidi}$^\textrm{\scriptsize 29}$,    
\AtlasOrcid[0000-0002-8279-9324]{O.S.~AbouZeid}$^\textrm{\scriptsize 40}$,    
\AtlasOrcid{N.L.~Abraham}$^\textrm{\scriptsize 156}$,    
\AtlasOrcid[0000-0001-5329-6640]{H.~Abramowicz}$^\textrm{\scriptsize 161}$
% ...
\bigskip
\\

% ...
$^{29}$Physics Department, Brookhaven National Laboratory, Upton NY; United States of America.\\
% ...
$^{36}$CERN, Geneva; Switzerland.\\
% ...
$^{40}$Niels Bohr Institute, University of Copenhagen, Copenhagen; Denmark.\\
% ...
$^{53}$II. Physikalisches Institut, Georg-August-Universit\"{a}t G\"ottingen, G\"ottingen; Germany.\\
% ...
$^{93}$Department of Physics, Royal Holloway University of London, Egham; United Kingdom.\\
% ...
$^{101}$CPPM, Aix-Marseille Universit\'e, CNRS/IN2P3, Marseille; France.\\
$^{102}$Department of Physics, University of Massachusetts, Amherst MA; United States of America.\\
% ...
$^{127}$Homer L. Dodge Department of Physics and Astronomy, University of Oklahoma, Norman OK; United States of America.\\
% ...
$^{156}$Department of Physics and Astronomy, University of Sussex, Brighton; United Kingdom.\\
% ...
$^{161}$Raymond and Beverly Sackler School of Physics and Astronomy, Tel Aviv University, Tel Aviv; Israel.\\
% ...

\end{flushleft}


\end{document}